\documentclass[11pt]{article}

\usepackage{mathrsfs}
\usepackage{dsfont}
\usepackage{amssymb}
\usepackage{float}
\usepackage{enumitem}
\usepackage{tikz}
\usepackage{chet}

\usetikzlibrary{decorations.markings,arrows,calc}

\tikzset{>=latex,
         ->/.style={decoration={markings,mark=at position 1 with
           {\arrow[scale=1.5]{>}}}, postaction={decorate}}}
\tikzset{->-/.style={decoration={markings,mark=at position 0.5 with
           {\arrow[scale=1.5]{>}}},postaction={decorate}}}
\tikzset{-<-/.style={decoration={markings,mark=at position 0.5 with
           {\arrow[scale=1.5]{<}}},postaction={decorate}}}
\tikzset{cross/.style={path picture={
      \draw[black]
            (path picture bounding box.south east) --
            (path picture bounding box.north west)
            (path picture bounding box.south west) --
            (path picture bounding box.north east);}}}

\def\Fermilab{Theoretical Physics Department, Fermilab, P.O. Box 500, Batavia, IL 60510, USA}
\def\ChicagoEFI{Enrico Fermi Institute, University of Chicago, Chicago, IL, 60637, USA}
\def\ChicagoKICP{Kavli Institute for Cosmological Physics, University of Chicago, Chicago, IL, 60637, USA}
\def\Barcelona{Institut de F\'isica d'Altes Energies (IFAE), The Barcelona Institute of Science and Technology (BIST),
Campus UAB, 08193 Bellaterra (Barcelona) Spain}
\def\Northwestern{Department of Physics and Astronomy, Northwestern University, Evanston, IL 60208, USA}

\preprint{FERMILAB-PUB-19-395-T}
\title{\vspace{-1.5cm}\LARGE Dark CP Violation and Gauged Lepton/Baryon Number for Electroweak Baryogenesis}
\author{{\large Marcela Carena$^{1,2,3}$, Mariano Quir\'os$^{4}$, Yue Zhang$^{1,5}$}}
\affiliation{
$^1$\Fermilab\\
$^2$\ChicagoEFI\\
$^3$\ChicagoKICP\\
$^4$\Barcelona\\
$^5$\Northwestern}

\abstract{
We explore the generation of the baryon asymmetry in an extension of the Standard Model where the lepton number is promoted to a $U(1)_\ell$ gauge symmetry with an associated  $Z^\prime$ gauge boson. This is based on a novel electroweak baryogenesis mechanism first proposed by us in Ref.~\cite{Carena:2018cjh}. Extra fermionic degrees of freedom - including a fermionic dark matter $\chi$ - are introduced in the dark sector for anomaly cancellation.  Lepton number is spontaneously broken at high scale and the effective theory, containing the Standard Model, the $Z^\prime$, the fermionic dark matter, and an additional complex scalar field $S$, violates CP in the dark sector. The complex scalar field couples to the Higgs portal and is essential in enabling  a strong first order phase transition. Dark CP violation is diffused in front of the bubble walls and creates a chiral asymmetry for $\chi$, which in turn creates a chemical potential for the Standard Model leptons. Weak sphalerons are then in charge of transforming the net lepton charge asymmetry into net baryon number. We explore the model phenomenology related to the leptophilic $Z^\prime$, the dark matter candidate, the Higgs boson and the additional scalar, as well as implications for electric dipole moments. We also discuss the case when baryon number $U(1)_B$ is promoted to a gauge symmetry, and discuss electroweak baryogenesis and its corresponding  phenomenology.}

\begin{document}

\maketitle

\tableofcontents

\section{Introduction}

The origin of the cosmic baryon asymmetry is a fascinating mystery for particle physics and cosmology. Electroweak baryogenesis (EWBG)~\cite{Kuzmin:1985mm, Cohen:1990py, Farrar:1993sp, Huet:1995mm, Huet:1995sh, Riotto:1995hh, Carena:1997gx, Carena:2000id, Cline:2000nw, Carena:2002ss, Lee:2004we, Cline:2006ts} is an elegant possibility and predicts new physics beyond the Standard Model (SM), near the electroweak scale, to trigger a sufficiently strong first-order electroweak phase transition (EWPT) and  source enough CP violation. If the new particles responsible for CP violation are charged under the SM, they will also contribute to the electric dipole moments (EDM) at low energies~\cite{Fromme:2006cm, Cirigliano:2006dg, Li:2010ax}. Models which belong to this class, including two Higgs-doublet models and supersymmetric models, are progressively receiving stronger and stronger constraints from the improved EDM measurements in recent years~\cite{Andreev:2018ayy, Baron:2013eja,Griffith:2009zz,Baker:2006ts}, specially after the discovery of the Higgs boson~\cite{Shu:2013uua, Ipek:2013iba, Jung:2013hka, Abe:2013qla, Inoue:2014nva, Cheung:2014oaa, Bian:2014zka, Chen:2015gaa, Fuyuto:2015ida, Jiang:2015cwa, Blinov:2015sna, Balazs:2016yvi, Bian:2016zba, Chen:2017com, Cesarotti:2018huy, Egana-Ugrinovic:2018fpy, Ramsey-Musolf:2017tgh}. This gives a strong motivation to study EWBG in models with a dark sector where SM gauge singlet particles source the required CP violation. 
\textit{The main challenge of such realizations is to find an efficient mechanism to transfer the CP violation from the dark sector to the visible sector in the early universe,
while still keeping contributions to EDMs sufficiently suppressed today.} 

To this respect, an interesting scenario of dark sector CP violation was presented in Ref.~\cite{Cline:2017qpe}, where a Yukawa interaction between a dark fermion and the SM fermion doublets is responsible for communicating CP violation into the visible sector. Such a realization, however, leads to two-loop level contributions to EDMs. In turn, suppressing such contributions to the EDMs requires a finely-tuned restoration of a global symmetry after the EWPT.

\begin{figure}[h]
\centerline{\includegraphics[width=0.65\textwidth]{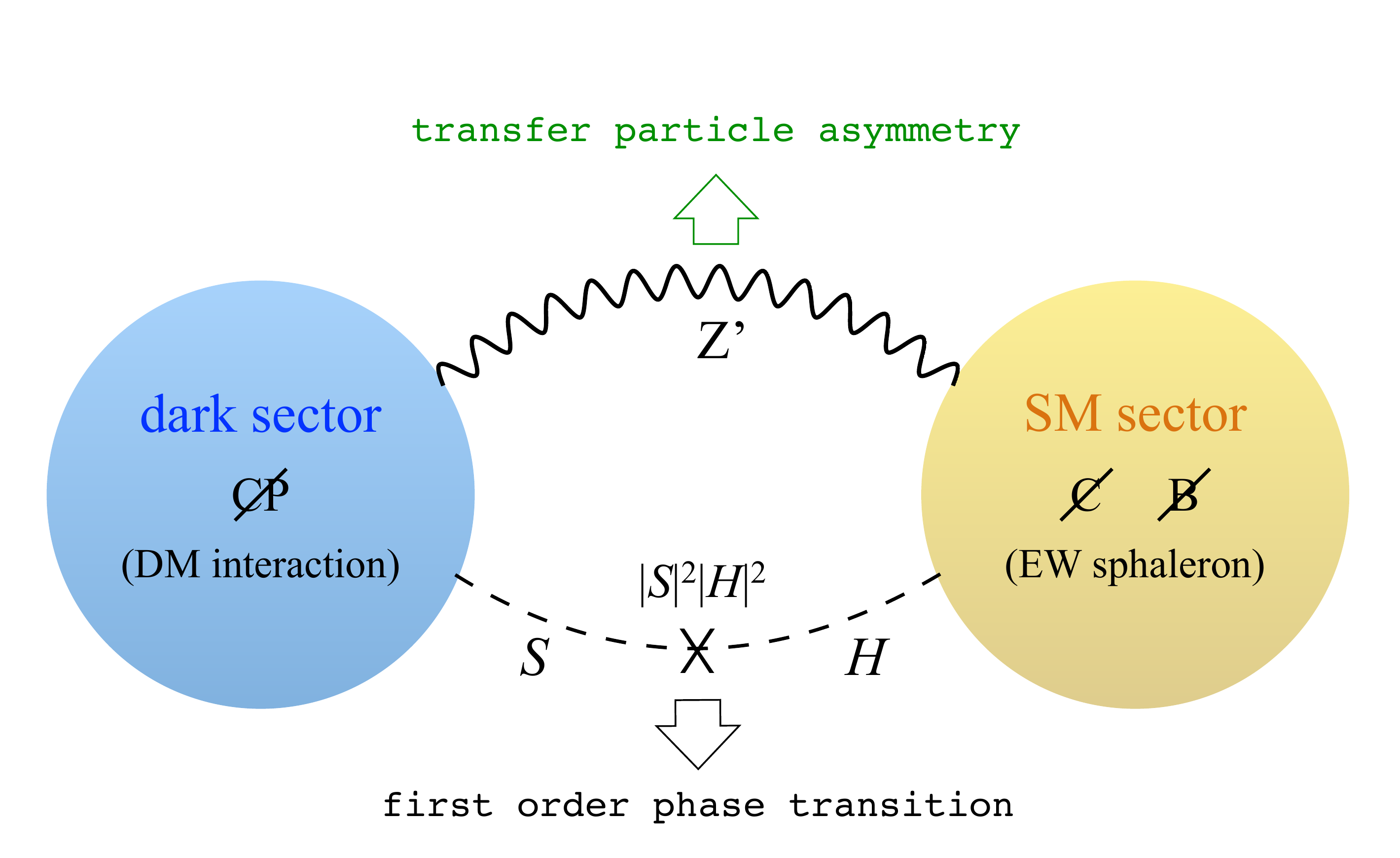}}
\caption{\it A schematic picture showing our model setup and the role played by each part in our proposed EWBG mechanism.}\label{SchematicPlot}
\end{figure}

In a recent short article~\cite{Carena:2018cjh}, we presented the basic idea of a new EWBG mechanism in which the role of messenger of the CP asymmetry can be played by a $Z^\prime$ gauge boson that couples to both the SM and the dark sector. The low-energy effective theory is a dark sector model containing a Dirac fermion $\chi$ (charged under the $Z^\prime$) with a CP violating coupling to a complex scalar field $S$. During a first-order phase transition, in the electroweak and the dark sectors involving both the Higgs field and the scalar $S$, a chiral-charge asymmetry in $\chi$ particles is first created. Through the time-like component of the $Z^\prime$ background (which is CP odd, and also CPT odd), the $\chi$ asymmetry leads to a chemical potential for all SM leptons. If the $Z^\prime$ is sufficiently light, it mediates a long range force that extends into the region outside the bubble wall with unbroken electroweak symmetry. This chemical potential then biases the sphaleron processes and generates a net baryon asymmetry inside the bubbles. After the EWPT is completed, the $Z^\prime$ background relaxes to zero and the dark CP violation becomes secluded from the SM sector. A schematic plot of this setup is shown in Fig.~\ref{SchematicPlot}.

There are several distinct features of this model.
\begin{itemize}

\item The $Z^\prime$ gauge boson needs to be light, not much heavier than the electroweak scale, and not too weakly coupled to the SM leptons, for generating sufficient baryon asymmetry. Therefore, the existence of a light leptophilic $Z'$ serves as a smoking-gun of the proposed EWBG mechanism, and provides a well-motivated target for various experimental searches, as we shall discuss below. 

\item Given that the CP violating interactions in the dark sector only involve SM gauge singlets, it follows that, in the absence of any Yukawa couplings involving both SM and the dark sector particles, the two-loop Barr-Zee type contributions to EDM~\cite{Barr:1990vd, Gunion:1990ce} are forbidden. Indeed, we will show that in this framework, the leading contribution to EDMs must appear at least at the three-loop level, which is much less constrained by current EDM results. This point is diagrammatically illustrated in Fig.~\ref{fig:noBZ}.

\begin{minipage}{\linewidth}
\begin{figure}[H]
\centerline{\includegraphics[width=0.75\textwidth]{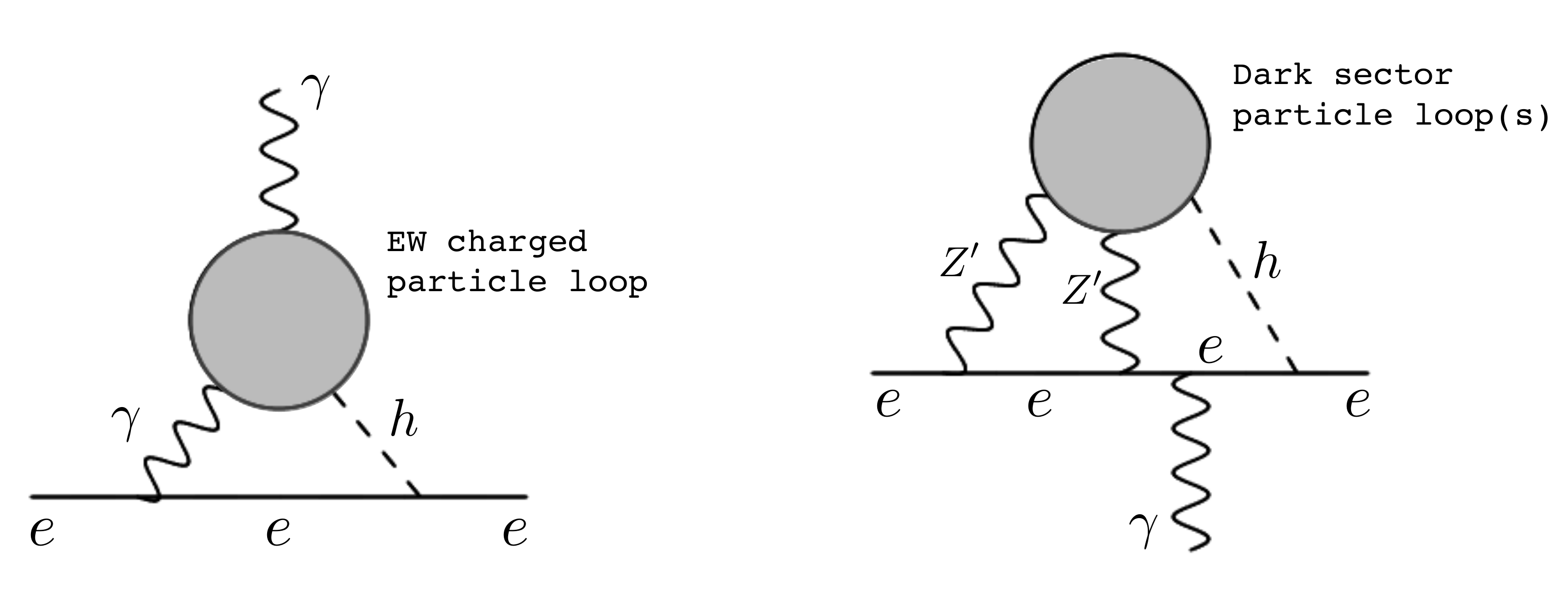}}
\caption{\it Representative diagrams showing the loop generated electron EDM in two classes of models, where CP violation occurs through the interactions from electroweak charged particles (left panel) or SM gauge singlets that couple to the $Z'$ (right panel). The gray blobs represent the loop generated $h F_{\mu\nu}F^{\mu\nu}$ and $h Z'_{\mu\nu}Z'^{\mu\nu}$ effective vertices in the two cases, respectively. In the former case, the contribution to EDMs can occur at two-loop level via the Barr-Zee type diagrams.
In the latter case, the contribution to EDMs must arise at more than two-loop level.}\label{fig:noBZ}
\end{figure}
\end{minipage}

\item The particle $\chi$ in this model could serve as the dark matter candidate, as we will show in detail in this paper.

\item The simple model we have just discussed can be embedded in an ultraviolet (UV) complete theory with gauged lepton number $U(1)_\ell$, whose gauge boson is $Z^\prime_\mu$, for the two interesting benchmark cases, where $\ell=L_e+L_\mu+L_\tau$ and $\ell=L_\mu+L_\tau$, which require the introduction of extra fermion fields (anomalons) to cancel the gauge anomalies. Below the spontaneous lepton number breaking scale, when part of the anomalon fields have been integrated out, the low energy effective theory is composed of the SM and a secluded dark sector. The two sectors are connected through the $Z^\prime$, which will transfer the CP asymmetry to the observable sector, and the Higgs portal interaction, responsible for inducing a first-order electroweak phase transition.

\end{itemize}

It is worth noticing that, for our EWBG mechanism to work, the vector current that couples to  the $Z'$ boson in the effective theory must be anomalous with respect to the 
SM $SU(2)_L$ gauge symmetry at the time of the EWPT. This is achieved by (Boltzmann) decoupling the heavy anomalons from the thermal plasma, such that only the SM fields are kept populated at the critical temperature of the EWPT~\footnote{In other words, while heavy anomalons protect the gauge theory at zero temperature from gauge anomalies, through the remaining Wess-Zumino terms~\cite{Wess:1971yu}, their abundance is Boltzmann suppressed at finite temperature so that they decouple from the thermal bath.}. The effect of the anomalous current is to generate a non-vanishing chemical potential, that triggers the electroweak sphaleron processes to create a net baryon asymmetry. The above observation implies that our mechanism will not work, for example, if the $Z'$ is the gauge boson of the $U(1)_{B-L}$ symmetry (anomaly free in the presence of right-handed neutrinos), the hypercharge $U(1)_Y$, or linear combinations thereof. The $U(1)_\ell$ lepton number symmetry we consider is anomaly free at high energy scales, but it becomes anomalous after the spontaneous breaking of the $U(1)_\ell$ gauge symmetry takes place and some of the new fermions - otherwise responsible for anomaly cancellation - are integrated out from the thermal plasma. The effective theory below the mass of the heavy anomalons is perfectly consistent, as gauge invariance is restored by the introduction of the Wess-Zumino terms~\cite{Wess:1971yu}. This is at the core of what makes our baryogenesis mechanism feasible. Similarly, our baryogenesis idea could also work for the gauged $U(1)_B$ baryon number symmetry, which is also known to be anomalous with respect to the SM.

The content of this paper is organized as follows. In Sec.~\ref{sec:UVcompletition}  we present our EWBG model, making explicit the structure of the extended dark fermion  and scalar sectors that interact with the SM particles through the $U(1)_\ell$ $Z^\prime$ gauge boson and the Higgs portal. In Sec.~\ref{EWZprimeBaryogenesis}, we discuss the necessary steps for the first order phase transition to occur, and the source of CP violation in the dark sector, as well as how the latter induces the actual mechanism of baryogenesis in the SM at the electroweak scale. In Secs.~\ref{pheno} and \ref{sec:G2} we concentrate on the phenomenological aspects of our model and its possible signatures in current and near future experiments, for the cases where $\ell=L_e+L_\mu+L_\tau$ and $\ell=L_\mu+L_\tau$, respectively. This includes the leptophilic $Z^\prime$ searches, dark matter $\chi$ direct detection searches, conditions for thermal freeze out, bounds from EDM's, and collider searches for dark scalar(s). We comment on the case of gauged $U(1)_B$ baryon number in Sec.~\ref{sec:GB}. We reserve Sec.~\ref{conclusions} for our conclusions and provide some details of the calculation of the lepton asymmetry in Apps.~\ref{AppendixA} and \ref{AppendixB}.

\section{A Model with Gauged Lepton Number}\label{sec:UVcompletition}

As the starting point, we consider an extension of the SM with gauged lepton number symmetry $U(1)_\ell$. Its gauge boson is called $Z^\prime$ and its gauge coupling $g^\prime$~\footnote{Not to be confused with the SM hypercharge $U(1)_Y$ gauge coupling, $g_Y$.}. There are various choices to define the lepton number, $\ell$. The most obvious choice is $\ell=L_e+L_\mu+L_\tau$ where all three lepton flavors are gauged universally. However, our baryogenesis mechanism will also work if only a reduced number of lepton flavors are gauged, {\it e.g.}~$\ell=L_\mu+L_\tau$. In the following discussion, we will keep the number of lepton flavors charged under the $U(1)_\ell$ as a free parameter, $N_g$, where $N_g=3\, (2)$ in the case $\ell=L_e+L_\mu+L_\tau\, (\ell=L_\mu+L_\tau)$.

Because the $U(1)_\ell$ symmetry in the SM is anomalous with respect to $SU(2)_L\times U(1)_Y$, additional fermions (so called anomalons) must be introduced for anomaly cancellation. A minimal set of new fermion content~\cite{FileviezPerez:2010gw, Duerr:2013dza, Schwaller:2013hqa} is given in Tab.~\ref{tab:table}, where $\mathtt{q}$ is an arbitrary real number. This is the UV complete framework we shall consider.

The right-handed neutrinos $\nu_R^{i},\ (i=1,\dots, N_g)$ could pair up with the active neutrinos $\nu_L^i$ in the SM, so that in this minimal setup the observed neutrino masses are Dirac~\footnote{The possibility of Majorana neutrinos will be considered in Sec.~\ref{sec:nuCosmo}.}. To pair up the other extra fermions and give them vector-like masses (with respect to the SM gauge symmetries), a complex scalar $\Phi$ is introduced carrying lepton number $N_g$. The vacuum expectation value (VEV) of $\Phi$, $v_\Phi$, spontaneously breaks the $U(1)_\ell$, giving mass to the $Z^\prime$ gauge boson, as
\begin{equation}\label{eq:Z'mass}
M_{Z'} = \sqrt{2} N_g g' v_\Phi \ ,
\end{equation}
and to the new fermions via the following Yukawa terms
\begin{equation}
\left( c_L \bar L''_R L'_L  + c_e \bar e''_L e'_R + c_\chi \bar \chi_L \chi_R \rule{0mm}{3.5mm}\right) \Phi + {\rm h.c.} \ .
\end{equation}
Hereafter, for simplicity, we will ignore the Yukawa couplings between lepton doublets and singlets with the Higgs boson (which would lead to subleading entries in the fermion mass matrix), as well as the potential Yukawa coupling between the SM leptons and some of the new leptons (only allowed for specific choices of $\mathtt{q}$, for example, $\mathtt{q}=1$), which also helps to suppress new sources of lepton flavor violation~\cite{Altmannshofer:2016oaq}.

\begin{table}[t]
\centering
\begin{tabular}{||c|c|c|c|c||}
\hline\hline
Particle & $SU(3)_c$ & $SU(2)_L$ & $U(1)_Y$ & $U(1)_\ell$ \\
\hline
$\nu_R^{i}$ & 1 & 1 & 0 & 1 \\
$L'_L=(\nu_L', e_L')^T$ & 1 & 2 & -1/2 & $\mathtt{q}$ \\
$e'_R$ & 1 & 1 & -1 & $\mathtt{q}$ \\
$\chi_R$ & 1 & 1 & 0 & $\mathtt{q}$ \\
$L''_R=(\nu_R'', e_R'')^T$ & 1 & 2 & -1/2 & $\mathtt{q}+N_g$ \\
$e''_L$ & 1 & 1 & -1 & $\mathtt{q}+N_g$ \\
$\chi_L$ & 1 & 1 & 0 & $\mathtt{q}+N_g$ \\
\hline\hline
\end{tabular}
\caption{\it Fermion content (anomalons), and its quantum numbers, in the anomaly free model with gauged $U(1)_\ell$ symmetry. $\mathtt{q}$ is a free (real) parameter.}
\label{tab:table}
\end{table}

Because $L'_L, L''_R, e'_R, e''_L$ contain fermions charged under the SM gauge group, which are constrained by the existing LEP and LHC searches, we will assume $v_\Phi$ to be well above the TeV scale and $c_L, c_e$ to be of order one, rendering these particles sufficiently heavy.  As a result, these particles could be integrated out at energy scales and temperatures of order of the $U(1)_\ell$ breaking scale.

For our baryogenesis mechanism to work, we will assume both the $g'$ and $c_\chi$ parameters to be small, so that the $Z'$ boson, as well as the $\chi_L, \chi_R$ fermions have masses around, or even below, the electroweak scale. In the forthcoming discussions, we will also show that $\chi$ qualifies to be the dark matter candidate.

After integrating out the $L'_L, L''_R, e'_R, e''_L$ fermions, which play a role in the anomaly cancellation mechanism, the $U(1)_\ell$ current involving only light degrees of freedom becomes anomalous at lower energy. As it is well known, integrating out the anomalon fields leads to the introduction of the Wess-Zumino (WZ) term~\cite{Wess:1971yu}, which is necessary for restoring the SM gauge invariance when calculating the triangle diagrams in the effective theory~\footnote{A manifestation of the non-decoupling properties of fields which acquire their masses only through a spontaneously breaking mechanism.}. However, the coefficient of the WZ term is not fixed but depends on the convention, {\it i.e.}~the momentum routings, and such convention needs to be respected when calculating the triangle diagrams~\cite{Bardeen:1984pm}. In particular, in the convention of ``covariant anomaly'', the coefficient of the WZ term vanishes~\cite{Rosenberg:1962pp}.  Observe, however, that in the baryogenesis mechanism discussed in this work all the relevant processes occur at tree level, and therefore issues of gauge invariance and appropriate loop momentum convention do not play a role, since they would only matter in one-loop processes involving the $Z'$ (see, {\it e.g.},~\cite{Fox:2018ldq}).

In addition to the above particle content, baryogenesis requires the presence of another complex (SM singlet) scalar $S$, which also carries lepton number $N_g$. We assume that $S$ is much lighter than $\Phi$, and its VEV $v_S$ evolves, together with that of the Higgs field, during the electroweak phase transition. In contrast, the VEV $v_\Phi$ of $\Phi$ remains constant as the universe evolves in the proximity of the electroweak phase transition, since at these scales the field $\Phi$ is decoupled. In the presence of the $S$ field, one can write down a Yukawa term that gives an additional mass to the fermion $\chi$.
It takes the form
\begin{eqnarray}\label{darkYukawa}
\bar \chi_L ( m_0 + \lambda_c S ) \chi_R + {\rm h.c.} \ ,
\end{eqnarray}
where the first term is given by $m_0=c_\chi v_\Phi$ and $\lambda_c$ is a (complex) Yukawa coupling. As a result, the mass of $\chi$ changes with the $S$ field profile during the electroweak phase transition, and, if the relative phase between $m_0$ and $\lambda_c S$ is physical, it will serve as a source of CP violation in our baryogenesis  mechanism.

To summarize, our assumptions lead to a low-energy effective theory below the $U(1)_\ell$ breaking scale ($v_\Phi$), which contains the SM fields plus the following new fields
\begin{equation}
\boxed{Z^\prime_\mu, \quad S, \quad \chi_L, \quad \chi_R \ }\ .
\end{equation}
Among them, $S$ and $ \chi_{L,R}$ are SM gauge singlets and belong to the dark sector. There  are two possible portals for them to interact with the SM sector. 

One way is through the leptonic $Z'$ portal,
\begin{eqnarray}\label{Z'portal}
\mathcal L\supset g' Z'_\mu \left[ (\mathtt{q}+N_g) \bar \chi_L \gamma^\mu \chi_L +  \mathtt{q}\, \bar \chi_R \gamma^\mu \chi_R + \bar L_L \gamma^\mu L_L + \bar \ell_R \gamma^\mu \ell_R \right] \ ,
\end{eqnarray}
where $L_L$ represents the SM left-handed lepton doublets and $\ell_R$ represents the SM right-handed charged leptons. Here, after integrating out the heavy $L'_L, L''_R, e'_R, e''_L$ fermion fields,  the $Z'$ couples to an anomalous current with respect to the SM gauge symmetries, in particular the $SU(2)_L$, which governs the lepton/baryon number violating sphaleron processes. This will be the key ingredient of our baryogenesis mechanism, which makes use of the $Z'$ field background, as we shall discuss in the following section. 

Another way, which is the other key ingredient in our baryogenesis mechanism, is the Higgs portal  interaction between $S$ and $H$, 
\begin{eqnarray}\label{Higgsportal}
\mathcal L=-\lambda_{SH} |S|^2 |H|^2 \ ,
\end{eqnarray}
that will be responsible for triggering a sufficiently strong-first order electroweak phase transition.

\section{Electroweak Baryogenesis Mediated by the $Z'$ Boson}\label{EWZprimeBaryogenesis}

In this section we will consider how the different ingredients play their roles for successful electroweak baryogenesis. We will discuss successively the out of equilibrium condition in the phase transition, the new source of CP violation (CPV), and the generation of the baryon asymmetry.

\subsection{\bf The Phase Transition(s)}\label{history}

We will consider a first-order electroweak phase transition during which the Higgs VEV turns on, while the VEV of the $S$ field varies at the same time. Such a scenario can be realized through the following steps in the history of our universe.

\begin{enumerate}

\item At very high temperatures, all symmetries are restored.

\item As the universe cools down to the temperature $T_\Phi\sim v_\Phi$, the $\Phi$ field acquires its VEV, $\langle\Phi\rangle=v_\Phi$, and the lepton number symmetry is broken. The nature of this phase transition is not relevant here, but the breaking of lepton number may possibly proceed by a second order phase-transition.

\item As the universe further cools down to a temperature $T_S$ not far above the electroweak scale $T_{EW}$, the $S$ field first develops a VEV, $\langle S\rangle\neq 0$, when its mass squared term (including the thermal corrections) becomes negative, while the Higgs VEV remains zero, $\langle H\rangle=0$. The transition to this step could be a simple crossover or just a second order phase transition. 

\item At the critical temperature near the electroweak scale,  $T_c$, a new minimum of the potential with $\langle H\rangle \neq 0, \langle S\rangle \simeq 0$ emerges that turns into the true minimum (replacing the former one with $\langle S\rangle \neq 0, \langle H\rangle =0$). This process must involve a first-order phase transition requiring the presence of a barrier between both minima. The universe tunnels from one vacuum to the other via bubble nucleation.

\end{enumerate}

\begin{figure}[h]
\centerline{\includegraphics[width=10cm]{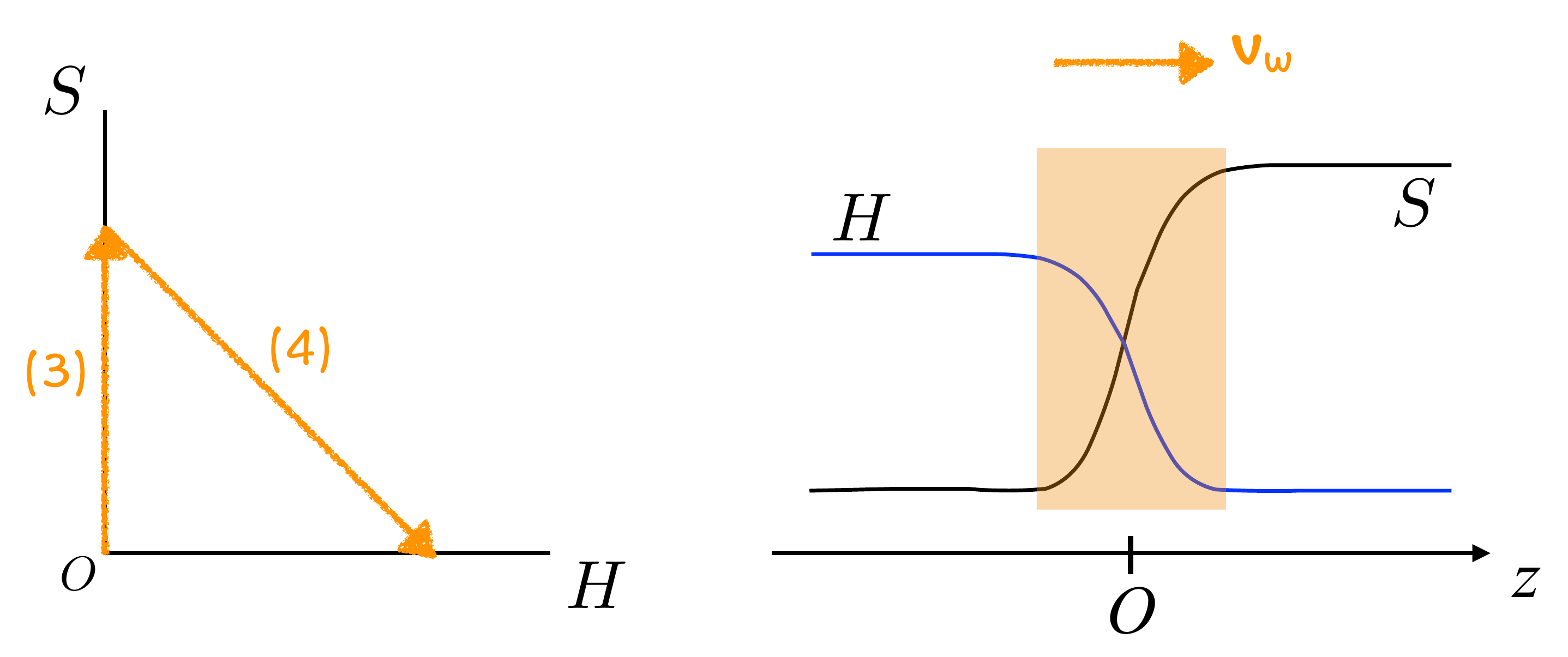}}
\caption{\it Schematic plot of the phase transitions. The left plot shows the change of $S$ and $H$ VEVs during the steps 3 and 4 discussed in the text.
The right plot, shows their VEV profiles in front of and behind the expanding bubble wall (shadowed region) during the electroweak phase transition in step 4. The bubble interior is for $z<0$.}\label{fig:EWPT}
\end{figure}

A schematic picture of the phase transitions in steps 3 and 4 is depicted in Fig.~\ref{fig:EWPT} (left panel). It has been shown~\cite{Espinosa:2011ax, Patel:2013zla, Cheung:2013dca, Curtin:2014jma, Cline:2017qpe, deVries:2017ncy} that the above evolutions could be realized dynamically by the interplay among the terms in the scalar potential describing the Higgs and the new scalar field $S$. At zero temperature, the scalar potential reads as~\footnote{As the field $\Phi$ is integrated out at the electroweak scale, the presence of the Higgs portal terms $\lambda_{\Phi H}|\Phi|^2|H|^2$ and $\lambda_{\Phi S}|\Phi|^2|S|^2$ in (\ref{scalarpotential}) would amount to a simple redefinition of the mass terms for $|H|^2$ and $|S|^2$, thus not changing the general conclusion which follows. Of course in that case we would have to face a little hierarchy problem, arising from the fact that $v_\Phi\gg v,v_S$, which can be mitigated e.g.~by assuming $\lambda_{\Phi H},\lambda_{\Phi S}\ll 1$.}
\begin{eqnarray}\label{scalarpotential}
V(H, S) = \lambda_H (|H|^2 - v^2)^2 + \lambda_S (|S|^2 - v_S^2)^2 + \lambda_{SH} |S|^2 |H|^2 \ .
\end{eqnarray}
The conditions for $H = v, S=0$ to be the global minimum are
\begin{eqnarray}
\lambda_H v^4 > \lambda_S v_S^4, \ \ \ \ \ \lambda_{SH} v^2 > 2 \lambda_S v_S^2 \ .
\label{conditions}
\end{eqnarray}

At high temperatures, both $H$ and $S$ receive thermal corrections to their quadratic terms, $a_H T^2 |H|^2$ and $a_ST^2 |S|^2$, with $a_{H,S}>0$. Thus, at very large $T$, the potential will be minimized for $\langle H\rangle=\langle S\rangle=0$ (steps 1 and 2). Given that the Higgs field couples to more degrees of freedom than $S$, it follows that $a_H>a_S$, and it is always possible to find an intermediate temperature where the Higgs quadratic term is positive, while the $S$ quadratic term is negative (step 3), thus triggering a minimum with $\langle S\rangle\neq 0, \langle H\rangle=0$. At lower temperatures, however, the Higgs quadratic term will also turn negative.
This implies that there should be a critical temperature where the two minima, ($\langle S\rangle\neq 0, \langle H\rangle=0$) and ($\langle H\rangle \neq 0, \langle S\rangle=0$), are degenerate allowing for step 4 to occur. The Higgs portal interaction $\lambda_{SH} |S|^2 |H|^2$ in Eq.~(\ref{scalarpotential}) [or Eq.~(\ref{Higgsportal})], which is a cross quartic term, could then provide a tree-level temperature-dependent barrier that separates the two minima allowing for a first-order phase transition. As this phenomenon depends on the particular values of the potential parameters, we will just assume hereafter that they are such that they provide a strong enough first order phase transition. Detailed model analyses can be found in Refs.~\cite{Espinosa:2011ax, Patel:2013zla, Cheung:2013dca, Curtin:2014jma, Cline:2017qpe}.

\subsection{\bf The Source of CP Violation}\label{sec:CPVsource}

The scalar potential, and the $\chi$-$S$ Yukawa coupling terms introduced so far [see Eqs.~(\ref{scalarpotential}) and (\ref{darkYukawa})], do not violate CP yet. This is because the scalar potential (\ref{scalarpotential}) is only a function of $|S|$ and, as a result, we are allowed to redefine the argument of $S$ to remove the relative phase between $m_0$ and $\lambda_c S$ in (\ref{darkYukawa}). Moreover, any overall phase of the $\chi$ mass term can be further removed by redefining the phases of $\chi_L$ and  $\chi_R$ fields. Hence any CP violation effect in the Yukawa terms can be absorbed by field redefinitions, leaving no physical effect during the phase transition.

In order to accommodate a physical CP violating effect, which is a necessary condition for baryogenesis, one option is to introduce terms in the potential depending on $S$, which will hinder the redefinition of $\arg(S)$. The general form of these terms is
\begin{equation}\label{deltaV}
\delta V(S) = \rho_S S + \mu_S^2 S^2 + \lambda_{3S} |S|^2 S + {\rm h.c.} \ . 
\end{equation}
Naively, these terms violate the $U(1)_\ell$ gauged symmetry and are forbidden in the UV complete theory. However, in this model, one can write renormalizable, $U(1)_\ell$ invariant terms involving $\Phi$ and $S$, as
\begin{equation}\label{deltaVorigin}
\delta V(\Phi,S) = \left(\mu_{\Phi S}^2 + \lambda_{\Phi S} |\Phi|^2 \right) \Phi^* S + \lambda_{\Phi S}' \Phi^{*2} S^2 + \lambda_{\Phi S}'' |S|^2 \Phi^* S + {\rm h.c.} \ . 
\end{equation}
Clearly, after $\Phi$ develops its VEV and the $U(1)_\ell$ symmetry is spontaneously broken, Eq.~(\ref{deltaVorigin}) can generate (\ref{deltaV}), leaving the coefficients $\rho_S, \mu_S, \lambda_{3S}$ complex in general. In this discussion, we neglect the back reaction of $\delta V$ on the VEV of the $\Phi$ field, which is a higher order effect in the small $v_S/v_\Phi$ expansion.

In the following, for simplicity, we present in more detail the case where only $\mu_S$ is non-zero. We could first use the freedom of field redefinition to make the parameters $m_0$ and $\mu_S^2$ real and positive, but $\lambda_c$ will  in general remain as a complex parameter. In this case, $\delta V(S) = 2 \mu_S^2 |S|^2 \cos [2\arg(S)]$ is the only term in the potential for $\arg(S)$. It is always minimized for $\arg(S) = \pi/2$, such that
\begin{equation}\label{deltaV2}
\delta V(S) = - 2 \mu_S^2 |S|^2 \ .
\end{equation}

The physical source of CP violation arises from the $\chi$ mass term, $M_\chi \bar\chi_L \chi_R + M_\chi^* \bar\chi_R \chi_L$, where
\begin{equation}\label{physicalCP}
M_\chi = m_0 + \lambda e^{i\theta} |S| \ .
\end{equation}
Here we make the phase of the second term explicit, with $\theta = \arg(\lambda_c)+\pi/2$ and $\lambda\equiv |\lambda_c|$. During a first-order electroweak phase transition, in the presence of a bubble wall, the magnitude of $|S|$ is space-time dependent, hence having used the freedom to make $m_0$ real, the phase of $M_\chi$ is not removable. As will be discussed in the following subsection, this phase modifies the dispersion relations of $\chi_{L,R}$, and their anti-particles, in a CP violating way~\cite{Cline:2000nw}, and provides the key source of CP violation for baryogenesis.

When minimizing the potential, we can combine Eq.~(\ref{deltaV2}) with (\ref{scalarpotential}) and repeat the discussions in Sec.~\ref{history}, which still hold with the replacement
\begin{equation}
v_S^2 \to v_S^2 + \frac{\mu_S^2}{\lambda_S} \ ,
\label{shift}
\end{equation}
provided conditions (\ref{conditions}) hold after the shift (\ref{shift}). A special feature of considering only a non-zero $\mu_S$ in Eq.~(\ref{deltaV}) is that, after the electroweak phase transition, the VEV of $S$ can relax to zero, and the mass of $\chi$ today is uniquely determined by $m_0$.

Alternatively, if the tadpole term $\rho_S S$ is turned on in (\ref{deltaV}), one can still derive the physical CP violating phase similar to (\ref{physicalCP}), but
the VEV of $S$ after the phase transition will remain non-zero. The impact of a non-zero $S$ VEV will only be of relevance for the contributions to EDM's, as will be discussed in Sec.~\ref{sec:EDM}. So in many of our subsequent discussions we will assume, unless explicit mention, that $\rho_S=0$.

\subsection{\bf The Baryogenesis Mechanism}\label{sec:diffusion}

In this subsection, we discuss the microscopic particle physics processes for our baryogenesis mechanism to work. All of them happen near the expanding bubble wall, during a first order electroweak phase transition (step 4 of the early universe history described in Sec.~\ref{history}), when the universe tunnels from the electroweak symmetric vacuum to the broken one via bubble nucleation. Such a phase transition involves the simultaneous changes in the SM Higgs field and the scalar field $S$. We first rewrite the $\chi$ mass term (\ref{physicalCP}) with explicit spatial coordinate dependence (labeled by $z$) in the rest frame of the bubble wall
\begin{equation}\label{eq:MChi}
M_\chi(z) =m_0 + \lambda e^{i\theta} |S(z)| \ ,
\end{equation}
where $z$ is the distance from the bubble wall, as shown in Fig.~\ref{fig:EWPT} (right panel). The $z>0\ (z<0)$ region is the electroweak symmetric (broken) phase located outside (inside) the bubble. Our discussion here is in the basis where $(m_0, \lambda,\, \theta)$ are all real parameters. 
We will parametrize the profile of $|S(z)|$ taking the form
\begin{eqnarray}\label{BPF}
|S(z)| = s_0 \left[1 + \kappa \tanh (z/L_w) \rule{0mm}{3mm}\right]/2 \ ,
\end{eqnarray}
where $s_0 (1+\kappa)/2$ is the value of $|S|$ in the electroweak symmetric phase ($z/L_\omega\to\infty$), and $s_0 (1-\kappa)/2$ parametrizes its value after the completion of the phase transition ($z/L_\omega\to-\infty$). The bubble wall width and velocity are denoted as $L_\omega$ and $v_\omega$, respectively. Here, we shall focus on the special case $\kappa=1$ where, after the phase transition (corresponding to $z\ll0$), the VEV of the $S$ field completely turns off. This can be realized in the presence of the $\mu_S^2 S^2$ term in Eq.~(\ref{deltaV}) as discussed above. We expect the qualitative features of our results to hold when the other terms in $\delta V(S)$ are turned on, so that $\kappa \neq1$.

The phase transition relevant quantities, including the wall width $L_\omega$, the wall velocity $v_\omega$, the scalar field profile across the bubble wall, as well as the critical and nucleation temperatures, $T_c$ and $T_n$,\footnote{$T_c$ is defined as the temperature at which the $H=0$ and $H=v(T_c)$ minima are degenerate, whereas $T_n$ is the temperature at which the phase transition occurs. respectively, are all calculable as functions of the model parameters  (see e.g. Ref.~\cite{Dorsch:2018pat}). The main goal of this work, however, is to present a new baryogenesis mechanism, hence we leave a detailed study of the strong first order phase transition, and in particular the precise calculation of the value of $T_n$ and the value of the Higgs field at $T_n$, $v(T_n)$, for a future publication.
The detailed analysis of the precise requirements on the model parameters for the phase transition is a straightforward task, that however involves computational intense calculations.
In the present work, we assume that the model parameters are such that $v(T_n)/T_n\gtrsim 1$, and we scan over a 
generous range of $T_n$ values, as well as over other relevant model parameters, including $L_\omega$ and $v_\omega$, as shown in Eq.~(\ref{eq:scanrange}).}

We define the particle chiral asymmetries in the dark sector as~\cite{Cline:2000nw,Cline:2006ts}, at the nucleation temperature,
\begin{eqnarray}
\begin{split}
&\xi_{\chi_L}(z) = \frac{3}{T_n^3} \left( n_{\chi_L} - n_{\chi_L^c}\right)  \ , \\
&\xi_{\chi_R}(z) = \frac{3}{T_n^3} \left( n_{\chi_R} - n_{\chi_R^c}\right) \ ,
\label{eq:chiLR}
\end{split}
\end{eqnarray}
where $T_n$ is the temperature when bubbles emerge, $n_{\chi_{L,R}}$ the number density of chiral asymmetry, and $\xi_{\chi_{L,R}}T_n\equiv \mu_{\chi_{L,R}}$  defines the corresponding chemical potentials. The Yukawa interaction of $\chi_{L, R}$ with the $S$ background violates CP but preserves a global symmetry $U(1)_\chi$, whose current is defined as $J^\mu_\chi = \bar\chi_L \gamma^\mu \chi_L + \bar\chi_R \gamma^\mu \chi_R$. As a result, although nonzero values for  $\xi_{\chi_L}$ and  $\xi_{\chi_R}$ can be generated by CP violation in the dark sector, the sum $\xi_{\chi_L}(z) +\xi_{\chi_R} (z)$ vanishes. The space-time dependence in the absolute value of the $\chi$ mass, $|M_\chi(z)|$, and its phase, $\arg(M_\chi)$, near the bubble wall play an important role by modifying the dispersion relations of $\chi_{L,R}$ particles and their antiparticles in a CP violating way. This affects the phase space distribution of these particles. The resulting chiral asymmetries evolve according to the diffusion equation
\begin{eqnarray}\label{eq:diffusion}
\begin{split}
&- D \xi_{\chi_L}'' - v_\omega \xi_{\chi_L}' + \Gamma_m (\xi_{\chi_L} - \xi_{\chi_R}) = S_{\rm CPV} \ , \\
\end{split}
\end{eqnarray}
where $(^\prime)$ means derivative with respect to $z$. The diffusion constant $D$ is given by $D = {\langle v^2 \rangle}/({3\Gamma_m})$, with $\Gamma_m \sim \lambda^2 T_n/(4\pi)$, $v$ is the particle velocity in the bubble wall rest frame, and $\langle \rangle$ is the thermal average over the Fermi-Dirac distribution function $f_i(p)$ ($i=\chi_{L},\chi_R$) in the rest frame of the bubble wall,
\begin{eqnarray}\label{eq:BoltzmannDist}
f_i(p) = \frac{1}{e^{ (E + v_\omega p_z-\mu_i)/T}+1} \ ,
\end{eqnarray}
where $\mu_i$ is the chemical potential. The corresponding number density for $\chi_L, \chi_R$ is defined as
\begin{equation}
n_i = \frac{2}{(2 \pi)^3} \int d^3p f_i(p) \ .
\end{equation}
The CP violating source term can be calculated using Refs.~\cite{Cline:2000nw,Cline:2006ts} as,
\begin{eqnarray}
S_{\rm CPV} &=& \frac{v_\omega}{\Gamma_m T_n} \left\langle \frac{v_z}{2E^2} \right\rangle \left[ |M_\chi(z)|^2 (\arg M_\chi(z))' \rule{0mm}{4mm}\right]^{\prime\prime} \nonumber \\
&=& \frac{v_\omega}{\Gamma_m T_n} \left\langle \frac{v_z}{2E^2} \right\rangle \frac{m_0 s_0 \lambda \left[ -2 + \cosh \left( \frac{2z}{L_\omega}  \right) \right] \sin\theta}{L_\omega^3 \cosh^4 \left(\frac{z}{L_\omega} \right)} \ ,
\end{eqnarray}
where $E^2=p^2+|M_\chi(z)|^2$.

Clearly, in Eq.~(\ref{eq:diffusion}), the source term $S_{\rm CPV}$ must be nonzero in order to generate nonzero asymmetries in the $\chi_{L, R}$ particle numbers, which are proportional to $\xi_{\chi_{L,R}}$, respectively. This requires a nonzero value of $(\arg M_\chi(z))'$, {\it i.e.}~the phase of the $\chi$ mass must \textit{not be} a constant --- it has to vary in together with the $S$ VEV along the $z$ direction. A quick glance at the form of the $\chi$ mass term in Eq.~(\ref{eq:MChi}) shows that $m_0$ has to be different from zero. We will come back to this point near the end of this section when discussing the numerical calculation of the baryon asymmetry and the scan over the parameter space.

\begin{figure}[htb!]
\centerline{\includegraphics[width=1\textwidth]{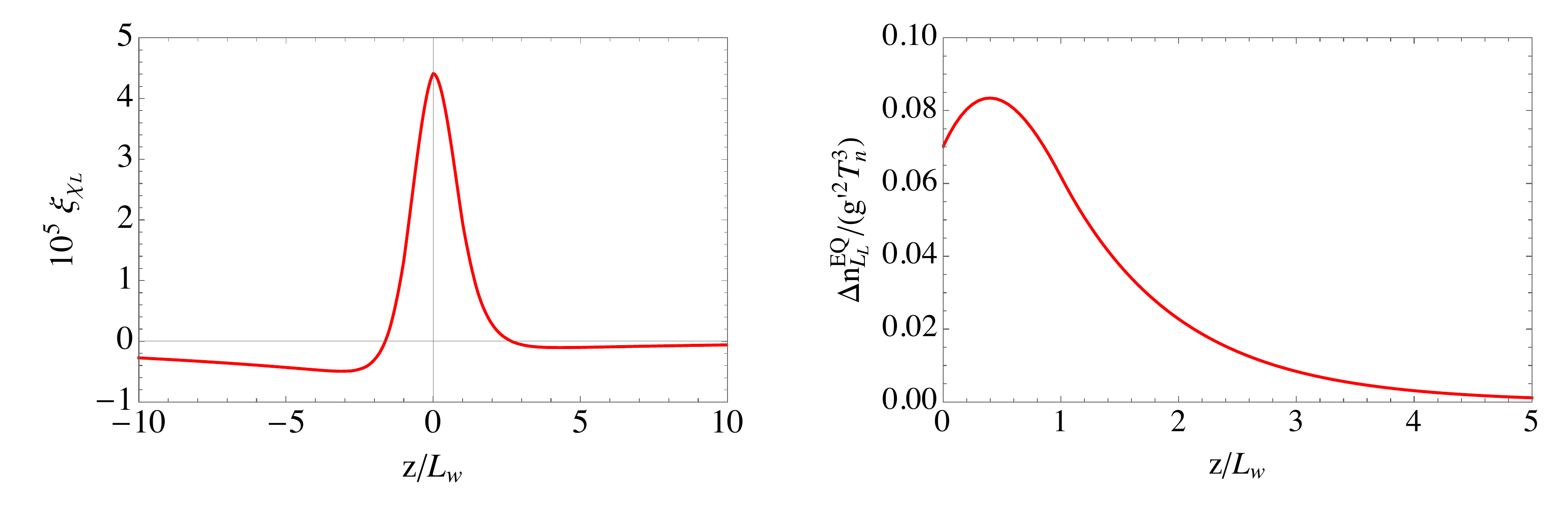}}
\caption{\it Left panel: Chiral charge asymmetry in $\chi_L$ (opposite for $\chi_R$) particles around the bubble wall, with parameters
$m_0=s_0=T_n=100\,{\rm GeV}$, $M_{Z'}=1\,$GeV, $\lambda=0.3$, $\theta=\pi/3$, $L_\omega=5/T_n$, $v_\omega=0.1$. Right panel: $\Delta n_{L_L}^{\rm EQ}(z)/g'^2T_n^3$ for the same values of the parameters. For this plot we only show the result in the region $z>0$ because it corresponds to the range of integral in Eq.~(\ref{eq:LeptonAsymmetry}), or (\ref{eq:A4}). 
}\label{asymmetry}
\end{figure}

The solution to the above diffusion equation is formally given by
\begin{eqnarray}
\xi_{\chi_L}(z) = \int_{-\infty}^{\infty} d z_0 \ G(z-z_0)\ S_{\rm CPV}(z_0) \ ,
\end{eqnarray}
where the Green's function $G(z)$ satisfies the equation
\begin{equation}
-D G^{\prime\prime}(z)-v_\omega G^\prime(z)+2 \Gamma_m G(z)=\delta(z)\ .
\end{equation}
The solution, continuous at the origin, is given by
\begin{eqnarray}
\begin{split}
&G(z) = \frac{D^{-1}}{k_+ - k_-} \left\{
\begin{array}{ll}
e^{-k_+z}, & z\geq0\\
e^{-k_-z}, & z<0
\end{array}
\right., \ \ \ \ \ 
k_{\pm} = \frac{v_\omega}{2D} \left( 1 \pm \sqrt{1+\frac{8\,\Gamma_m D}{v_\omega^2}} \right) \ .
\end{split}
\end{eqnarray}
In the left panel of Fig.~\ref{asymmetry}, we show the chiral asymmetry distribution of $\chi_L$ as a function of the $z$ coordinate, for a given set of model and phase transition parameters.

Unlike in the usual electroweak baryogenesis scenarios, here the particle chiral charge asymmetry is generated in the dark sector through the $\chi$ particle, which is an $SU(2)_L$ singlet and thus does not couple to the electroweak sphalerons. Moreover, for general values of $\mathtt{q}$, the gauge symmetry $U(1)_\ell$ forbids any renormalizable operators through which the asymmetries in $\chi$ might be directly shared with the SM fermions that carry the $SU(2)_L$ charge~\footnote{As explained in the introduction, this aspect serves as a major difference between our work and that in Ref.~\cite{Cline:2017qpe}. In our case, a new way of transferring the $\chi$ particle chiral charge asymmetry to the visible sector is presented.}. We here make the observation that, thanks to the leptonic $Z^\prime$ portal, which couples to both $\chi$ and the SM leptons, the CP violating effect in the dark sector can be transferred in a novel  way to the observable sector.

The main point here is that $\chi_L$ and $\chi_R$ carry different $U(1)_\ell$ charges ($\mathtt{q}+N_g$ and $\mathtt{q}$ respectively)~\footnote{Note their charges are not chosen by hand but, instead, required by the anomaly cancellation conditions discussed in Sec.~\ref{sec:UVcompletition} and Tab.~\ref{tab:table}.}. Consequently, the above chiral asymmetries imply a net $U(1)_\ell$ charge density near the bubble wall as,
\begin{equation}\label{nl}
\begin{split}
\rho_\ell (z) &= (\mathtt{q}+N_g) \left[n_{\chi_L} - n_{\chi_L^c}\right] + \mathtt{q} \left[n_{\chi_R} - n_{\chi_R^c}\right] = \frac{1}{3} N_g T_n^3 \, \xi_{\chi_L}(z) \ ,
\end{split}
\end{equation}
where use has been made of Eq.~(\ref{eq:chiLR}). The existence of this net $U(1)_\ell$ charge density yields a Coulomb background of the $Z'$ potential, $\langle Z'_0 \rangle$. In the approximation of very large bubbles, this lepton number potential could be calculated in cylindrical coordinates as,
\begin{equation}\label{eq:Zbackground}
\left\langle Z'_0(z)  \rule{0mm}{3.5mm}\right\rangle
= \frac{g' }{2M_{Z'}} \int_{-\infty}^\infty dz_1\ \rho_\ell(z_1)\ \exp\left[-M_{Z'} |z-z_1|  \rule{0mm}{3.5mm}\right]\ ,
\end{equation}
where we neglect the impact of $|S(z)|$ on the mass of $Z^\prime$, which is mainly set by the value of $v_\Phi$ at a much higher scale.

The background of the vector field $Z^\prime$ breaks the Lorentz symmetry and thus is a CPT violating effect, which is also odd under the CP transformation. It retains certain similarities to the spontaneous baryogenesis mechanism~\cite{Cohen:1991iu, Kolb:1990vq} (also with gravitational baryogenesis~\cite{Davoudiasl:2004gf, Davoudiasl:2013pda}), where a time-dependent (CPT violating) scalar field couples to the vector current of a particle, and serves as its chemical potential~\footnote{Notice that the VEV of $Z^\prime_0$ vanishes after the electroweak phase transition, as its value stems from the asymmetry in $\chi_{L,R}$ particles, which vanishes when $\arg(M_{\chi})$ becomes a constant and the source of CP violation $S_{CPV}$ vanishes. Therefore at zero temperature our model does not contain any violation of Lorentz symmetry.}. 
In our model, we use the time-like component of the $Z^\prime_\mu$ gauge boson, whose CP and CPT violating background is generated due to the microscopic interaction processes between the dark sector particles and the bubble wall described above. The $Z'_0$ background couples to the SM lepton current (see Eq.~(\ref{Z'portal})). As we shall see, given that this current is anomalous with respect to the SM $SU(2)_L$ gauge symmetry, it could bias the sphaleron process to work in one direction. The $Z'_0$ background then yields a ``chemical potential'' for the SM leptons,
\begin{eqnarray}\label{chemical}
\mu_{L_L}(z) = \mu_{\ell_R}(z) = g' \left\langle Z'_0(z)  \rule{0mm}{3.5mm}\right\rangle \ .
\end{eqnarray}
The thermal equilibrium asymmetry in SM lepton number would then be given by (considering left-handed lepton doublets)
\begin{eqnarray}\label{eq:ThermalLeptonAsymmetry}
\Delta n_{L_L}^{\rm EQ}(z) = \frac{2N_gT_n^2}{3}  \mu_{L_L}(z) = \frac{2g' N_gT_n^2}{3} \left\langle Z'_0(z)  \rule{0mm}{3.5mm}\right\rangle \ .
\end{eqnarray}
We show in the right panel of Fig.~\ref{asymmetry} the spatial distribution of $\Delta n_{L_L}^{\rm EQ}(z)$ for a given set of model and phase transition parameters.
It is worth mentioning that the profiles $\Delta n_{L_L}^{\rm EQ}(z)$ and $\left\langle Z'_0(z) \right\rangle$ depend on our assumption of the bubble profile, Eq.~(\ref{BPF}).

In the presence of the electroweak sphaleron processes, which can change the lepton number, the actual SM lepton number asymmetry will evolve toward its equilibrium value. This evolution is governed by the following rate equation, 
\begin{eqnarray}\label{EWsph}
\begin{split}
\frac{\partial \Delta n_{L_L}(z, t)}{\partial t} &= \Gamma_{\rm sph}(z-v_\omega t)  \left[ \Delta n_{L_L}^{\rm EQ}(z-v_\omega t) - \Delta n_{L_L}(z,t) \rule{0mm}{3.5mm}\right] \ ,
\end{split}
\end{eqnarray}
where $\Gamma_{\rm sph}$ is the rate for the sphaleron process at the nucleation temperature $T_n$. The second term on the right-hand side of Eq.~(\ref{EWsph}) represents the washout term, which would drive the asymmetry to zero if the sphaleron processes did not go out of equilibrium quickly enough. Assuming a strong first-order electroweak phase transition, where the condition $v_n/T_n\gtrsim 1$ is fulfilled ($v_n$ is the Higgs VEV at the nucleation temperature $T_n$), a good approximation for $\Gamma_{\rm sph}$ is that it is unsuppressed at any point $z$ outside the bubble wall, but becomes exponentially suppressed after the bubble wall has passed through taking this point to the bubble interior, {\it i.e.}
\begin{eqnarray}\label{RatePhase} 
\Gamma_{\rm sph}(z-v_\omega t) = \left\{\begin{array}{ll} 
\Gamma_0&:\ t < z/v_\omega \\
\Gamma_0 e^{-M_{\rm sph}/T_c} &:\ t > z/v_\omega \\
\end{array} \right. \ .
\end{eqnarray}
In Eq.~(\ref{RatePhase}),  $\Gamma_0\simeq 120\,\alpha_w^5 T_n\simeq 10^{-6}T_n$~\cite{Bodeker:1999gx}, and $M_{\rm sph} = {4\pi v_n} B/{g_2}$ is the sphaleron mass in the broken phase, where $B$ is a fudge factor~\cite{Kuzmin:1985mm} which depends on the Higgs mass, and the weak coupling $g_2$.  In the SM, for the experimental value of the Higgs mass it turns out that $B\simeq 1.96$.
As discussed in detail in \cite{Moore:1998swa, Zhou:2019uzq}, the sphaleron rate in the presence of an additional singlet
depends on the parameters in the $V(S, H)$ potential, and could be calculated once this parameter dependence of the first order phase transition is worked out.

The solution to the rate equation takes the form~\cite{Cline:2000nw}
\begin{equation}\label{eq:LeptonAsymmetry}
\Delta n_{L_L} =\frac{\Gamma_0}{v_\omega} \int^{\infty}_0 dz \,
\Delta n_{L_L}^{\rm EQ}(z)e^{-\Gamma_0 z/v_\omega} \ .
\end{equation}
We refer the reader to App.~\ref{app:solve} for more details on obtaining this result.
At this point it is important to realize that the final lepton number density, as given by Eq.~(\ref{eq:LeptonAsymmetry}), is non-vanishing as a consequence of the fact that the effective theory at the scale of electroweak baryogenesis has an anomalous lepton number. Had we not integrated out any anomalon propagating in the UV theory, the final lepton number density $\Delta n_L$ would have been zero. This statement is proven in detail in App.~\ref{AppendixB}. See also~\cite{Carena:2018cjh}.

Because the sphaleron processes preserve $B-L$, equal asymmetries will be generated for baryon and lepton numbers, $\Delta n_{B}= \Delta n_{L_L}$. The entropy density of the universe at the EW scale is $s\simeq ({2\pi^2}) g_{*} T_c^3/{45}$, where $g_*\simeq g_B+(7/8)g_F\simeq \mathcal O(100)$ is the effective number of degrees of freedom at the EW phase transition. The final generated baryon-to-entropy ratio is then
\begin{eqnarray}
\eta_B = \frac{\Delta n_B}{s} \ .
\end{eqnarray}

\begin{figure}[t]
\centerline{\includegraphics[width=0.7\textwidth]{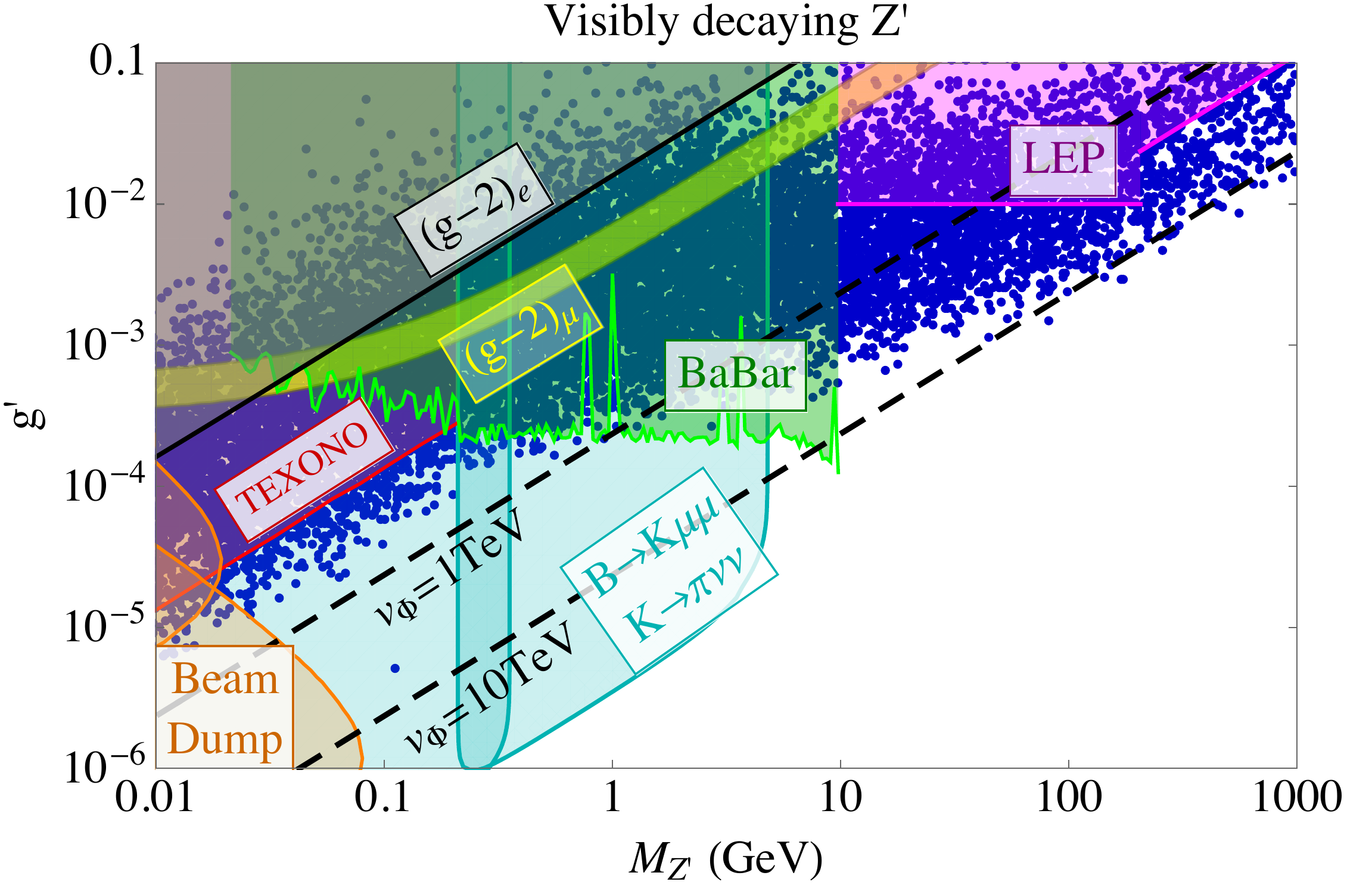}}
\caption{\it
The parameter space of our model (assuming $N_g=3$) that could generate the observed baryon asymmetry of the universe is covered by the blue points, in the $g^\prime$ versus $M_{Z^\prime}$ parameter space. The colorful shaded regions have been excluded by the existing constraints from LEP, BaBar, electron $g-2$, beam dump, and neutrino-electron scattering experiments, 
as well as the measurement of flavor changing $K\to \pi$, $B\to K$ decay rates.
The yellow band is the favored region for explaining the muon $g-2$ anomaly. The black dashed line corresponds to the VEV $v_\Phi$ equal to 1, 10\,TeV. We consider in the parameter scanning the condition $m_0>M_{Z'}/2$.}\label{fig:workingpoints}
\end{figure}

The dark blue points in Fig.~\ref{fig:workingpoints} show the working parameter space where the observed baryon asymmetry~\cite{Aghanim:2018eyx} 
\begin{eqnarray}
\eta_B \simeq 0.9\times10^{-10} 
\end{eqnarray} 
can be generated. They are obtained by scanning over all the model parameters in the following ranges,
\begin{eqnarray}\label{eq:scanrange}
\begin{split}
&M_{Z'}, \ m_0\in (10^{-3}, 10^3)\,{\rm GeV}, \ \ \ s_0,\ T_n \in (100, 500)\,{\rm GeV}, \ \ \ \lambda \in (10^{-2}, 1), \\ 
&g'\in (10^{-6},0.1),\quad \theta \in (-\pi/2, \pi/2), \ \ \ L_w \in (1/T_n, 10/T_n), \ \ \  v_\omega \in (0.05,0.5) \ .
\end{split}
\end{eqnarray} 
Here, the parameter $m_0$ is the mass of the $\chi$ particle, assuming $S$ has no VEV today.

We display, in Fig.~\ref{fig:workingpoints}, the baryogenesis viable points in the $g^\prime$ versus $M_{Z^\prime}$ plane assuming $N_g=3$ (the case $N_g=2$ will be independently exhibited in Sec.~\ref{sec:G2}), where the mass parameters satisfy the relation $m_0>M_{Z'}/2$. The result shows that the smaller the $Z^\prime$ mass, the smaller the value of $g^\prime$ in the allowed region. In particular, with $M_{Z'}$ around 100\,MeV, the gauge coupling $g'$ should be as small as $10^{-5}$. This feature is expected from the value of the $Z^{\prime}_0$ background during baryogenesis, calculated in Eq.~(\ref{eq:Zbackground}), where parametrically the final baryon asymmetry is proportional to $\sim g'^2/M_{Z'}^2$. In this case, $m_0>M_{Z'}/2$, the $Z^\prime$ boson is kinematically forbidden to decay into $\chi\bar\chi$. If created in the laboratory, it will decay into SM particles. This is a visible decay, and in the next section we will confront these points with the existing, and near-future, $Z^\prime$ searches.
It is worth pointing out that the values of $g'$ of interest for successful baryogenesis are much smaller than 1, thus the back reaction of $Z'$ particles on the bubble wall is negligible.

On the other hand, we find that the resulting points with $m_0 < M_{Z'}/2$ exhibit a different $g'$ versus $M_{Z'}$ correlation behavior. In particular, we find that when the $Z'$ is light (well below the electroweak scale), $m_0$ is thus small and the required values of $g'$ for successful baryogenesis are much larger (with $g'>10^{-3}$ everywhere).  This could be understood from the explicit expression for the source of CP violation for the baryogenesis mechanism $S_{CPV}$. As discussed in the paragraph below Eq.~(\ref{eq:BoltzmannDist}), the relevant CP violation source is proportional to the gradient of $\arg(M_\chi)$ along the $z$ direction, where the VEV of $S$ changes. Clearly, if the $m_0$ term is very small, $\arg(M_\chi)$ remains approximately $\theta$, and $(\arg(M_\chi))'$ would be suppressed. To compensate for this suppression, larger values of $g'$ are needed. In this case, we find that the experimental constraints from invisibly decaying $Z'$ searches~\cite{Banerjee:2016tad, Lees:2017lec} are already strong enough to exclude almost the entire viable parameter space for baryogenesis. Therefore, we will not consider this case any further.

\section{Phenomenology} \label{pheno}

In this section, we will discuss the phenomenological consequences of the above described baryogenesis mechanism. We will show that generating the observed baryon asymmetry in the model has a strong impact on the $Z^\prime$ boson search, on the physics of $\chi$ as the dark matter candidate, and on the electric dipole moments, as well as on possible LHC signals of the Higgs boson and the dark Higgs $S$.

Throughout the discussions in this section, we will assume the parameter $N_g$, the number of lepton flavors charged under the $U(1)_\ell$, to be equal to 3. We will comment on the differences in phenomenology if only two lepton flavors are gauged, {\it e.g.}~$L_\mu+L_\tau$, in the upcoming Sec.~\ref{sec:G2}.

\subsection{\bf Searches for the Leptophilic $Z^\prime$}\label{sec:Z'search}

First of all, let us recall that the presence of the $Z^\prime$ boson is the key for the success of our electroweak baryogenesis mechanism. It needs to develop a CP (and CPT) violating background during the electroweak phase transition, which permits to transfer the CP violating effect from the dark sector to the SM leptons. In order to generate sufficient final baryon asymmetry, which is proportional to $g'^2/M_{Z^\prime}^2$, the gauge boson $Z^\prime$ cannot be too heavy and the coupling $g^\prime$ should not be too small, as shown in Fig.~\ref{fig:workingpoints}.

At the same time, since the $Z^\prime$ is the gauge boson for the lepton number symmetry, it couples to the SM charged leptons and neutrinos. Such a new vector particle has been directly searched for at $e^+e^-$ colliders, such as LEP (both through resonances~\cite{Tanabashi:2018oca} and contact interactions~\cite{Alcaraz:2006mx}) and BaBar~\cite{Lees:2014xha}, as well as at electron beam dump experiments~\cite{Bjorken:2009mm}, and neutrino experiments that are sensitive to neutrino-electron interactions (such as TEXONO)~\cite{Bauer:2018onh}.  The $Z'$ could also be exchanged at the loop level and contribute to the anomalous magnetic moments of charged leptons~\cite{Davoudiasl:2012qa}. Many of these constraints are similar to, and could be translated from, the limits on dark photons~\cite{Bauer:2018onh, Battaglieri:2017aum}. Because the $Z^\prime$ now mainly couples to charged-leptons and neutrinos, we re-evaluate its branching ratios based on the following partial decay widths
\begin{eqnarray}
\begin{split}
& \Gamma_{Z'\to \ell\bar\ell} = \frac{g'^2}{12\pi} M_{Z'} \left( 1 + \frac{2m_\ell^2}{M_{Z'}^2} \right)\sqrt{ 1- \frac{4m_\ell^2}{M_{Z'}^2}} \ , \\
& \Gamma_{Z'\to \nu\bar\nu} = 3\times \frac{g'^2}{24\pi} M_{Z'} \ , \\
\end{split}
\end{eqnarray} 
where $\ell=e,\mu,\tau$. We neglect the $Z'$ decay into right-handed neutrinos, assuming it is kinematically forbidden~\footnote{The origin of right-handed neutrino masses will be addressed in Sec.~\ref{sec:nuCosmo}.}. Because the $Z^\prime$ boson in this model is hadrophobic, the constraints from meson decays ($\pi^0$, $J/\Psi$, $\Upsilon$) into $Z^\prime$ only apply through loop level processes~\cite{Bauer:2018onh}.


Moreover, because the $Z'$ couples to an anomalous current with respect to $SU(2)_L^2$ in the low energy theory, it makes important contributions to flavor-changing meson decays such as $K\to \pi Z'$ and $B\to K Z'$ through the Wess-Zumino term which occurs at two loop level~\cite{Dror:2017ehi,Dror:2017nsg}~\footnote{We thank Jeff Dror for pointing out to us the results in Ref.~\cite{Dror:2017ehi,Dror:2017nsg}.}. For very light $Z'$, these decays are mainly into the longitudinal component of the $Z'$ boson and the corresponding rates are enhanced by $1/M_{Z'}^2$. 
The $Z'$ boson will then decay into charged lepton pairs or neutrinos. Following Ref.~\cite{Dror:2017ehi,Dror:2017nsg}, we find that with these final states stringent limits can be set on the gauge coupling $g'$.

The existing experimental constraints on a leptophilic $Z^\prime$ are summarized in Fig.~\ref{fig:workingpoints} for the $N_g=3$ model. 
These limits, altogether, set a lower bound on the $Z^\prime$ mass of around 10\,GeV. 
A prospective Higgs factories~\cite{dEnterria:2016fpc, CEPC-SPPCStudyGroup:2015csa} could explore regions with larger $Z^\prime$ masses.

In addition, the gauge coupling $g'$ is indirectly constrained by requiring the anomalon fields for the $U(1)_\ell$ symmetry to be sufficiently heavy.
As discussed in Sec.~\ref{sec:UVcompletition}, the gauged $U(1)_\ell$ symmetry is broken by the VEV of a scalar field $\Phi$ above the electroweak scale. The same VEV will define the mass of the anomalon fields, as a function of their Yukawa coupling.
To secure the anomalon fields are already decoupled during the electroweak phase transition, while avoiding the Yukawa couplings to be in the strongly coupled regime, values of
$v_\Phi$ above a few times the electroweak scale are required.
In Fig.~\ref{fig:workingpoints}, we show 
indicative values of $v_\Phi=1$ and 10 TeV, respectively, where we have used the $Z^\prime$ mass given in Eq.~(\ref{eq:Z'mass}).
This shows that most of the experimentally-allowed, EWBG-favored solutions are in the region of $M_{Z'}$ above 10\,GeV.

Finally we wish to comment that, in general, there is a kinetic mixing between the hypercharge gauge boson $B_\mu$ and the new gauge boson $Z_\mu^\prime$, as
\begin{equation}
\mathcal L_{kin}=-\frac{1}{2} c(\mu) F_{Y}^{\mu\nu}F^\prime_{\mu\nu}  \ ,
\end{equation} 
where the coefficient $c(\mu)$ receives renormalization at loop level. Its one-loop beta function takes the form~\cite{Carone:1995pu}
\begin{equation}\label{uvRGE}
\frac{\partial c(\mu)}{\partial\log\mu}=\frac{g_Y g^\prime}{12\pi^2} \textrm{Tr} (YL) \ .
\end{equation}
In the complete UV theory considered here, we have that $\textrm{Tr}(YL)=-4(\mathtt{q}+3)$. There is a special case, $\mathtt{q}=-3$, where the kinetic mixing parameter $c(\mu)$ does not run at energies above the $U(1)_\ell$ symmetry breaking scale, $v_\Phi$. For $\mu<v_\Phi$, after integrating out the anomalon fields $L_L', L_R'', e_L'', e_R'$, $\textrm{Tr}(YL)=-6$ in the effective theory. This implies that even if we set $c_{UV}=0$ at high scale as the boundary condition, it will be generated at low energies as
\begin{equation}
c(M_Z)\simeq c_{\rm UV} + \frac{g_Y g^\prime}{2\pi^2}\log\frac{v_\Phi}{M_Z} \ ,
\end{equation}
where we are assuming that the masses of $L_L', L_R'', e_L'', e_R'$ are all of order $v_\Phi$, and compute the value of $c$ at the $M_{Z}$ mass scale where it is measured at the LEP experiment.
A non-zero kinetic mixing between $B_\mu$ and $Z^\prime_\mu$ generates, after electroweak breaking, a mixing between $Z_\mu$ and $Z_\mu^\prime$. This could impact LEP observables including the $Z$ boson mass (the $\rho$ parameter), the $Z$ hadronic width, and the forward-backward asymmetries in leptonic $Z$ decays. The analysis in~\cite{Carone:1995pu} finds that $c(M_{Z})$ is constrained to be less than the percent level, with a much stronger constraint in the region where $Z$ and $Z'$ are nearly degenerate~\cite{Hook:2010tw, Dobrescu:2014fca}. Compared to the EWBG favored region for $g'$ in Fig.~\ref{fig:workingpoints}, we find it easy to satisfy these constraints provided $c_{\rm UV}$ is small enough.

To summarize, after taking into account all the above constraints, the mass window of the $Z^\prime$ for our baryogenesis mechanism to work is $10\,{\rm GeV}<M_{Z'}<\mathcal O({\rm TeV})$.


\subsection{\bf Neutrino Cosmology}\label{sec:nuCosmo}

It is worth commenting on the neutrino sector of the gauged $U(1)_\ell$ model, and implications of cosmological measurements on additional neutrino degrees of freedom, $\Delta N_{eff}$~\cite{Aghanim:2018eyx}. 

As discussed in Sec.~\ref{sec:UVcompletition}, within the minimal setup, the neutrino mass is Dirac, generated by the Yukawa coupling between the SM active neutrinos $\nu_{L_i}$ and the right-handed ones, $\nu_{R_j}$. In the early universe, at sufficiently high temperatures, the $U(1)_\ell$ gauge interaction could thermalize all $\nu_{R_i}$, and make a contribution to $\Delta N_{eff}$~\cite{Barger:2003zh, Ghosh:2010hy, Hamann:2011ge}. To avoid an excessive contribution to $\Delta N_{eff}$, one option is to make the $U(1)_\ell$ interaction to decouple early enough, preferably above the QCD phase transition temperature, $T_{\rm QCD} \sim 100\,$MeV. The relevant process for thermalizing the $\nu_R$'s is, $\ell_{\rm SM} \bar \ell_{\rm SM} \to \nu_{R_i} \bar \nu_{R_i}$, through the $s$-channel $Z'$ exchange, where $\ell_{\rm SM}=e, \mu, \nu_{L_i}$ are the SM relativistic species, around the $T_{\rm QCD}$ temperature. The corresponding annihilation cross section times relative velocity is
\begin{equation}
\sigma v (\ell_{\rm SM} \bar \ell_{\rm SM} \to \nu_{R_i} \bar \nu_{R_i}) = \frac{g_{\ell_{\rm SM}} g'^4 s_{\rm CM}}{48 \pi (s_{\rm CM} - M_{Z'}^2)^2} \ ,
\end{equation}
where $g_{e}=g_\mu=2g_{\nu_{L_i}} = 2$, and $s_{\rm CM}$ is the center-of-mass energy squared of the annihilation, of order $T^2$. The thermal averaged annihilation rate per particle $\ell_{\rm SM}$, given by $n_{\ell_{\rm SM}} \sigma v (\ell_{\rm SM} \bar \ell_{\rm SM} \to \nu_{R_i} \bar \nu_{R_i})$, goes as $T^5$, in the heavy $Z'$ limit, $M_{Z'}\gg T$. In this case, decoupling $\nu_R$'s no later than $T_{\rm QCD}$ amounts to requiring the annihilation rate to be less than the Hubble expansion rate at $T_{\rm QCD}$. This in turn implies that 
\begin{equation}\label{condition1}
v_\Phi \gtrsim 10\,{\rm TeV}, \ \ \ \ \ {\rm for}\ \ M_{Z'}\gg T_{\rm QCD} \ .
\end{equation}
On the other hand, if $M_{Z'} \ll T_{\rm QCD}$, the thermal averaged annihilation rate will scale as $T$ until the temperature falls below the $Z'$ mass. In this case, requiring that $\nu_{R_i}$ never reaches thermal equilibrium implies that
\begin{equation}\label{condition2}
g' \lesssim 10^{-5} \left( \frac{M_{Z'}}{{1\, \rm MeV}} \right)^{1/4}, \ \ \ \ \ {\rm for}\ \ M_{Z'}\ll T_{\rm QCD} \ .
\end{equation}
Satisfying conditions (\ref{condition1}) and (\ref{condition2}) imposes strong constraints on the EWBG viable parameter space found in Fig.~\ref{fig:workingpoints}.
See~\cite{Abazajian:2019oqj} for a recent calculation in a similar context.

The viable alternative option for neutrino mass is to implement the seesaw mechanism by giving Majorana masses to $\nu_{R_i}$.  If all the $\nu_{R_i}$ are heavier than $\sim 500\,$MeV, they will decay before the big-bang nucleosynthesis and will have no effect in $\Delta N_{eff}$~\cite{Berryman:2017twh}. However this option requires extending the scalar sector of the model by introducing an extra SM singlet $\Phi^\prime$ with lepton number $L=2$, which couples to the right-handed neutrinos as 
\begin{equation}
\sum_{\alpha,\beta = e,\mu,\tau}Y'_{\alpha\beta}\bar \Phi^\prime\bar\nu_{R\,\alpha}^c\nu_{R\,\beta} + {\rm h.c.}
\end{equation}
For large enough values of $v_{\Phi'}\gtrsim100\,$GeV, the new scalars from the $\Phi'$ field could kinematically evade searches at LEP. Note, on the other hand, that we need $v_{\Phi'}\ll v_\Phi \sim$\,TeV in order to not perturb the results of this paper on electroweak baryogenesis. 
However, this will require a more detailed study of the effects on the nature of the electroweak phase transition.
The experimental search for heavy Majorana neutrinos is of great phenomenological interest~\cite{Atre:2009rg}, especially as the $U(1)_\ell$ gauge interaction here opens a new production channel for them. We will investigate this exciting opportunity in a future work.

\subsection{\bf $\chi$ as Dark Matter}\label{sec:DMsearch}

As mentioned earlier, in this model, the particle $\chi$ from the dark sector could be a dark matter candidate, since there is a $\mathbb Z_2$ symmetry in the Lagrangian ($\chi\to -\chi$) allowing it to be stable.

\begin{figure}[h]
\centerline{\includegraphics[width=0.7\textwidth]{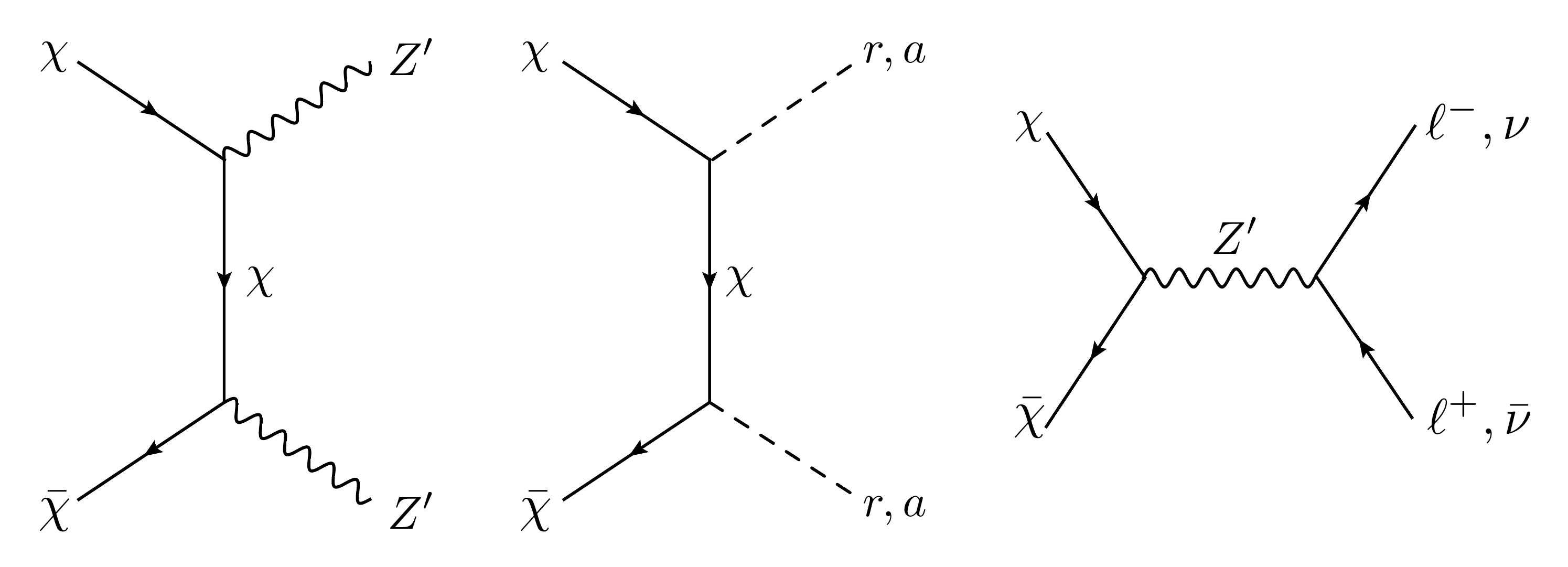}}
\centerline{\includegraphics[width=0.5\textwidth]{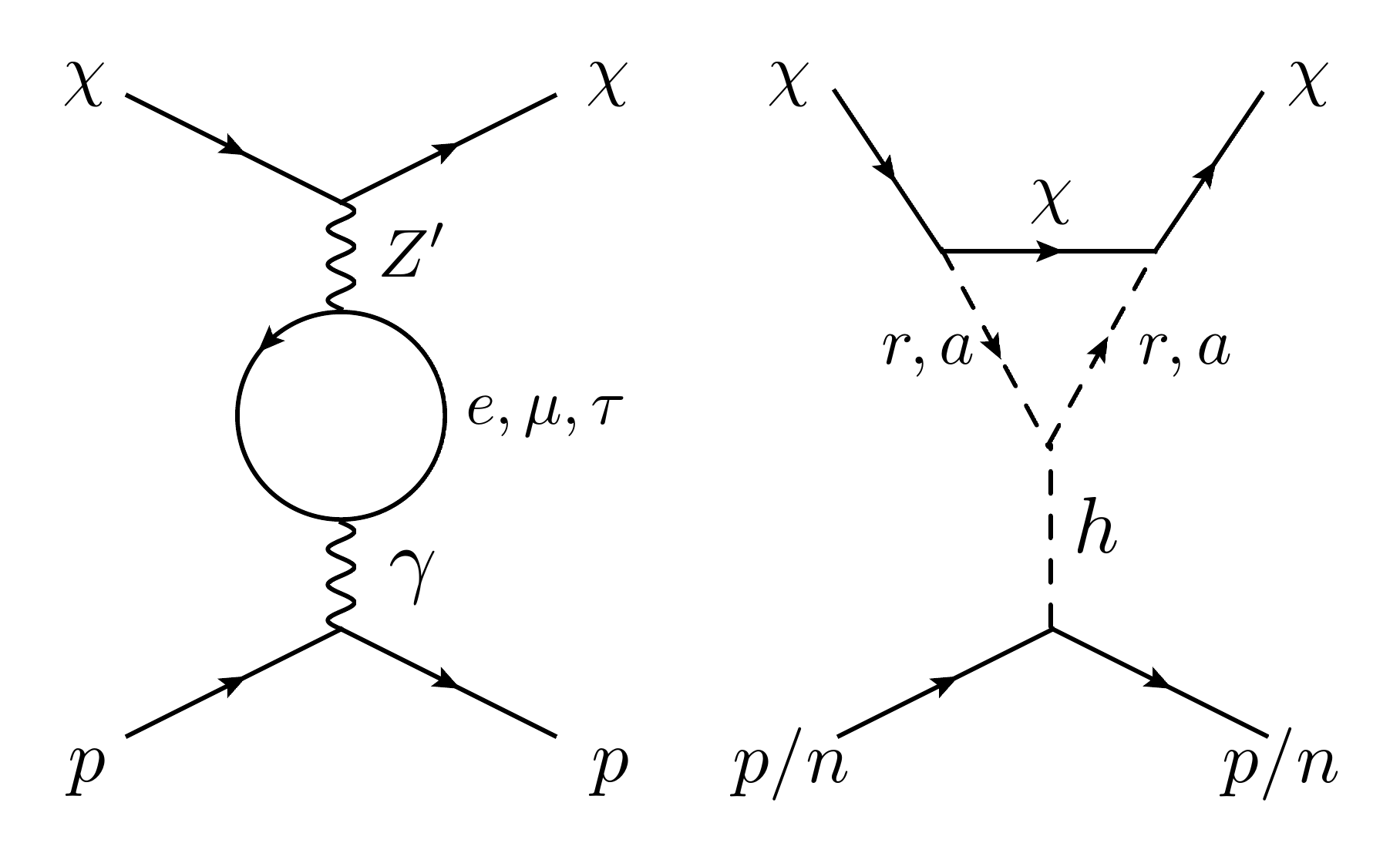}}
\caption{\it Feynman diagrams for dark matter thermal freeze out (first row) and direct detection (second row) in the model we consider. Time flows from left to right.}\label{fig:DM}
\end{figure}

\subsubsection{The Thermal Relic Density}\label{DMpheno}

If the VEV of $S$ relaxes to zero after the electroweak phase transition, the mass of $\chi$ will be given by $m_0$. From the above baryogenesis analysis point of view, we find that $\chi$ is favored to be heavier than $Z^\prime$ (see Fig.~\ref{fig:workingpoints} and corresponding discussions). In the following, we will consider all the possible annihilation channels, as shown in the first row of Fig.~\ref{fig:DM}, that will contribute to the dark matter relic density.

\begin{figure}[h]
\centerline{\includegraphics[width=1\textwidth]{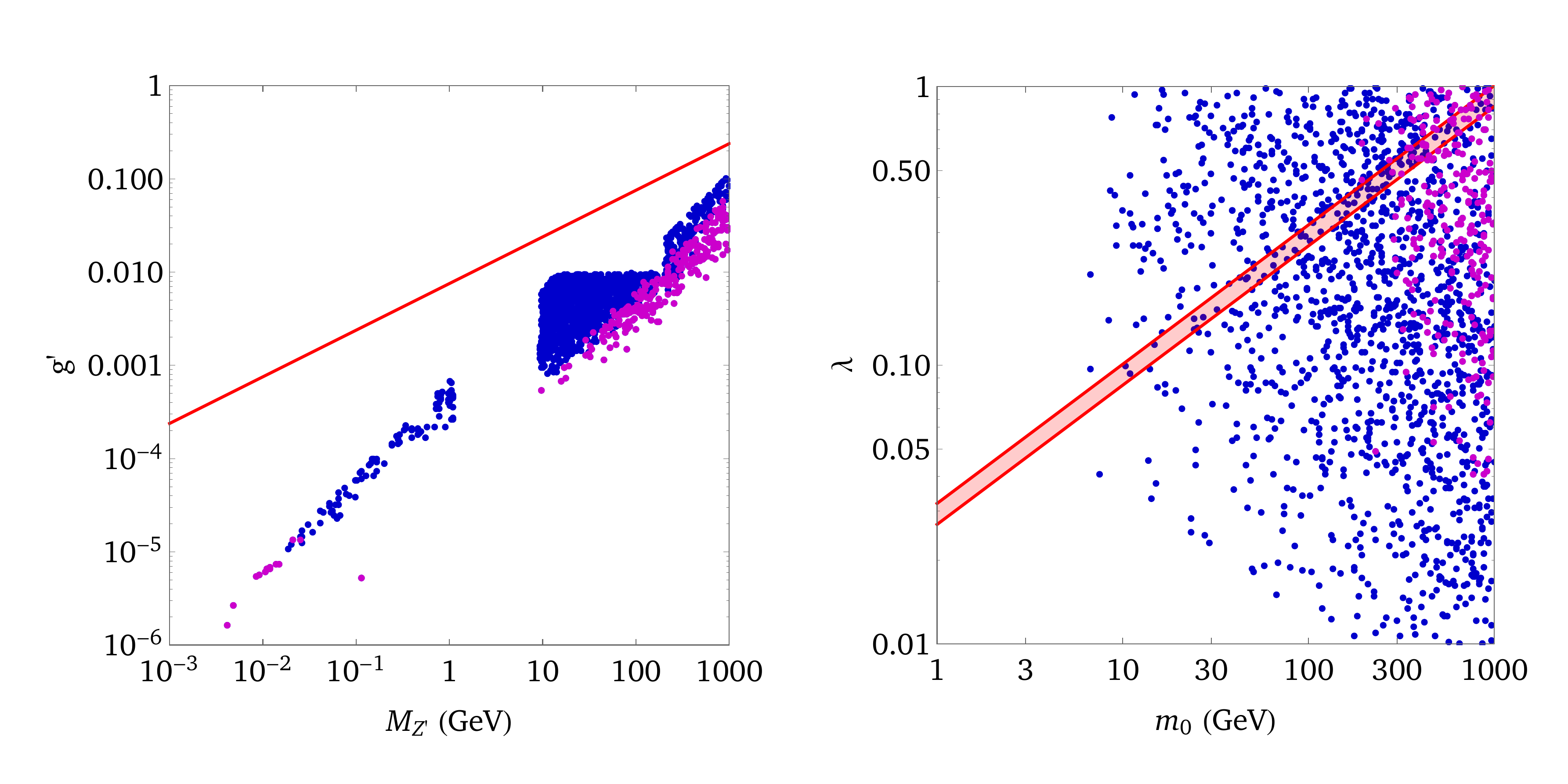}}
\caption{\it Confronting the electroweak baryogenesis favored parameter space (shown by the blue and magenta points) with dark matter observables, assuming the $\chi$ particle, which sources CP violation in baryogenesis, is also the dark matter candidate. All the blue and magenta points in the plots satisfy the constraints on the $Z^\prime$ boson shown in Fig.~\ref{fig:workingpoints}. The magenta points are consistent with both the observed baryon asymmetry and the dark matter direct detection experiments, while the blue points fails to pass the latter constraint. In the left (right) panel, on the red curve (band), the $\chi$ particle could explain the correct relic density through the thermal freeze out mechanism via the annihilation channel $\chi\bar\chi\to Z^\prime Z^\prime$ ($\chi\bar\chi\to rr, aa, ra$).}\label{fig:correlation}
\end{figure}

Let us first consider the annihilation channel $\chi\bar\chi\to Z^\prime Z^\prime$ (upper-left diagram of Fig.~\ref{fig:DM}). The annihilation cross section is~\cite{Krovi:2018fdr}
\begin{eqnarray}\label{ChiChiBartoZpZp}
(\sigma v_{\rm rel})_{\chi\bar\chi\to Z'Z'} &&= \frac{g'^4}{64\pi M_{Z'}^2} \frac{\left( 1 - \frac{M_{Z'}^2}{m_0^2} \right)^{3/2}}{\left(1 - \frac{M_{Z'}^2}{2m_0^2}  \right)^{2}} 
\left[ 18 (2\mathtt{q}+3)^2 + \frac{M_{Z'}^2}{m_0^2} (2\mathtt{q}^2-9)(2\mathtt{q}^2+12\mathtt{q}+9) \right] \nonumber\\
&&\xrightarrow[]{m_0\gg M_{Z'}} \frac{9g'^4 (2\mathtt{q}+3)^2}{32\pi M_{Z'}^2} \ ,
\end{eqnarray} 
where $ v_{\rm rel}$ is the relative velocity between $\chi$ and $\bar\chi$ particles before the annihilation, and in the last step we take the limit that $m_0\gg M_{Z^\prime}$. Requiring that $\chi$ obtains the observed relic abundance~\cite{Aghanim:2018eyx} through this annihilation mechanism, we get
\begin{eqnarray}
g' \simeq \sqrt{\frac{M_{Z'}}{5.9\,{\rm TeV}\times|2\mathtt{q}+3|}} \ .
\end{eqnarray} 
This relation is shown by the red curve in Fig.~\ref{fig:correlation} (left panel), for a particular value of $\mathtt{q}=-3$ (similar results hold for other values of $\mathtt{q}$, as long as $\mathtt{q}$ is of order one).
Comparing with the blue and magenta dots, which are the phenomenologically allowed points for successful baryogenesis (surviving the various constraints in Fig.~\ref{fig:workingpoints}), we find these values of $g^\prime$ are too small to account for the correct dark matter relic density this way, unless the dark matter charge $\mathtt{q}$ value is unnaturally large. Hence, we need larger contributions to the dark matter annihilation cross section from additional channels.

Next, we consider the $s$-channel $Z'$ exchange, as shown by Fig.~\ref{fig:DM} (upper-right diagram), where $\chi\bar\chi$ annihilate into SM charged leptons and neutrinos. The corresponding cross section is (assuming the limit $m_0\gg M_{Z'}$),
\begin{eqnarray}
(\sigma v_{\rm rel})_{\chi\bar\chi\to \ell^+\ell^-, \nu\bar\nu} = \frac{9g'^4(2\mathtt{q}+3)^2}{128\pi m_0^2} \ .
\end{eqnarray} 
Comparing this expression with Eq.~(\ref{ChiChiBartoZpZp}), we find that $(\sigma v_{\rm rel})_{\chi\bar\chi\to \ell^+\ell^-, \nu\bar\nu}$ is not sufficiently large, since it is
parametrically smaller than $(\sigma v_{\rm rel})_{\chi\bar\chi\to Z'Z'}$, for $m_0\gg M_{Z'}$. The latter having an enhancement factor, $m_0^2/M_{Z'}^2$, which arises from $\chi\bar\chi$ mainly annihilating into the longitudinal component of the $Z'$ boson.

Finally, we consider the dark matter annihilation into the dark scalar $S$.  Here we first derive the dark scalar spectrum and its couplings to the dark matter $\chi$. The most general scalar potential of $S$ is given by the sum of Eqs.~(\ref{scalarpotential}) and (\ref{deltaV}). We will focus on the case where in (\ref{deltaV}) only the quadratic term $\mu_S^2 S^2 +{\rm h.c.}$ is present, and the VEV of $S$ relaxes to zero when the dark matter freezes out (which typically occurs at temperatures below the electroweak phase transition). In this case, CP can be violated in the dark sector as explained in Sec.~\ref{sec:CPVsource}. We can first redefine the phases of $S$ and $\chi_{L,R}$ fields so that $m_0$ and $\mu_S$ are real parameters, but the $\lambda_c$ coupling in Eq.~(\ref{darkYukawa}) remains complex in general. As before, we rewrite $\lambda_c= \lambda e^{i\theta_\lambda}$ with $\lambda$ and $\theta_\lambda$ being real parameters. In this basis, the complex scalar $S$ is separated into its real and imaginary parts $S = (r + ia)/\sqrt2$, where $r$ and $a$ are the physical mass eigenstates, with respective masses
\begin{eqnarray}\label{eq:ramasses}
\begin{split}
&M_r^2 = \lambda_{SH} v^2 - 2 \lambda_S v_S^2 + 2\mu_S^2 \ ,\\
&M_a^2 = \lambda_{SH} v^2 - 2 \lambda_S v_S^2 - 2\mu_S^2 \ .
\end{split}
\end{eqnarray} 
Conditions~(\ref{conditions}) and (\ref{shift}) guarantee that both $M_r^2$ and $M_a^2$ are positive. Clearly, the presence of the $\mu_S^2 S^2 +{\rm h.c.}$ potential term breaks the degeneracy between $r$ and $a$, $M_r\neq M_a$. It is then straightforward to rewrite the Yukawa interaction, Eq.~(\ref{darkYukawa}), into those between $r, a$ and the fermion $\chi$, which takes the form
\begin{eqnarray}\label{eq4.8}
\begin{split}
\mathcal{L}_{\rm dark\,Yukawa} & = \lambda e^{i \theta_\lambda} \bar \chi_L \chi_R S + {\rm h.c.}\\
&=\frac{ r}{\sqrt{2}} \left( \lambda \cos\theta_\lambda \bar\chi \chi + \lambda \sin\theta_\lambda \bar\chi i\gamma_5 \chi \right) + \frac{a}{\sqrt{2}} \left( - \lambda \sin\theta_\lambda \bar\chi \chi + \lambda \cos\theta_\lambda \bar\chi i\gamma_5 \chi \right) \ .
\end{split}
\end{eqnarray} 

With these interactions, we calculate the cross sections for $\chi\bar\chi$ annihilating into $rr$, $aa$ and $ra$. The corresponding Feynman diagrams are shown in Fig.~\ref{fig:DM} (upper-middle diagram). The sum of these annihilation cross sections is
\begin{eqnarray}\label{DominantXsec}
(\sigma v_{\rm rel})_{\chi\bar\chi\to rr} +(\sigma v_{\rm rel})_{\chi\bar\chi\to aa} +(\sigma v_{\rm rel})_{\chi\bar\chi\to ra} \simeq \frac{\lambda^4 \left( 3- \cos4\theta_\lambda \rule{0mm}{3.5mm}\right)}{256\pi m_0^2} \ ,
\end{eqnarray} 
where we assume that the final state particles $r$ and $a$ are much lighter than $\chi$. Obtaining the correct relic density for $\chi$ through this channel then requires $\lambda$ to lie within the window
\begin{eqnarray}
\sqrt{\frac{m_0}{1.4\,{\rm TeV}}} < \lambda < \sqrt{\frac{m_0}{1.0\,{\rm TeV}}} \ ,
\label{eq:band}
\end{eqnarray} 
for $0<\theta_\lambda<2\pi$. This relation is derived by assuming the ${\chi\bar\chi\to Z^\prime Z^\prime}$ and $\chi\bar\chi\to \ell^+\ell^-, \nu\bar\nu$ annihilation cross sections discussed above are much smaller than the one in Eq.~(\ref{DominantXsec}), and thus negligible when accounting for the total value of the thermal relic density. Region (\ref{eq:band}) is shown by the red band in Fig.~\ref{fig:correlation} (right panel).  Again, the blue/magenta dots are the phenomenologically viable points obtained from the baryogenesis scan, and now shown in the $\lambda$ versus $m_0$ parameter space. This comparison makes it clear that there exists a viable region in the parameter space where both \textit{successful electroweak baryogenesis and correct dark matter relic density} are achievable. 
The favored region of dark matter mass is around a few hundred GeV.

\subsubsection{The Direct Detection}

Direct detection of dark matter in this model could occur through $Z^\prime$ exchange. However, because the $Z^\prime$ is the gauge boson for lepton number, it does not directly couple to nucleons, implying that the dark matter-nucleon scattering should occur through loop of charged leptons which effectively act as a kinetic mixing between the $Z^\prime$ and the photon, as shown in Fig.~\ref{fig:DM} (lower-left diagram). The corresponding spin-independent cross section for this process is~\cite{Fox:2011fx},
\begin{eqnarray}
\sigma_{\chi p\to\chi p} = \frac{16\alpha^2 \alpha'^2 (\mathtt{q}+3/2)^2 \mu_p^2}{81 \pi \left( q^2 - M_{Z'}^2\right)^2} \left[ \sum_{\ell=e,\mu,\tau} f(q^2, m_\ell) \right]^2 \ ,
\end{eqnarray} 
where $\alpha' = g'^2/(4\pi)$, $\mu_p = m_0 m_p/(m_0+m_p)$ is the reduced mass of the dark matter and target nucleus system ($m_p$ is the proton mass), and
\begin{eqnarray}
f(q^2, m_\ell) = \frac{1}{q^2} \left[ 5q^2 + 12 m_\ell^2 + 6(q^2+2m_\ell^2) \sqrt{1-\frac{4m_\ell^2}{q^2}} {\rm arccoth} \left( \sqrt{1-\frac{4m_\ell^2}{q^2}} \right) +3 q^2 \log\frac{\Lambda^2}{m_\ell^2}  \right] \ , \nonumber\\
\end{eqnarray} 
where $\Lambda$ is the cutoff scale corresponding to the renormalization of the effective $Z'-\gamma$ kinetic mixing. We set $\Lambda=1\,$TeV in our calculation, and assume $\mathtt{q}\sim O(1)$. The typical square momentum transfer of the scattering is of order $q^2 = -4 \mu^2 v^2$, where $v\simeq10^{-3}$ is the typical halo dark matter velocity~\footnote{In the case of Xenon nucleus target, we have $\mu=m_0 m_{\rm Xe}/(m_0+m_{\rm Xe})$, with $m_{\rm Xe} \sim 130 m_p$.}. In Fig.~\ref{fig:correlation}, the points in magenta are compatible with the present dark matter direct detection constraints~\cite{Aprile:2018dbl, Cui:2017nnn, Akerib:2016vxi}, and can generate the observed baryon asymmetry in the universe.

In addition, the dark matter direct detection could also be mediated by the scalar $S$ (or equivalently the $r,a$ mass eigenstates) and the Higgs boson exchange. If $S$ has no VEV today, the dark matter scattering is a loop level process, as shown in Fig.~\ref{fig:DM} (lower-right diagram). In this case, the cross section arises from a loop suppressed Higgs portal interaction and is sufficiently small and can be neglected~\cite{Chowdhury:2011ga, Cline:2013gha}. On the other hand, if $S$ were to have a nonzero VEV, it would mix with the Higgs boson and the dark matter scattering would occur at tree level. In such a case, the direct detection constraints could become important depending on the mass of $S$ and the size of its mixing with the Higgs boson~\cite{Wise:2014jva, Zhang:2015era, Kouvaris:2014uoa}.

\subsection{\bf Implications for Electric Dipole Moments}\label{sec:EDM}

We will comment here on the implications of our baryogenesis model for the electric dipole moment experiments.  It is generically expected that the CP violating interaction between $S$ and $\chi$, required for successful baryogenesis, will propagate at loop level to the Standard Model sector, giving rise to EDMs.

\begin{figure}[h]
\centerline{\includegraphics[width=0.65\textwidth]{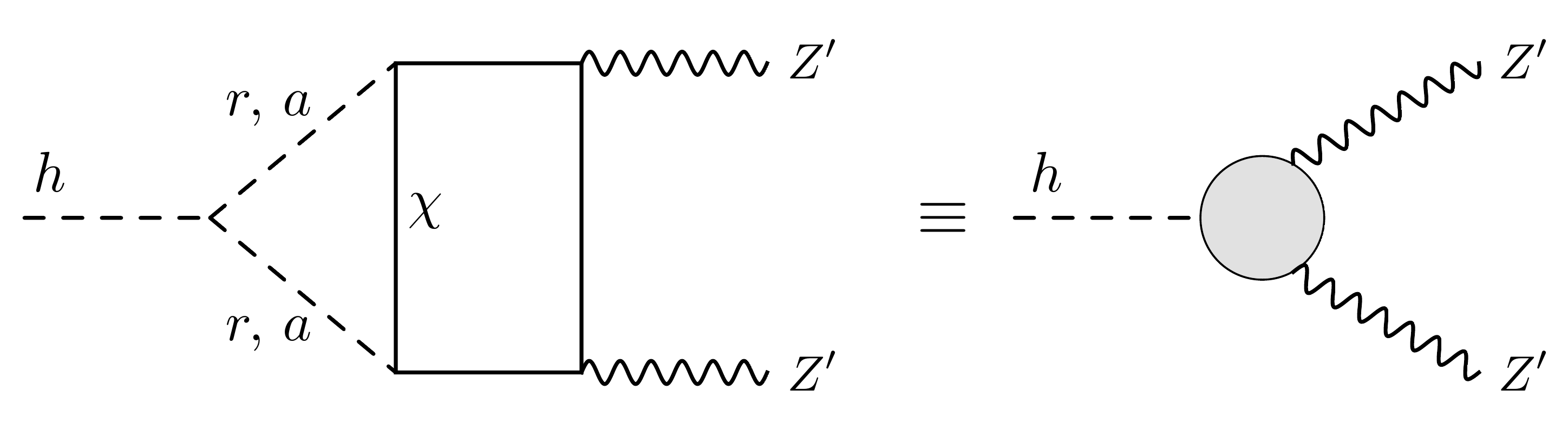}}
\caption{\it Two-loop generated $h Z^\prime_{\mu\nu}\tilde Z^{\prime\mu\nu}$ vertex.}\label{fig:EDM-part1}
\end{figure}

The relevant interaction and mass terms for CP violation in the dark sector are given in Eqs.~(\ref{darkYukawa}) and (\ref{deltaV}). We first consider the case where the VEV of $S$ at zero temperature is zero and only the $\mu_S^2 S^2 +{\rm h.c.}$ term is present in Eq.~(\ref{deltaV}). As explained in Sec.~\ref{DMpheno}, the complex scalar $S$ splits into its real and imaginary parts, yielding the physical mass eigenstates, $r$ and $a$, respectively, and their interactions with dark matter are given by Eq.~(\ref{eq4.8}). If $\theta_\lambda\neq0$, the $r$ and $a$ fields couple to both scalar ($\bar\chi\chi$) and pseudoscalar ($\bar\chi i\gamma_5\chi$) operators involving the $\chi$ fields. At the same time, they also couple to the SM Higgs boson through the Higgs portal interaction, Eq.~(\ref{Higgsportal}),
\begin{eqnarray}\label{eq4.9}
\lambda_{SH} |S|^2 |H|^2 \supset \frac{\lambda_{SH} v}{2} h (r^2 + a^2)\ .
\end{eqnarray}
Then Eqs.~(\ref{eq4.8}) and (\ref{eq4.9}) allow us to derive a CP violating Higgs-$Z^\prime$ operator, of the form $h Z^\prime_{\mu\nu}\tilde Z^{\prime\mu\nu}$, at two loop level, as shown in Fig.~\ref{fig:EDM-part1}. Out of the two vertices where the dark scalars ($r$ or $a$) are attached to the $\chi$ loop, one of them needs to be the scalar coupling in Eq.~(\ref{eq4.8}) and the other the pseudoscalar coupling, so that CP can be violated. The resulting coefficient of the $h Z^\prime_{\mu\nu}\tilde Z^{\prime\mu\nu}$ operator will be proportional to $\lambda^2 \sin\theta_\lambda \cos\theta_\lambda$.

It is worth noting that the non-degeneracy between $r$ and $a$ is the key for the coefficient of this operator to be nonzero, otherwise  the coupling structure in (\ref{eq4.8}) would lead to a complete cancellation between the two diagrams involving $r$ and $a$, respectively. This cancellation could also be understood from a symmetry argument.  Based on the discussions in Sec.~\ref{sec:CPVsource}, if the $\delta V$ potential (containing $\mu_S^2S^2$ term) vanishes, thus leading to degenerate $r$ and $a$ fields, there is no CP violation in the dark sector  --- all the parameters can be made real by field redefinitions --- and there is no contribution to any CP violating operators.

\begin{figure}[h]
\centerline{\includegraphics[width=0.39\textwidth]{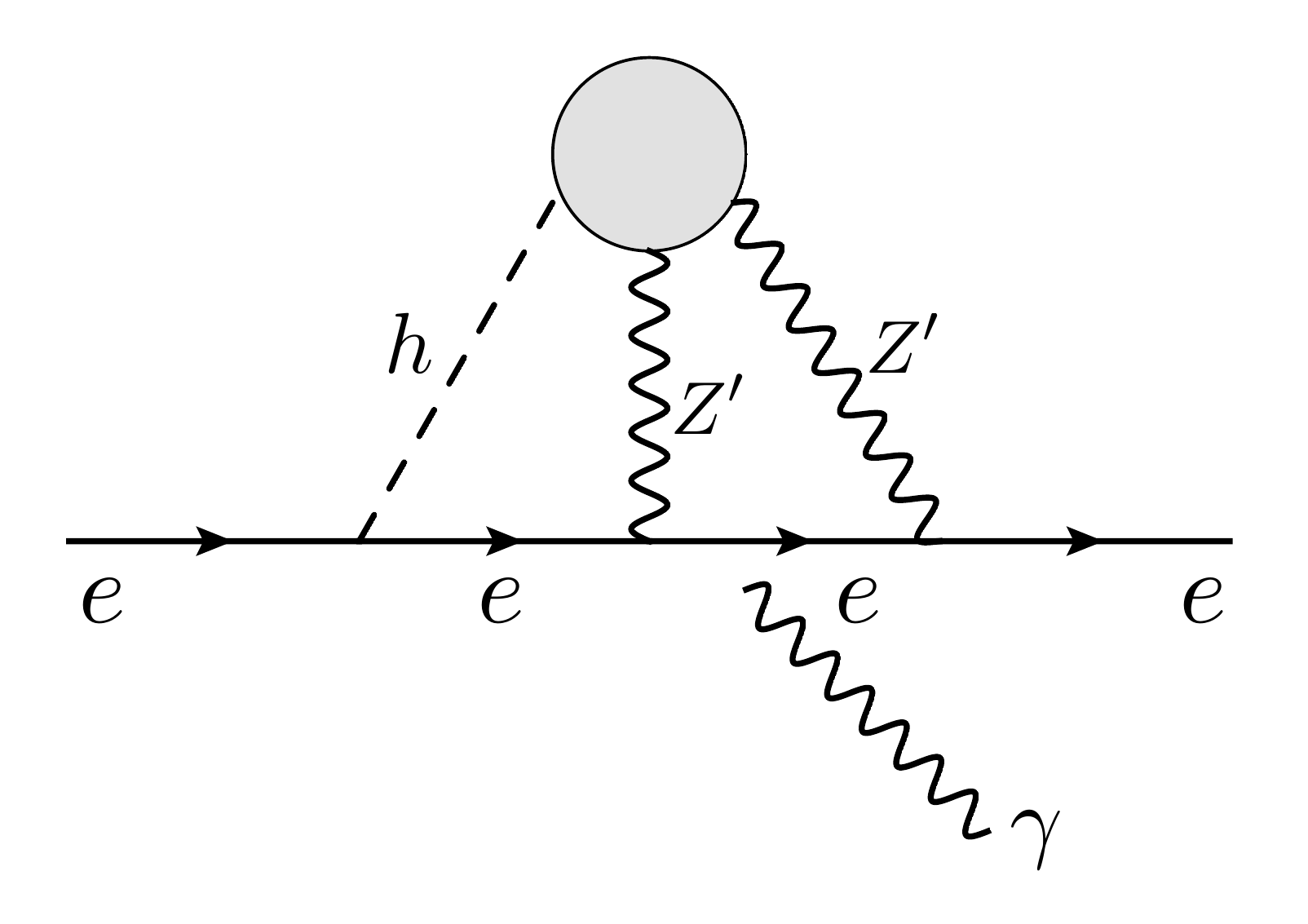}}
\caption{\it Two-loop generated electron EDM, from the $h Z^\prime_{\mu\nu} \tilde Z^{\prime\mu\nu}$ vertex (represented by the gray blob).
In our model, the $h Z^\prime_{\mu\nu} \tilde Z^{\prime\mu\nu}$ is generated at two loop level, see Fig.~\ref{fig:EDM-part1}.
The photon must be radiated from one of the internal propagators and that has to be an electron propagator because everybody else is electrically neutral.}\label{fig:EDM-part2}
\end{figure}

In the presence of dark sector CP violation, when the contribution to the $h Z^\prime_{\mu\nu}\tilde Z^{\prime\mu\nu}$ operator is nonzero, we could use it to further generate the EDM for the electron, at the price of another two loops, as shown in Fig.~\ref{fig:EDM-part2}. Unlike the Barr-Zee type diagrams for EDMs~\cite{Barr:1990vd}, 
here we must attach both $Z^\prime$s to the electron line and the external photon to either of the internal electron propagators. 

By simple power counting, the resulting electron EDM is
\begin{eqnarray}\label{eq:4.10}
d_e \sim \frac{e\, G_F m_e}{(16\pi^2)^4} (\lambda_{SH} \lambda^2 g'^4 \mathtt{q}^2) \sin (2\theta_{\lambda}) \lesssim 10^{-30}  (\lambda_{SH} \lambda^2 g'^4 \mathtt{q}^2) \sin (2\theta_{\lambda}) \, e\,{\rm cm} \ .
\end{eqnarray} 
This estimate is valid assuming that the $r$ and $a$ mass difference is around the electroweak scale. With the factor $(\lambda_{SH} \lambda^2 g^{\prime 4} \mathtt{q}^2)<1$,
the resulting electron EDM is well below the current upper bound on $d_e$, which comes from the ACME experiment~\cite{Andreev:2018ayy}: $d_e<1.1\times 10^{-29} \, e\,{\rm cm}$.  As mentioned in the introduction, this is an appealing feature of our model for electroweak baryogenesis which, unlike many others, is safe from the EDM constraints, even if the CP phase is of order one.

Finally, we comment on the case where the VEV of $S$ at zero temperature is non-zero. In this case, from the Higgs portal interaction, Eq.~(\ref{Higgsportal}), there is a direct mixing between $r$ and $h$ fields. As a result, the $h Z^\prime_{\mu\nu}\tilde Z^{\prime\mu\nu}$ vertex could be generated by replacing the scalar loop in Fig.~\ref{fig:EDM-part1} by the $r-h$ mixing, with only one $r$ attached to the fermion loop via the pseudoscalar coupling, which becomes a one-loop diagram.
The contribution to the electron EDM in this case reduces to three loops,
\begin{eqnarray}
d_e \sim 10^{-28} \left(\lambda_{SH} v v_S/M_r^2 \right) \sin \theta_{\lambda} \, e\,{\rm cm} \ ,
\end{eqnarray} 
where the factor $(\lambda_{SH} v v_S/M_r^2)$ is the mixing between $r$ and $h$. The Higgs boson rate measurements at the LHC requires this mixing must be less than $\lesssim 20\%$~\cite{Carena:2018vpt, Khachatryan:2016vau}. This implies that $d_e \lesssim 10^{-29} \sin\theta_{\lambda} \, e\,{\rm cm}$, allowing the predicted EDM to be closer to the current upper bound and giving a prospect for future electron EDM searches.

\subsection{\bf Possible LHC Signals of the Dark Scalar(s)}\label{sec:LHC}

\begin{figure}[h]
\centerline{\includegraphics[width=0.9\textwidth]{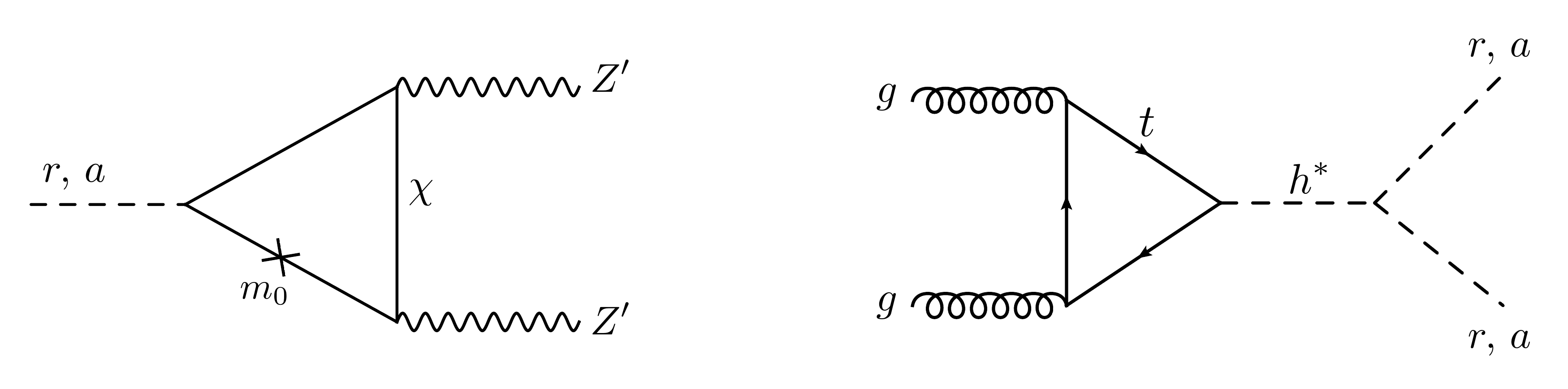}}
\centerline{\includegraphics[width=0.56\textwidth]{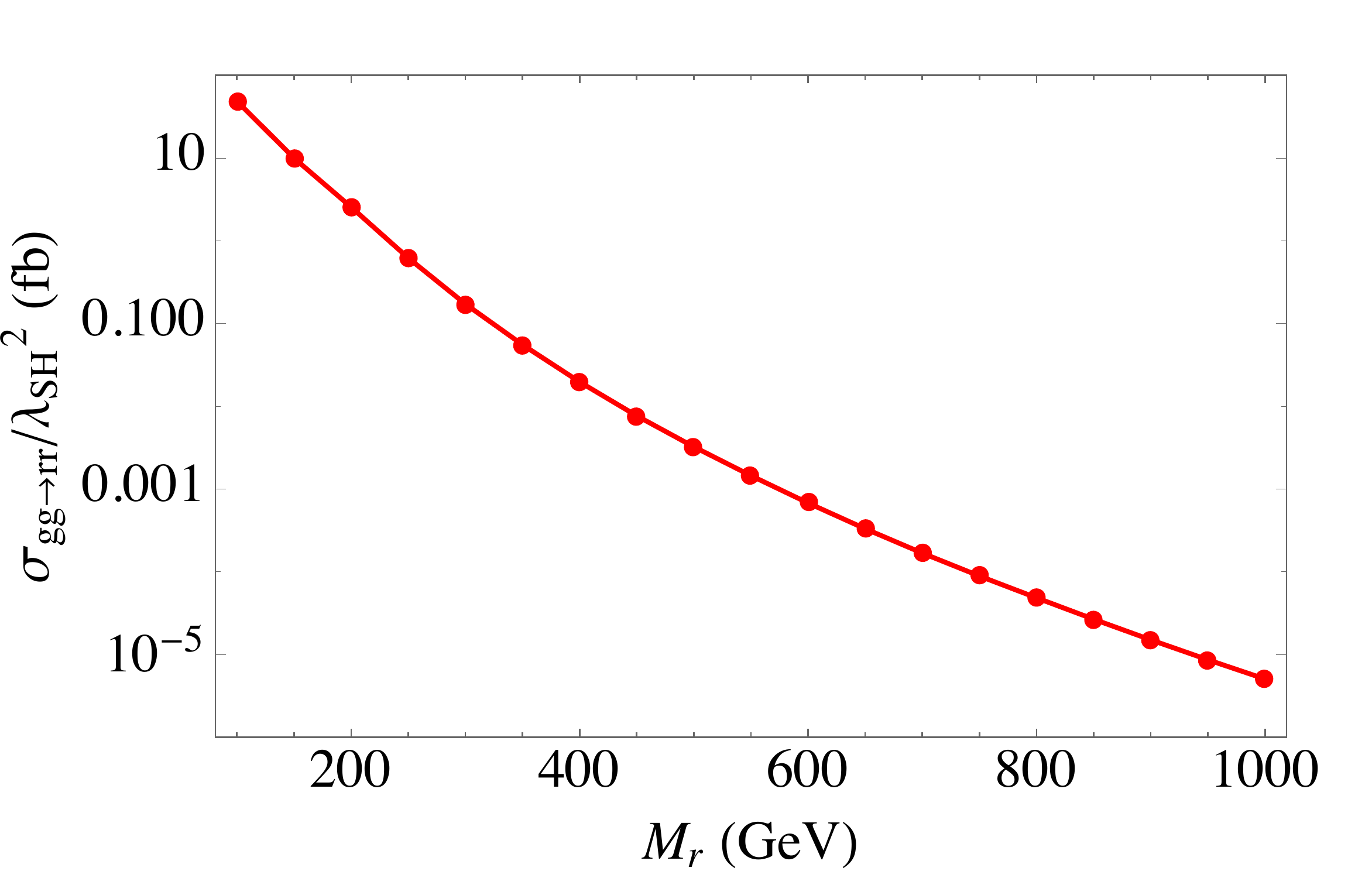}}
\caption{\it Feynman diagrams for the loop induced decay of $r, a$ into two $Z^\prime$ bosons (upper left) and the production process $gg\to rr$ (or $aa$) via an off-shell Higgs boson (upper right).
The cross section for the latter at $\sqrt{s}=13$ TeV LHC is shown in the lower panel.}\label{fig:Sdecay}
\end{figure}

In this subsection, we comment on the possible collider signals of the new scalar $S$ in our model. Unlike the electroweak phase transition discussion, where only the $S$ field background is relevant, here we consider the $S$ excitations, being produced as particles. As mentioned in Sec.~\ref{sec:EDM}, the physical states from the $S$  field are its real, $r$, and imaginary, $a$, parts, which have different masses. Their interactions with $\chi$ are given by Eq.~(\ref{eq4.8}), thus, if kinematically allowed, they could dominantly decay into $\chi\bar\chi$. However, as discussed in Sec.~\ref{sec:DMsearch}, for the dark matter $\chi$ to freeze out effectively we need $r$ and $a$ to be lighter than $\chi$. In this case, they have to decay via a loop of $\chi$ into a pair of $Z^\prime$ bosons, as shown in Fig.~\ref{fig:Sdecay} (upper left panel)~\footnote{A similar diagram makes in the standard model the ``golden channel" decay $h\to \gamma\gamma$ via a top quark loop.}. This could lead to a potentially interesting signature because the $Z^\prime$ boson, which is typically lighter than $\chi$ (necessary for successful baryogenesis), has to decay into SM charged leptons or neutrinos. Each decaying $r$ or $a$ could then produce as many as four charged leptons.

There is important information about the model in these charged lepton decay products. First, each pair of the charged leptons sit on the $Z^\prime$ resonance, so their invariant masses all line up in the same energy bin corresponding to the $Z^\prime$ mass. Moreover, because $r$ (and $a$) has both CP even and odd couplings with $\chi$, the effective operators for its decay (after integrating out $\chi$ in the loop) are $r Z^\prime_{\mu\nu}Z^{\prime\mu\nu}$ and $r Z^\prime_{\mu\nu}\tilde Z^{\prime\mu\nu}$. The interference of the two decay amplitudes allows us to probe CP violating observables in the final state charged lepton angular distributions, in analogy to using the ``golden-channel'' of the Higgs decay to probe CP violation~\cite{Accomando:2006ga, Chen:2014gka}.

For the production of the new scalars $r, a$, we resort to the Higgs portal interaction, Eq.~(\ref{Higgsportal}) or (\ref{eq4.9}). If the $S$ field has no VEV today, there is a $\mathbb Z_2$ symmetry at this vertex which requires that $r$ or $a$ must be pair produced. This may occur at the LHC, or a prospective future hadron collider, through  the gluon fusion process that creates an off-shell Higgs boson,  which later on splits into two $r$ (or $a$) particles, as shown in Fig.~\ref{fig:Sdecay} (upper right panel). The corresponding production cross section at the LHC is shown in the lower panel of Fig.~\ref{fig:Sdecay}. 
Quantitatively, $\sigma_{gg\to rr,aa}\sim 10\lambda_{SH}^2$ fb ($\sim 0.1\lambda_{SH}^2$ fb) for $M_{r,a}\simeq 150$ GeV (for $M_{r,a}\simeq 300$ GeV).
After the decays of the $r$ (or $a$) scalars, the final state could contain as many as 4 pairs of charged leptons, which would provide  a very striking signal. A recent analysis~\cite{Izaguirre:2018atq} has shown that the multi-lepton final state data from the LHC~\cite{Sirunyan:2017lae} could already set  useful limits on dark sector models. Comparing the production cross section shown in Fig.~\ref{fig:Sdecay} with the limits derived in~\cite{Izaguirre:2018atq}, we find that the existing LHC data could already cover the region where the dark scalar ($r$ or $a$) is lighter $\sim 200\,$GeV for $\lambda_{SH} \sim \mathcal{O}(1)$.

Finally, we comment on the case where $S$ has a nonzero VEV today. A nonzero VEV of $S$ allows $r$-Higgs boson mixing implying that, in addition to the above pair production mode, $r$ may be  singly produced through mixing via   the gluon fusion channel.  There are two possibilities to consider: a) the Higgs boson is produced off-shell and subsequently mixes with $r$, that is produced on shell as a new resonance and decays to a  $Z^\prime$ pair at tree level, leading to 4 leptons in the final state. This is an interesting signature to be explored.
The new $r$ resonance can also decay to SM final states, but this will be further suppressed by an additional $r-h$ mixing factor. b) The Higgs boson can be produced on shell and its decays can be modified through its mixing with $r$.
Importantly,  this  has a direct impact on precision measurements of the SM-like Higgs boson,
by modifying the Higgs couplings to SM particles, allowing for Higgs exotic decays, and affecting the di-Higgs production  rate. In particular, the current bound~\cite{Aaboud:2018fvk} on Higgs exotic decay $h\to 2Z'\to 4\ell$ is consistent with an order one $r-h$ mixing, for $g'\lesssim 10^{-2}$ and $v_S\lesssim 100\,$GeV. This region of parameter space is just below the  LEP bound shown in Fig. \ref{fig:workingpoints} and is an interesting benchmark for future collider searches.

%

\section{The Case of Gauged $L_\mu+L_\tau$}\label{sec:G2}

In this section, we consider another incarnation of the gauged $U(1)_\ell$ model where only two lepton flavors  are gauged, $\ell=L_\mu+L_\tau$,   $N_g=2$. We will comment on the differences and similarities for  the EWBG predictions, as well as the phenomenological implications  between this two flavor case and the previously studied three flavor case with 
 $\ell=L_e+L_\mu+L_\tau$.

\begin{figure}[htb]
\centerline{\includegraphics[width=0.7\textwidth]{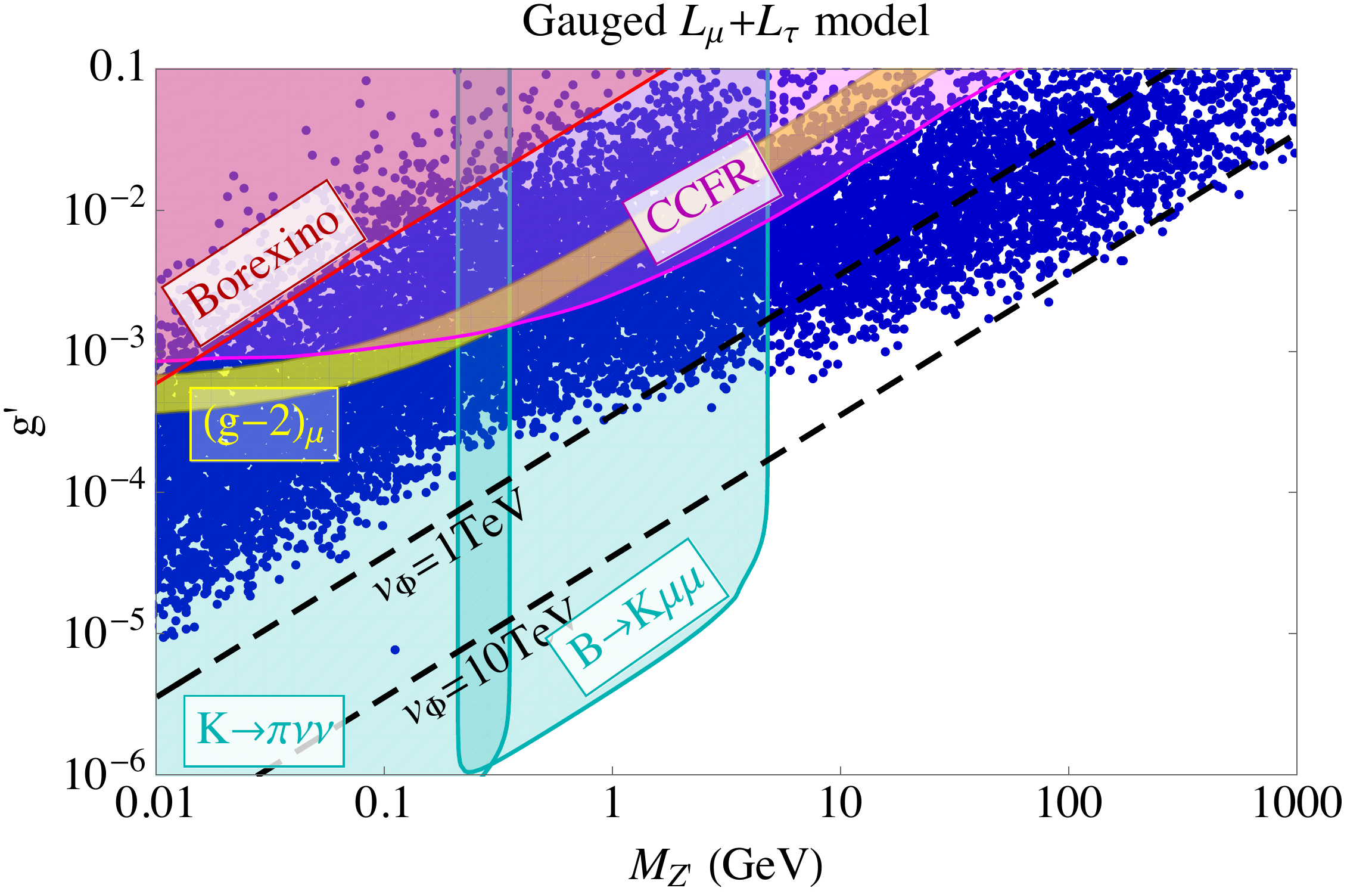}}
\caption{\it
Scanned points (blue) in the $g^\prime$ -- $M_{Z^\prime}$ plane, compatible with the observed baryon asymmetry of the universe assuming $N_g=2$ . The colorful shaded regions have been excluded by the existing constraints from the CCFR, Borexino experiments, and the $K\to \pi \nu\bar\nu$ and $B\to K\mu\mu$ decay rate measurements, respectively. The yellow band is the favored region for explaining the muon $g-2$ anomaly. The black dashed lines correspond to $v_\Phi$ equal 1 and  10\,TeV; two indicative values related to the anomalon masses
that need to be above the electroweak scale.
}\label{fig:workingpoints2G}
\end{figure}

The previous  discussion on our recently proposed EWBG mechanism in Sec.~\ref{EWZprimeBaryogenesis} has assumed a generic value of $N_g$. The parametric dependence of the final baryon asymmetry to entropy ratio is given by
\begin{equation}
\eta_B = \frac{\Delta n_{B}}{s} \propto \frac{g'^2 N_g^2 T_c^3 L_\omega \alpha_W^5}{M_{Z'}^2 v_\omega} \ ,
\end{equation}
from where one observes that, for a fixed value of $M_{Z'}$,  it scales as $g'^2 N_g^2$  {\it i.e.}~the favored values of $g'$ in the $N_g=2$ case will be $\sim 1.5$ times larger than those in the $N_g=3$ case. In Fig.~\ref{fig:workingpoints2G}, the blue points show the EWBG favored region of parameter space in the $g'$ versus $M_{Z'}$ plane, obtained by scanning  over the model and phase transition parameters given by  Eq.~(\ref{eq:scanrange}). This figure is the analogous to  Fig.~\ref{fig:workingpoints} for the $N_g=2$ case.

Experimentally, the gauged $L_\mu+L_\tau$ model is interesting because the $Z'$ does not couple to electrons at tree level. This helps to avoid most constraints discussed in Sec.~\ref{sec:Z'search}.
There are, however, relevant constraints from neutrino trident production (CCFR)~\cite{Altmannshofer:2014pba} and loop-induced solar-neutrino-electron scattering (Borexino)~\cite{Araki:2017wyg, Bauer:2018onh} which exclude the correspondingly labeled shaded regions in Fig.~\ref{fig:workingpoints2G}. In this model, the Borexino experiment stands out to be the most important neutrino scattering experiment because the solar neutrino contains a $\nu_\mu$ component. 
Like the $L_e+L_\mu+L_\tau$ case, for small $M_{Z'}$, this model is also strongly constrained by flavor-changing meson decays due to the anomalous $Z'WW$ coupling~\cite{Dror:2017ehi,Dror:2017nsg}.
The measurement of $K\to \pi \nu\bar\nu$ and $B\to K\mu\mu$ decay rates have already excluded the cyan shaded region in Fig.~\ref{fig:workingpoints2G}.
A prospective high-energy electron-positron collider could probe the viable region of $Z'$ masses via the multi-muon searches, similar to limit set by BaBar (not shown in the Fig.~\ref{fig:workingpoints2G} because it is superseded by CCFR.)~\cite{TheBABAR:2016rlg}.


In view of neutrino cosmology, the gauged $L_\mu+L_\tau$ model has an attractive aspect where the scalars $\Phi$ and $S$ both carry $U(1)_\ell$ charge 2. This allows them to directly give Majorana masses to the right-handed neutrinos, which is necessary for being consistent with the $\Delta N_{eff}$ bound in cosmology and keeping the $Z'$ sufficiently light, as discussed in Sec.~\ref{sec:nuCosmo}. However, with the minimal particle content given in Table~\ref{tab:table}, the gauged $L_\mu+L_\tau$ model cannot generate realistic active neutrino masses and mixings. This is mainly because the electron neutrino in this model is not charged under the $U(1)_\ell$, which forbids it to mix with the $\mu$ and $\tau$ flavors unless a charge one  scalar (named $S'$) under $U(1)_\ell$, with a non-vanishing VEV, is introduced. The relevant Yukawa interactions, and Majorana mass terms, accounting for realistic neutrino masses and mixings take the form
\begin{equation}
Y_{\nu}^{ee} \bar L_e \tilde H \nu_{Re} + \sum_{\alpha,\beta=\mu,\tau} Y_{\nu}^{\alpha\beta} \bar L_\alpha \tilde H \nu_{R\beta} + M_{ee} \bar \nu_{Re}^c \nu_{Re}  + \sum_{\alpha =\mu,\tau} Y''_{e\alpha} S' \bar\nu_{Re}^c \nu_{R\beta} + \sum_{\alpha,\beta=\mu,\tau} Y'_{\alpha\beta} \Phi \bar \nu_{R\alpha}^c \nu_{R\beta} + {\rm h.c.} \ ,
\end{equation}
where we also have to introduce an electron flavored right-handed neutrino $\nu_{Re}$ which is a $U(1)_\ell$ singlet and can have a bare Majorana mass $M_{ee}$. 

The dark matter phenomenology in the gauged $L_\mu+L_\tau$ model is similar to that discussed in Sec.~\ref{sec:DMsearch}, except that there could be an additional annihilation channel $\chi\bar\chi\to \nu_R\nu_R$ through an $s$-channel $\Phi$ or $S$ exchange, if kinematically allowed, as their $U(1)_\ell$ quantum numbers match for $N_g=2$. These new annihilation channels introduce additional model dependence in the relic density calculations.

Finally, the contribution to electron EDM in the gauged $L_\mu+L_\tau$ model is suppressed compared to the gauged $L_e+L_\mu+L_\tau$ case, by the absence of $Z'$-electron coupling.

\section{The Case of Gauged Baryon Number $B$}\label{sec:GB}

In this section we will comment on an alternative $U(1)$ extension of the Standard Model where the new electroweak baryogenesis mechanism proposed in this work could also work. Here we will consider gauging the baryon number, $U(1)_B$, instead of the lepton number, under which the SM quarks carry charge 1/3 but leptons are neutral. An interesting observation is that the same new fermion content as in Tab.~\ref{tab:table} could also cancel all $U(1)_B$ gauge anomalies, where the $L'_L$, $e'_R$, $\chi_R$, $L''_R$, $e''_L$, $\chi_L$ fields carry, under $U(1)_B$, the same charges assigned in Tab.~\ref{tab:table}, Ref.~\cite{Duerr:2013dza}\footnote{ We keep the same notation as for $U(1)_l$, in spite of the fact that these new states carry baryon number. Observe they are all color singlets.}. On the other hand, the right-handed neutrinos $\nu_{R}^i$, are now neutral under $U(1)_B$ and they are just introduced for the purpose of giving mass to the neutrinos. An immediate consequence of this setup is that, without participating in the new $U(1)_B$ interactions, the $\nu_{R}^i$'s will not be thermalized in the early universe. Therefore, unlike the $U(1)_\ell$  case, the Dirac neutrino mass scenario is consistent with the cosmological constraints on $\Delta N_{eff}$ in the gauged $U(1)_B$ model.

\begin{figure}[ht]
\centerline{\includegraphics[width=0.7\textwidth]{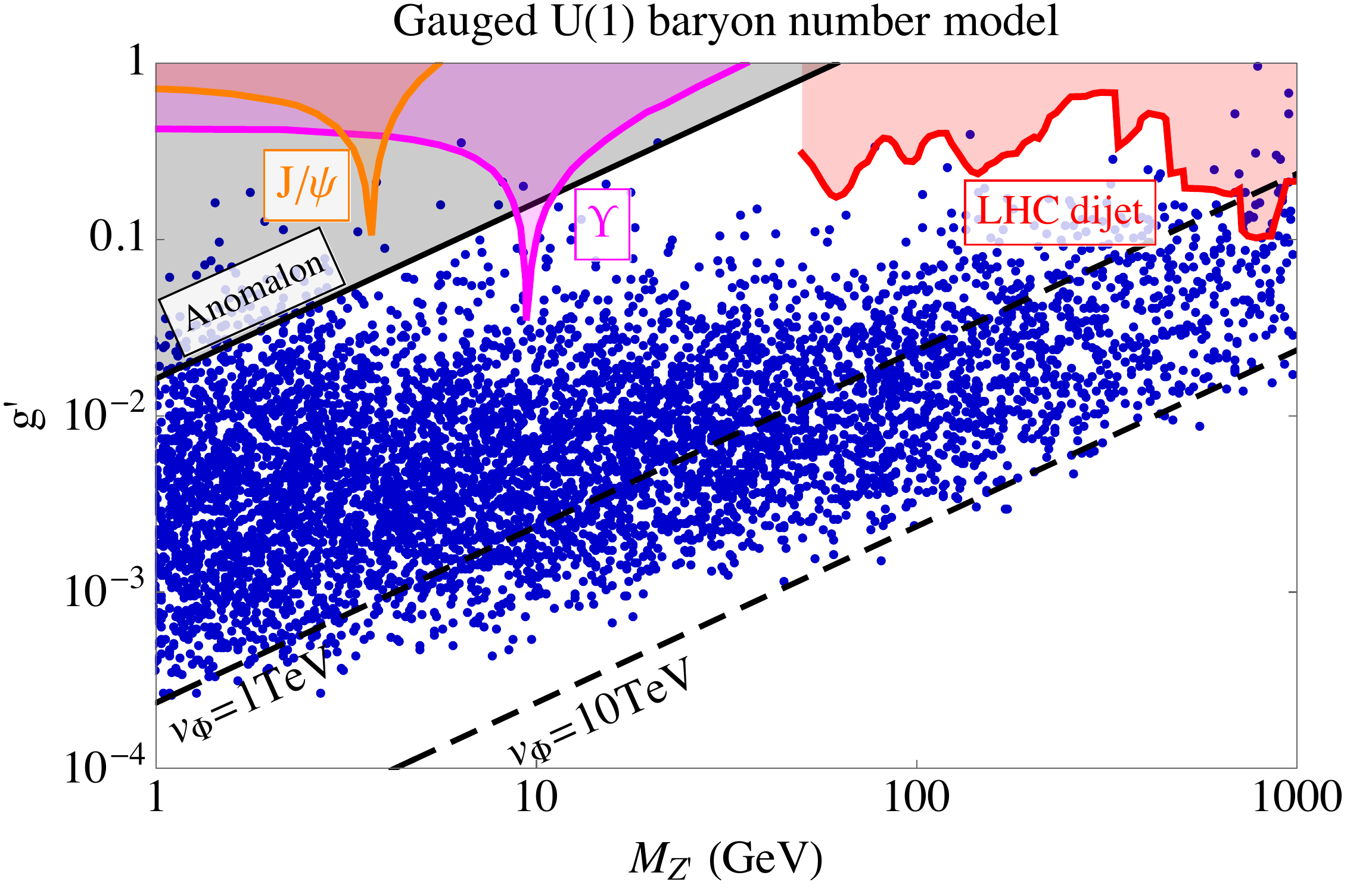}}
\caption{\it The parameter space of the gauged $U(1)_B$ model that could generate the observed baryon asymmetry of the universe (blue points), in the $g^\prime$ -- $M_{Z^\prime}$ plane. The colorful shaded regions have been excluded by the existing constraints from LHC dijet searches (red), hadronic width of $\Upsilon$ (magenta) and $J/Psi$ (Orange). The gray shaded region is the minimally excluded region by the LEP bound on electric charged anomalon fields, assuming their Yukawa couplings with the VEV $v_\Phi$ is near the perturbative limit $\sqrt{4\pi}$. The black dashed lines correspond to  $v_\Phi$ equal to 1, 10\,TeV, 
}\label{fig:workingpointsB}
\end{figure}

For electroweak baryogenesis, the baryonic $Z'_0$ background could still be generated from the $\chi$-bubble-wall interaction, which now serves as the baryon number chemical potential for the SM quarks, instead of leptons as in the $U(1)_\ell$ models. As a result, the Boltzmann equation Eq.~(\ref{EWsph}) will become directly one for the baryon asymmetry, with the replacement $\Delta n_L \to \Delta n_B$, the thermal equilibrium asymmetry $\Delta n_B^{\rm EQ}$ being identical to Eq.~(\ref{eq:ThermalLeptonAsymmetry}).  It is worth noting that the baryon charge factor 1/3 for quarks is now compensated by the number of colors. The existing constraints on the baryogenesis viable parameter space are shown in Fig.~\ref{fig:workingpointsB}. The baryogenesis viable parameter space in this model is the same as the blue points shown in Fig.~\ref{fig:workingpoints}, except for a different set of experimental constraints on the baryonic $Z'$~\cite{An:2012va, Dobrescu:2013coa, FileviezPerez:2018jmr}. In particular, the LHC constraints on the baryonic $Z'$-quark coupling is much weaker than the LEP constraint on leptophilic $Z'$~\cite{Sirunyan:2017nvi, ATLAS:2016xiv, Sirunyan:2016iap, Aaboud:2017yvp}. This allows a wider window for our EWBG mechanism to be successful. 


Because the $Z'$ in this case only couples to quarks, the dark sector CP violation will dominantly contribute to quark EDMs instead of the electron EDM, which are less severely constrained.

Like the gauged $U(1)_\ell$ model, here the dark fermion $\chi$ could still be a thermal dark matter candidate. Its annihilation channels are similar to those depicted in Fig.~\ref{fig:DM}, except that the annihilation final states will be quarks instead of leptons. On the other hand, direct detection constraints become much stronger because in the gauged $U(1)_B$ model the $Z'$ directly couples to quarks and the dark-matter-nucleon scattering now occurs at tree level.
For generic values of $\mathtt{q}$ of order one, the current direct detection limit on spin-independent dark-matter-nucleon scattering cross section implies $v_\Phi\gtrsim20\,$TeV. This constraint is in tension with most of the EWBG favored points in Fig.~\ref{fig:workingpointsB}. 
A possible way to alleviate this tension is to choose $\mathtt{q}=-3/2$ in which case the dark-matter-$Z^\prime$ coupling becomes an axial current interaction and the corresponding dark-matter-nucleon scattering is suppressed by the incoming dark matter velocity in the galactic halo.

Analogous to previous cases, because the $Z'$ couples to an anomalous current with respect to $SU(2)_L^2$ in the low energy theory, it makes contributions to flavor-changing meson decays such as $K\to \pi Z'$ and $B\to K Z'$ as shown in Ref.~\cite{Dror:2017ehi,Dror:2017nsg}. However, in the $U(1)_B$ model the  $Z'$ dominantly decays into quarks and antiquarks, while the decay into charged leptons could only occur through a $Z'\gamma$ kinetic mixing, and is subdominant if the kinetic mixing is generated at loop level.
As a result, the corresponding flavor-changing constraints are much weaker and do not appear in the range shown in Fig.~\ref{fig:workingpointsB}.
%


\section{Conclusion}\label{conclusions}

\label{sec:conclusion}

One of the main challenges to electroweak baryogenesis models is that the required amount of CP violation can be at odd with the improved limits on the electron and neutron electric dipole moments. In this work, we propose a model where electroweak baryogenesis is triggered by a CP violating dark sector. During the electroweak phase transition, the CP violating effect is transferred from the dark to the visible sector at tree level via the background of a $Z'_0$ gauge boson, whereas at zero temperature the transmission of  CPV effects could be suppressed up to four loop level. This mechanism helps to alleviate the otherwise severe EDM constraints on the viable baryogenesis parameter space.

The $U(1)_\ell$ model we have considered is based on a gauged lepton number symmetry, where the anomaly cancellation condition requires extending the SM sector with new fermions carrying lepton number. The lightest of these fermions plays the role of dark matter. After the spontaneous breaking of the gauged lepton number, once  all the new  fermion fields (the anomalons)  -~with the exception of the dark matter candidate~- are integrated out, the fermion content of the effective theory contains all SM fermions, right handed neutrinos and the dark matter.  The force carrier of the new gauge interaction, $Z^\prime$, couples to the lepton number current involving all fermions in the effective theory, which is anomalous with respect to $SU(2)_L$ -- a key ingredient for the baryogenesis mechanism to work.

To achieve a first order electroweak phase transition we introduce a SM singlet $S$ in the dark sector, 
 which couples to the Higgs boson portal and may allow for a two-step phase transition in the early universe. 
 Similar studies in the literature have shown that after  an initial transition from a trivial vacuum state $(v_S,0)$ at very high temperatures, it is possible to trigger a strong first order transition to the electroweak vacuum $(0,v)$,  thereby 
 creating the out-of-equilibrium condition necessary for baryogenesis. 
 A detailed analysis of the phase transition history and its relation to the proposed mechanism for electroweak baryogenesis will be presented elsewhere.

The role of the dark sector CP violation in our baryogenesis mechanism for the $U(1)_\ell$ model can be summarized in the following steps:
\begin{enumerate}
\item
CP is first violated in the dark sector, containing the $\chi_{L,R}$ fermions. Their mass term has an irreducible phase that becomes time-dependent only during the first-order electroweak phase transition, involving both the Higgs field and the dark scalar $S$, as described above.
\item
This time dependent CP violating mass generates particle chiral asymmetries for $\chi_{L,R}$ in the dark sector, which diffuse to the exterior of the bubble wall, where SM sphalerons are active.
\item
By model construction, $\chi_L$ and $\chi_R$ carry different $U(1)_\ell$ charges. As a result, their chiral asymmetries generate a net $U(1)_\ell$ charge density near the wall, that yields a Coulomb background for the $Z^\prime_0$ gauge field.
\item
Given that the gauge field $Z^\prime_0$ couples, not only to the dark sector leptons but also to the SM leptons, it generates a chemical potential for the SM leptons.
\item
In the presence of sphaleron processes, which are active outside the bubble, 
the SM lepton number asymmetry will evolve towards its equilibrium value set by the above chemical potential. 
\item
As sphalerons preserve $B-L$, that originally was zero,  they can change the generated SM lepton number into baryon number. Hence,  a baryon number asymmetry will be equally generated.
\item
Inside the bubbles the sphaleron processes  are suppressed, and  the baryon asymmetry generated  at the phase transition is not washed out. This process sets  the baryon asymmetry as an input for the initial condition in standard cosmology.
\end{enumerate}

As for the phenomenology of the present model, the contributions to EDM are highly suppressed, below the present experimental limits, and we do not expect to see a positive signal in the next generation of experiments.
Instead, one of our main predictions, in particular for the $U(1)_\ell$ model, is a leptophilic $Z'$ boson with mass below the TeV scale.  The lighter the $Z'$, the more weakly coupled it should be, as shown in Fig.~\ref{fig:workingpoints}.
It serves as a very well-motivated target for a number of searches at near future and prospective experiments, such as BELLE II, NA64 ($\mu$ mode) and SHiP, as well as a possible Higgs factory.

Accommodating a dark matter candidate within this new EWBG mechanism provides an additional handle in probing this idea. Concerning the fermion candidate $\chi$ to dark matter, we show that the annihilation cross sections involving the new force carrier $Z'$ are too small.  However the dark matter annihilation into the new scalar $S$ comes to the rescue, yielding the correct relic abundance via thermal freeze out.  Direct detection experiments also yield important information on the parameter space compatible with EWBG. The most relevant, straightforward  contribution comes from the $Z'$ exchange which, given the leptophilic nature of this new gauge boson in the  $U(1)_\ell$ model, implies that dark matter scattering occurs at loop level.  Future direct dark matter searches, with an improvement of about two order of magnitude over present bounds, will provide an important test of the viable parameter space in the $U(1)_\ell$ model of EWBG.

Finally,  there are novel collider signals from the new additional scalar $S$, which can be pair produced via an s-channel off-shell Higgs boson, or singly produced through mixing with the Higgs boson. The former, pair-production mode could lead to 8 charged lepton final states from the decays of the $Z'$s. The latter, single-production mode, instead, could yield 4 charged leptons. For both cases, one could reconstruct the $Z'$ mass from the invariant mass of the charged lepton pairs. 
The new scalar $S$ can also be virtually produced via mixing with the Higgs boson, altering the Higgs boson phenomenology.
Current bounds on the Higgs boson exotic decays still allow for a large region of parameter space compatible with our EWBG mechanism, and provide interesting opportunities for near-future searches in the Higgs decay to $Z'Z'$ when kinematically allowed.

Similar, corresponding, comments should apply to the $U(1)_B$ model after replacing $L$ by $B$ and leptons by quarks.
However, for the DM candidate $\chi$ in the $U(1)_B$ case, already present direct detection constraints make the scenario quite challenging.
Nevertheless observe that it is possible for $\chi$ to be only a fraction of the total dark matter in the universe. In that case, the direct detection bounds, as computed here for any of the models, would become less stringent.


 \subsection*{Acknowledgments}
We thank Zackaria Chacko, James Cline, Bogdan Dobrescu, Bhaskar Dutta, Pavel Fileviez Perez, Paddy Fox, Ian Low, David Morrissey and Tim Tait for useful discussions and correspondence. We are also grateful to Julian Heeck, Alexis Plasencia, and especially Jeff Dror, for very useful comments on the first version of this paper. This manuscript has been authored by Fermi Research Alliance, LLC under Contract No. DE-AC02-07CH11359 with the U.S. Department of Energy, Office of Science, Office of High Energy Physics.   The work of M.Q. is partly supported by Spanish MINEICO under Grant CICYT-FEDER-FPA2014-55613-P and FPA2017-88915-P, by the Government of Catalonia under Grant 2017SGR1069 and by the Severo Ochoa Excellence Program of MINEICO under Grant SEV-2016-0588.  The work of Y.Z. is partly supported by the DoE under contract number DE-SC0007859. M.C. and Y.Z. would like to thank the Aspen Center for Physics, which is supported by National Science Foundation grant PHY-1607611, where part of this work was performed, and Colegio De Fisica Fundamental E Interdisciplinaria De Las Americas (COFI) for a travel support during the completion of this work. M.Q. would like to thank the Department of Physics, University of Notre Dame, where part of this work was done, for hospitality.

\begin{appendix}

\section{Equation for the lepton asymmetry}\label{app:solve}
\label{AppendixA}

In this appendix we provide more details about solving the sphaleron rate equation~(\ref{EWsph}), to obtain the final lepton/baryon asymmetry. We first rewrite Eq.~(\ref{EWsph}) here,
\begin{eqnarray}
\begin{split}
\frac{d \Delta n_{L_L}(z,t)}{dt} &= \Gamma_{\rm sph}(z-v_\omega t)  \left[ \Delta n_{L_L}^{\rm EQ}(z-v_\omega t) - \Delta n_{L_L} (z,t)\rule{0mm}{3.5mm}\right] \ .
\end{split}
\label{dDeltandt}
\end{eqnarray}
where the sphaleron rate $\Gamma_{\rm sph}$ was given in Eq.~(\ref{RatePhase}) for the symmetric and broken phases.

Several remarks are in order here:
\begin{itemize}
\item We want to solve $\Delta n_L$ for a generic point at a distance $z$ from the moving bubble wall.  We assume that the bubble is formed at an initial time, that we arbitrarily fix to $t=0$. The bubble wall will pass through the point $z$ at time $t=z/v_\omega$, and turn on the Higgs VEV at this point. We are interested in its final value, {\it i.e.}~in principle at $t\to \infty$, after
the bubble wall has passed through and bubble nucleation took place.

\item The electroweak sphaleron rate is strongly suppressed in the broken phase for a strong first order phase transition, where the Higgs VEV at the tunneling (or nucleation) temperature is $v_n\gtrsim T_n$. This behavior follows since  $e^{-M_{\rm sph}/T_n}\ll1$, and hence $\Gamma_{\rm sph}$ at the broken phase is  negligible.  
At point $z$, instead,  the sphaleron process is active, and its rate is a constant, {\it i.e.}~$\Gamma_{\rm sph}(z-v_\omega t) = \Gamma_0\neq0$, for the time window $0\leq t\leq z/v_\omega$.

\item As calculated, and shown in the left panel of Fig.~\ref{asymmetry}, the source term is peaked, and localized, around the moving bubble wall. 
It is highly suppressed at large instantaneous distance from the wall, {\it i.e.}~for $z$ greater than a few times the wall width $L_\omega$.

\end{itemize}

To solve Eq.~(\ref{dDeltandt}), we first get rid of the damping term on the right-hand side with the redefinition
\begin{eqnarray}\label{redefine}
A(z,t) \equiv \Delta n_{L_L}(z,t) e^{\Gamma_0 t} \ .
\end{eqnarray}
The differential equation for $A(z,t)$ is then
\begin{eqnarray}
\frac{d A(z,t)}{dt} = e^{\Gamma_0 t} \Gamma_0 \Delta n_{L_L}^{\rm EQ}(z-v_\omega t) \ ,
\end{eqnarray} 
As explained in the second bullet above, this equation is only valid in the time window $0\leq t\leq z/v_\omega$, as for larger values of $t$, $\Gamma_{\rm sph}\simeq 0$. The solution for $A(z,t)$ could be obtained by simply integrating the right-hand side over time, and then we could use Eq.~(\ref{redefine}) to compute $\Delta n_{L_L}(z,t)$. For $t=z/v_\omega$, we have
\begin{eqnarray}\label{eq:A4}
\begin{split}
 \left. \Delta n_{L_L}(z,t)\right|_{t=z/v_\omega} &= \Gamma_0 \int_0^{z/v_\omega} dt' \, \Delta n_{L_L}^{\rm EQ}(z-v_\omega t')\, e^{\Gamma_0(t'-z/v_\omega)}  \\
& = \frac{\Gamma_0}{v_\omega} \int_0^z dy \, \Delta n_{L_L}^{\rm EQ}(y)\, e^{-\Gamma_0 y/v_\omega} \ ,
\end{split}
\end{eqnarray} 
where in the second step, we have changed the integration variable from $t'$ to $y=z-v_\omega t'$, the coordinate in the bubble wall center of mass frame.

Based on the above discussion, after the bubble wall passes through the point $z$, the Higgs VEV turns on, and the sphaleron process is highly suppressed.
Consequently, the quantity $\Delta n_{L_L}$ is conserved in the broken electroweak phase. In other words, the created baryon/lepton asymmetry freezes in, and we can derive that at $t\to \infty$,
\begin{eqnarray}
\Delta n_{L_L}(z)\equiv \Delta n_{L_L}(z,\infty) \simeq \frac{\Gamma_0}{v_\omega} \int^{z}_0 dy \, \Delta n_{L_L}^{\rm EQ}(y) \, e^{-\Gamma_0 y/v_\omega} \ .
\end{eqnarray}
Therefore we define the asymmetry of the final lepton density in the universe, integrating over all points $z$, as
\begin{eqnarray}
\Delta n_{L_L} =\int_0^\infty dz \frac{d \Delta n_{L_L}(z)}{dz}=\frac{\Gamma_0}{v_\omega} \int^{\infty}_0 dz \, \Delta n_{L_L}^{\rm EQ}(z) \, e^{-\Gamma_0 z/v_\omega} \ ,
\label{ecuacionfinal}
\end{eqnarray}
which is the result quoted in Eq.~(\ref{eq:LeptonAsymmetry}) in the main text. Notice that we are integrating over all points $z>0$, outside the bubble, as we are assuming that in the interior of the bubble, $z<0$, $\Gamma_{sph}\simeq 0$.

\section{The case of a non-anomalous $U(1)_\ell \otimes SU(2)_L^2$ effective theory}
\label{AppendixB}

Let us first consider the case where the masses of $L_L'$ and $L_R''$ doublet fields are much smaller than the critical temperature of EWPT, and
they are not integrated out. The fermionic current $\mathcal{J}^\mu$ that $Z'$ couples to takes then the form
\begin{equation}
\mathcal{J}^\mu = \sum_{i=1}^{N_g} \bar L_{L_{i}} \gamma^\mu L_{L_i} + \texttt{q} \bar L_L' \gamma^\mu L_L' + (\texttt{q}+N_g) \bar L_R'' \gamma^\mu L_R'' + \cdots \ ,
\end{equation}
where $L_{L_{i}} (i=1,2,3)$ are the SM lepton doublets, and the ellipsis represents the terms involving $SU(2)_L$ singlet fields.
The current $\mathcal{J}^\mu$ is non-anomalous with respect to the SM $SU(2)_L$, {\it i.e.},
\begin{equation}\label{anomalyfree}
\partial_\mu \mathcal{J}^\mu \propto {\rm tr} (\ell \tau^a \tau^b) W^a\widetilde W^b \propto \left[ N_g \times 1 +\texttt{q} - (\texttt{q}+N_g) \right] {\rm tr} (W\widetilde W) = 0 \ ,
\end{equation}
where $W$ ($\widetilde W$) is the $SU(2)_L$ field (dual field) strength, and the Pauli matrices $\tau^{a}$ are $SU(2)_L$ generators.

Next, we assume the $\left\langle Z'_0 \right\rangle$ background to be present during EWBG, still generated by the CP violating $\chi$-bubble-wall interaction, given by Eq.~(\ref{eq:Zbackground}). Through the gauge interactions, the $Z'_0$ background serves as chemical potential for the fields charged under it, and leads to the thermal equilibrium asymmetry in their number densities. Of particular interest to us are those for the $SU(2)_L$ doublets,
\begin{equation}\label{EQasymmetries}
\begin{split}
&\Delta n_{L_L}^{\rm EQ} = \hspace{0.3cm}  N_g \hspace{0.3cm} \times \hspace{0.3cm}1 \hspace{0.4cm}\times \hspace{0.4cm} \frac{2}{3} T_c^2 g' \left\langle Z'_0 \right\rangle \ , \\
&\Delta n_{L_L'}^{\rm EQ} = \hspace{0.4cm} 1 \hspace{0.5cm} \times \hspace{0.3cm} \mathtt{q} \hspace{0.4cm} \times \hspace{0.4cm} \frac{2}{3} T_c^2 g' \left\langle Z'_0 \right\rangle \ , \\
&\Delta n_{L_R''}^{\rm EQ} =\hspace{0.4cm} 1 \hspace{0.5cm} \times \hspace{0.3cm} (\mathtt{q}+N_g) \hspace{0.3cm}\times  \hspace{0.35cm}\frac{2}{3} T_c^2 g' \left\langle Z'_0 \right\rangle \ . \\
\end{split}
\end{equation}

In the context of EWBG, the electroweak sphaleron processes are responsible for changes in the lepton and baryon numbers in the universe. In the presence of $L_L', L_R''$ fields in the thermal bath, they will also participate. The actual changes in the particle asymmetries are tied to each other, and satisfy the following relations,
\begin{eqnarray}\label{sumrule}
\frac{\partial}{\partial t}\Delta n_{B_L} = \frac{\partial}{\partial t}\Delta n_{L_L} =3 \frac{\partial}{\partial t}\Delta n_{L_L'} = -3 \frac{\partial}{\partial t}\Delta n_{L_R''} \ ,
\end{eqnarray}
where $B_L$ denotes the baryon number in left-handed SM doublets. It is useful to define the ``effective total lepton asymmetry'' as
\begin{eqnarray}
\Delta n_{L, \rm eff}(z, t) \equiv \Delta n_{L_L}(z, t) + \Delta n_{L_L'}(z, t) - \Delta n_{L_R''}(z, t) \ ,
\end{eqnarray}
so that Eq.~(\ref{sumrule}) implies
\begin{eqnarray}\label{conservation}
\frac{\partial}{\partial t}\Delta n_{B_L}(z, t) = \frac{3}{5}\frac{\partial}{\partial t} \Delta n_{L, \rm eff}(z, t) \ .
\end{eqnarray}

The Boltzmann equation for $\Delta n_{L, \rm eff}(z, t)$ satisfy
\begin{eqnarray}
\frac{\partial}{\partial t}\Delta n_{L, \rm eff}(z, t) &&= \Gamma_{\rm sph}(z-v_\omega t) \left[ \Delta n_{L, \rm eff}^{\rm EQ}(z-v_\omega t) - \Delta n_{L, \rm eff}(z, t) \rule{0mm}{4mm}\right] \ ,\label{newLasymmetry}\\
\Delta n_{L, \rm eff}^{\rm EQ} &&= \Delta n_{L_L}^{\rm EQ} + \Delta n_{L_L'}^{\rm EQ} - \Delta n_{L_R''}^{\rm EQ} \ .\label{newEQLasymmetry}
\end{eqnarray}

Eq.~(\ref{EQasymmetries}) then implies that a cancellation occurs in Eq.~(\ref{newEQLasymmetry}), leading to $\Delta n_{L, \rm eff}^{\rm EQ}=0$.
In this case, the Boltzmann equation for $\Delta n_{L, \rm eff}(z, t)$ has no source term, and assuming the universe begins without any particle asymmetries, no $\Delta n_{L, \rm eff}$ will be generated. In turn Eq.~(\ref{conservation}) implies that the baryon asymmetry cannot be generated.

One should note that such a conclusion is drawn by assuming the $L_L', L_R''$ fields to be relativistic degrees of freedom in the thermal bath during the EWPT.
As pointed out in~\cite{Carena:2018cjh}, the above cancellation is closely related to Eq.~(\ref{anomalyfree}), the conservation of the current $\mathcal{J}^\mu$, with respect to $SU(2)_L$.

On the other hand, if $L_L', L_R''$ obtain a sufficiently large $U(1)_\ell$ symmetry breaking mass through the Yukawa coupling to the $\Phi$ field as discussed in the main text, their thermal number densities in Eq~(\ref{EQasymmetries}) will become Boltzmann suppressed. In this case, the above cancellation no longer occurs, and the proposed EWBG mechanism could be successful. In the limit when $L_L', L_R''$ are very heavy and integrated out, the current that $Z'$ couples to in the low energy theory becomes
\begin{equation}
J^\mu = \sum_{i=1}^{3} \bar L_{L_{i}} \gamma^\mu L_{L_i} + \cdots \ ,
\end{equation}
which is anomalous with respect to $SU(2)_L$. In summary, the created baryon asymmetry should be proportional to the non-conservation of the current $J^\mu$~\cite{Carena:2018cjh}, as previously stated.

\end{appendix}

\bibliography{References}

\begin{thebibliography}{100}
\ifx\href\asklfhas\newcommand{\href}[2]{#2}\fi
\ifx\arxivref\asklfhas\newcommand{\arxivref}[2]{\href{http://arxiv.org/abs/#1}{#2}}\fi
\ifx\doiref\asklfhas\newcommand{\doiref}[2]{\href{http://dx.doi.org/#1}{#2}}\fi
\parskip 0pt
\normalsize

\bibitem{Carena:2018cjh}
M.~Carena, M.~Quiros \& Y.~Zhang,
\textit{``{Electroweak Baryogenesis From Dark CP Violation}''},
\doiref{10.1103/PhysRevLett.122.201802}{Phys.~Rev.~Lett. \textbf{122}, 201802
  (2019)\ignorespaces}\ignorespaces,
\normalsize{\texttt{\arxivref{1811.09719}{arXiv:1811.09719}}}\ignorespaces
\bibitem{Kuzmin:1985mm}
V.~A. Kuzmin, V.~A. Rubakov \& M.~E. Shaposhnikov,
\textit{``{On the Anomalous Electroweak Baryon Number Nonconservation in the
  Early Universe}''},
\doiref{10.1016/0370-2693(85)91028-7}{Phys.~Lett. \textbf{155B}, 36
  (1985)\ignorespaces}\ignorespaces
\bibitem{Cohen:1990py}
A.~G. Cohen, D.~B. Kaplan \& A.~E. Nelson,
\textit{``{WEAK SCALE BARYOGENESIS}''},
\doiref{10.1016/0370-2693(90)90690-8}{Phys.~Lett. \textbf{B245}, 561
  (1990)\ignorespaces}\ignorespaces
\bibitem{Farrar:1993sp}
G.~R. Farrar \& M.~E. Shaposhnikov,
\textit{``{Baryon asymmetry of the universe in the minimal Standard Model}''},
\doiref{10.1103/PhysRevLett.71.210.2,
  10.1103/PhysRevLett.70.2833}{Phys.~Rev.~Lett. \textbf{70}, 2833
  (1993)\ignorespaces}\ignorespaces,
\normalsize{\texttt{\arxivref{hep-ph/9305274}{hep-ph/9305274}}}\ignorespaces,
[Erratum: Phys. Rev. Lett.71,210(1993)]\ignorespaces
\bibitem{Huet:1995mm}
P.~Huet \& A.~E. Nelson,
\textit{``{CP violation and electroweak baryogenesis in extensions of the
  standard model}''},
\doiref{10.1016/0370-2693(95)00674-A}{Phys.~Lett. \textbf{B355}, 229
  (1995)\ignorespaces}\ignorespaces,
\normalsize{\texttt{\arxivref{hep-ph/9504427}{hep-ph/9504427}}}\ignorespaces
\bibitem{Huet:1995sh}
P.~Huet \& A.~E. Nelson,
\textit{``{Electroweak baryogenesis in supersymmetric models}''},
\doiref{10.1103/PhysRevD.53.4578}{Phys.~Rev. \textbf{D53}, 4578
  (1996)\ignorespaces}\ignorespaces,
\normalsize{\texttt{\arxivref{hep-ph/9506477}{hep-ph/9506477}}}\ignorespaces
\bibitem{Riotto:1995hh}
A.~Riotto,
\textit{``{Towards a nonequilibrium quantum field theory approach to
  electroweak baryogenesis}''},
\doiref{10.1103/PhysRevD.53.5834}{Phys.~Rev. \textbf{D53}, 5834
  (1996)\ignorespaces}\ignorespaces,
\normalsize{\texttt{\arxivref{hep-ph/9510271}{hep-ph/9510271}}}\ignorespaces
\bibitem{Carena:1997gx}
M.~Carena, M.~Quiros, A.~Riotto, I.~Vilja \& C.~E.~M. Wagner,
\textit{``{Electroweak baryogenesis and low-energy supersymmetry}''},
\doiref{10.1016/S0550-3213(97)00412-4}{Nucl.~Phys. \textbf{B503}, 387
  (1997)\ignorespaces}\ignorespaces,
\normalsize{\texttt{\arxivref{hep-ph/9702409}{hep-ph/9702409}}}\ignorespaces
\bibitem{Carena:2000id}
M.~Carena, J.~M. Moreno, M.~Quiros, M.~Seco \& C.~E.~M. Wagner,
\textit{``{Supersymmetric CP violating currents and electroweak
  baryogenesis}''},
\doiref{10.1016/S0550-3213(01)00032-3}{Nucl.~Phys. \textbf{B599}, 158
  (2001)\ignorespaces}\ignorespaces,
\normalsize{\texttt{\arxivref{hep-ph/0011055}{hep-ph/0011055}}}\ignorespaces
\bibitem{Cline:2000nw}
J.~M. Cline, M.~Joyce \& K.~Kainulainen,
\textit{``{Supersymmetric electroweak baryogenesis}''},
\doiref{10.1088/1126-6708/2000/07/018}{JHEP \textbf{0007}, 018
  (2000)\ignorespaces}\ignorespaces,
\normalsize{\texttt{\arxivref{hep-ph/0006119}{hep-ph/0006119}}}\ignorespaces
\bibitem{Carena:2002ss}
M.~Carena, M.~Quiros, M.~Seco \& C.~E.~M. Wagner,
\textit{``{Improved results in supersymmetric electroweak baryogenesis}''},
\doiref{10.1016/S0550-3213(02)01065-9}{Nucl.~Phys. \textbf{B650}, 24
  (2003)\ignorespaces}\ignorespaces,
\normalsize{\texttt{\arxivref{hep-ph/0208043}{hep-ph/0208043}}}\ignorespaces
\bibitem{Lee:2004we}
C.~Lee, V.~Cirigliano \& M.~J. Ramsey-Musolf,
\textit{``{Resonant relaxation in electroweak baryogenesis}''},
\doiref{10.1103/PhysRevD.71.075010}{Phys.~Rev. \textbf{D71}, 075010
  (2005)\ignorespaces}\ignorespaces,
\normalsize{\texttt{\arxivref{hep-ph/0412354}{hep-ph/0412354}}}\ignorespaces
\bibitem{Cline:2006ts}
J.~M. Cline,
\textit{``{Baryogenesis}''},
in \textit{``{Les Houches Summer School - Session 86: Particle Physics and
  Cosmology: The Fabric of Spacetime Les Houches, France, July 31-August 25,
  2006}''}\bibitem{Fromme:2006cm}
L.~Fromme, S.~J. Huber \& M.~Seniuch,
\textit{``{Baryogenesis in the two-Higgs doublet model}''},
\doiref{10.1088/1126-6708/2006/11/038}{JHEP \textbf{0611}, 038
  (2006)\ignorespaces}\ignorespaces,
\normalsize{\texttt{\arxivref{hep-ph/0605242}{hep-ph/0605242}}}\ignorespaces
\bibitem{Cirigliano:2006dg}
V.~Cirigliano, S.~Profumo \& M.~J. Ramsey-Musolf,
\textit{``{Baryogenesis, Electric Dipole Moments and Dark Matter in the
  MSSM}''},
\doiref{10.1088/1126-6708/2006/07/002}{JHEP \textbf{0607}, 002
  (2006)\ignorespaces}\ignorespaces,
\normalsize{\texttt{\arxivref{hep-ph/0603246}{hep-ph/0603246}}}\ignorespaces
\bibitem{Li:2010ax}
Y.~Li, S.~Profumo \& M.~Ramsey-Musolf,
\textit{``{A Comprehensive Analysis of Electric Dipole Moment Constraints on
  CP-violating Phases in the MSSM}''},
\doiref{10.1007/JHEP08(2010)062}{JHEP \textbf{1008}, 062
  (2010)\ignorespaces}\ignorespaces,
\normalsize{\texttt{\arxivref{1006.1440}{arXiv:1006.1440}}}\ignorespaces
\bibitem{Andreev:2018ayy}
ACME Collaboration, V.~Andreev et~al.,
\textit{``{Improved limit on the electric dipole moment of the electron}''},
\doiref{10.1038/s41586-018-0599-8}{Nature \textbf{562}, 355
  (2018)\ignorespaces}\ignorespaces
\bibitem{Baron:2013eja}
ACME Collaboration, J.~Baron et~al.,
\textit{``{Order of Magnitude Smaller Limit on the Electric Dipole Moment of
  the Electron}''},
\doiref{10.1126/science.1248213}{Science \textbf{343}, 269
  (2014)\ignorespaces}\ignorespaces,
\normalsize{\texttt{\arxivref{1310.7534}{arXiv:1310.7534}}}\ignorespaces
\bibitem{Griffith:2009zz}
W.~C. Griffith, M.~D. Swallows, T.~H. Loftus, M.~V. Romalis, B.~R. Heckel \&
  E.~N. Fortson,
\textit{``{Improved Limit on the Permanent Electric Dipole Moment of
  Hg-199}''},
\doiref{10.1103/PhysRevLett.102.101601}{Phys.~Rev.~Lett. \textbf{102}, 101601
  (2009)\ignorespaces}\ignorespaces
\bibitem{Baker:2006ts}
C.~A. Baker et~al.,
\textit{``{An Improved experimental limit on the electric dipole moment of the
  neutron}''},
\doiref{10.1103/PhysRevLett.97.131801}{Phys.~Rev.~Lett. \textbf{97}, 131801
  (2006)\ignorespaces}\ignorespaces,
\normalsize{\texttt{\arxivref{hep-ex/0602020}{hep-ex/0602020}}}\ignorespaces
\bibitem{Shu:2013uua}
J.~Shu \& Y.~Zhang,
\textit{``{Impact of a CP Violating Higgs Sector: From LHC to Baryogenesis}''},
\doiref{10.1103/PhysRevLett.111.091801}{Phys.~Rev.~Lett. \textbf{111}, 091801
  (2013)\ignorespaces}\ignorespaces,
\normalsize{\texttt{\arxivref{1304.0773}{arXiv:1304.0773}}}\ignorespaces
\bibitem{Ipek:2013iba}
S.~Ipek,
\textit{``{Perturbative analysis of the electron electric dipole moment and CP
  violation in two-Higgs-doublet models}''},
\doiref{10.1103/PhysRevD.89.073012}{Phys.~Rev. \textbf{D89}, 073012
  (2014)\ignorespaces}\ignorespaces,
\normalsize{\texttt{\arxivref{1310.6790}{arXiv:1310.6790}}}\ignorespaces
\bibitem{Jung:2013hka}
M.~Jung \& A.~Pich,
\textit{``{Electric Dipole Moments in Two-Higgs-Doublet Models}''},
\doiref{10.1007/JHEP04(2014)076}{JHEP \textbf{1404}, 076
  (2014)\ignorespaces}\ignorespaces,
\normalsize{\texttt{\arxivref{1308.6283}{arXiv:1308.6283}}}\ignorespaces
\bibitem{Abe:2013qla}
T.~Abe, J.~Hisano, T.~Kitahara \& K.~Tobioka,
\textit{``{Gauge invariant Barr-Zee type contributions to fermionic EDMs in the
  two-Higgs doublet models}''},
\doiref{10.1007/JHEP01(2014)106, 10.1007/JHEP04(2016)161}{JHEP \textbf{1401},
  106 (2014)\ignorespaces}\ignorespaces,
\normalsize{\texttt{\arxivref{1311.4704}{arXiv:1311.4704}}}\ignorespaces,
[Erratum: JHEP04,161(2016)]\ignorespaces
\bibitem{Inoue:2014nva}
S.~Inoue, M.~J. Ramsey-Musolf \& Y.~Zhang,
\textit{``{CP-violating phenomenology of flavor conserving two Higgs doublet
  models}''},
\doiref{10.1103/PhysRevD.89.115023}{Phys.~Rev. \textbf{D89}, 115023
  (2014)\ignorespaces}\ignorespaces,
\normalsize{\texttt{\arxivref{1403.4257}{arXiv:1403.4257}}}\ignorespaces
\bibitem{Cheung:2014oaa}
K.~Cheung, J.~S. Lee, E.~Senaha \& P.-Y. Tseng,
\textit{``{Confronting Higgcision with Electric Dipole Moments}''},
\doiref{10.1007/JHEP06(2014)149}{JHEP \textbf{1406}, 149
  (2014)\ignorespaces}\ignorespaces,
\normalsize{\texttt{\arxivref{1403.4775}{arXiv:1403.4775}}}\ignorespaces
\bibitem{Bian:2014zka}
L.~Bian, T.~Liu \& J.~Shu,
\textit{``{Cancellations Between Two-Loop Contributions to the Electron
  Electric Dipole Moment with a CP-Violating Higgs Sector}''},
\doiref{10.1103/PhysRevLett.115.021801}{Phys.~Rev.~Lett. \textbf{115}, 021801
  (2015)\ignorespaces}\ignorespaces,
\normalsize{\texttt{\arxivref{1411.6695}{arXiv:1411.6695}}}\ignorespaces
\bibitem{Chen:2015gaa}
C.-Y. Chen, S.~Dawson \& Y.~Zhang,
\textit{``{Complementarity of LHC and EDMs for Exploring Higgs CP
  Violation}''},
\doiref{10.1007/JHEP06(2015)056}{JHEP \textbf{1506}, 056
  (2015)\ignorespaces}\ignorespaces,
\normalsize{\texttt{\arxivref{1503.01114}{arXiv:1503.01114}}}\ignorespaces
\bibitem{Fuyuto:2015ida}
K.~Fuyuto, J.~Hisano \& E.~Senaha,
\textit{``{Toward verification of electroweak baryogenesis by electric dipole
  moments}''},
\doiref{10.1016/j.physletb.2016.02.053}{Phys.~Lett. \textbf{B755}, 491
  (2016)\ignorespaces}\ignorespaces,
\normalsize{\texttt{\arxivref{1510.04485}{arXiv:1510.04485}}}\ignorespaces
\bibitem{Jiang:2015cwa}
M.~Jiang, L.~Bian, W.~Huang \& J.~Shu,
\textit{``{Impact of a complex singlet: Electroweak baryogenesis and dark
  matter}''},
\doiref{10.1103/PhysRevD.93.065032}{Phys.~Rev. \textbf{D93}, 065032
  (2016)\ignorespaces}\ignorespaces,
\normalsize{\texttt{\arxivref{1502.07574}{arXiv:1502.07574}}}\ignorespaces
\bibitem{Blinov:2015sna}
N.~Blinov, J.~Kozaczuk, D.~E. Morrissey \& C.~Tamarit,
\textit{``{Electroweak Baryogenesis from Exotic Electroweak Symmetry
  Breaking}''},
\doiref{10.1103/PhysRevD.92.035012}{Phys.~Rev. \textbf{D92}, 035012
  (2015)\ignorespaces}\ignorespaces,
\normalsize{\texttt{\arxivref{1504.05195}{arXiv:1504.05195}}}\ignorespaces
\bibitem{Balazs:2016yvi}
C.~Balazs, G.~White \& J.~Yue,
\textit{``{Effective field theory, electric dipole moments and electroweak
  baryogenesis}''},
\doiref{10.1007/JHEP03(2017)030}{JHEP \textbf{1703}, 030
  (2017)\ignorespaces}\ignorespaces,
\normalsize{\texttt{\arxivref{1612.01270}{arXiv:1612.01270}}}\ignorespaces
\bibitem{Bian:2016zba}
L.~Bian \& N.~Chen,
\textit{``{Cancellation mechanism in the predictions of electric dipole
  moments}''},
\doiref{10.1103/PhysRevD.95.115029}{Phys.~Rev. \textbf{D95}, 115029
  (2017)\ignorespaces}\ignorespaces,
\normalsize{\texttt{\arxivref{1608.07975}{arXiv:1608.07975}}}\ignorespaces
\bibitem{Chen:2017com}
C.-Y. Chen, H.-L. Li \& M.~Ramsey-Musolf,
\textit{``{CP-Violation in the Two Higgs Doublet Model: from the LHC to
  EDMs}''},
\doiref{10.1103/PhysRevD.97.015020}{Phys.~Rev. \textbf{D97}, 015020
  (2018)\ignorespaces}\ignorespaces,
\normalsize{\texttt{\arxivref{1708.00435}{arXiv:1708.00435}}}\ignorespaces
\bibitem{Cesarotti:2018huy}
C.~Cesarotti, Q.~Lu, Y.~Nakai, A.~Parikh \& M.~Reece,
\textit{``{Interpreting the Electron EDM Constraint}''},
\normalsize{\texttt{\arxivref{1810.07736}{arXiv:1810.07736}}}\ignorespaces
\bibitem{Egana-Ugrinovic:2018fpy}
D.~Egana-Ugrinovic \& S.~Thomas,
\textit{``{Higgs Boson Contributions to the Electron Electric Dipole
  Moment}''},
\normalsize{\texttt{\arxivref{1810.08631}{arXiv:1810.08631}}}\ignorespaces
\bibitem{Ramsey-Musolf:2017tgh}
M.~J. Ramsey-Musolf, P.~Winslow \& G.~White,
\textit{``{Color Breaking Baryogenesis}''},
\doiref{10.1103/PhysRevD.97.123509}{Phys.~Rev. \textbf{D97}, 123509
  (2018)\ignorespaces}\ignorespaces,
\normalsize{\texttt{\arxivref{1708.07511}{arXiv:1708.07511}}}\ignorespaces
\bibitem{Cline:2017qpe}
J.~M. Cline, K.~Kainulainen \& D.~Tucker-Smith,
\textit{``{Electroweak baryogenesis from a dark sector}''},
\doiref{10.1103/PhysRevD.95.115006}{Phys.~Rev. \textbf{D95}, 115006
  (2017)\ignorespaces}\ignorespaces,
\normalsize{\texttt{\arxivref{1702.08909}{arXiv:1702.08909}}}\ignorespaces
\bibitem{Barr:1990vd}
S.~M. Barr \& A.~Zee,
\textit{``{Electric Dipole Moment of the Electron and of the Neutron}''},
\doiref{10.1103/PhysRevLett.65.2920,
  10.1103/PhysRevLett.65.21}{Phys.~Rev.~Lett. \textbf{65}, 21
  (1990)\ignorespaces}\ignorespaces,
[Erratum: Phys. Rev. Lett.65,2920(1990)]\ignorespaces
\bibitem{Gunion:1990ce}
J.~F. Gunion \& R.~Vega,
\textit{``{The Electron electric dipole moment for a CP violating neutral Higgs
  sector}''},
\doiref{10.1016/0370-2693(90)90246-3}{Phys.~Lett. \textbf{B251}, 157
  (1990)\ignorespaces}\ignorespaces
\bibitem{Wess:1971yu}
J.~Wess \& B.~Zumino,
\textit{``{Consequences of anomalous Ward identities}''},
\doiref{10.1016/0370-2693(71)90582-X}{Phys.~Lett. \textbf{37B}, 95
  (1971)\ignorespaces}\ignorespaces
\bibitem{FileviezPerez:2010gw}
P.~Fileviez~Perez \& M.~B. Wise,
\textit{``{Baryon and lepton number as local gauge symmetries}''},
\doiref{10.1103/PhysRevD.82.079901, 10.1103/PhysRevD.82.011901}{Phys.~Rev.
  \textbf{D82}, 011901 (2010)\ignorespaces}\ignorespaces,
\normalsize{\texttt{\arxivref{1002.1754}{arXiv:1002.1754}}}\ignorespaces,
[Erratum: Phys. Rev.D82,079901(2010)]\ignorespaces
\bibitem{Duerr:2013dza}
M.~Duerr, P.~Fileviez~Perez \& M.~B. Wise,
\textit{``{Gauge Theory for Baryon and Lepton Numbers with Leptoquarks}''},
\doiref{10.1103/PhysRevLett.110.231801}{Phys.~Rev.~Lett. \textbf{110}, 231801
  (2013)\ignorespaces}\ignorespaces,
\normalsize{\texttt{\arxivref{1304.0576}{arXiv:1304.0576}}}\ignorespaces
\bibitem{Schwaller:2013hqa}
P.~Schwaller, T.~M.~P. Tait \& R.~Vega-Morales,
\textit{``{Dark Matter and Vectorlike Leptons from Gauged Lepton Number}''},
\doiref{10.1103/PhysRevD.88.035001}{Phys.~Rev. \textbf{D88}, 035001
  (2013)\ignorespaces}\ignorespaces,
\normalsize{\texttt{\arxivref{1305.1108}{arXiv:1305.1108}}}\ignorespaces
\bibitem{Altmannshofer:2016oaq}
W.~Altmannshofer, M.~Carena \& A.~Crivellin,
\textit{``{$L_\mu - L_\tau$ theory of Higgs flavor violation and
  $(g-2)_\mu$}''},
\doiref{10.1103/PhysRevD.94.095026}{Phys.~Rev. \textbf{D94}, 095026
  (2016)\ignorespaces}\ignorespaces,
\normalsize{\texttt{\arxivref{1604.08221}{arXiv:1604.08221}}}\ignorespaces
\bibitem{Bardeen:1984pm}
W.~A. Bardeen \& B.~Zumino,
\textit{``{Consistent and Covariant Anomalies in Gauge and Gravitational
  Theories}''},
\doiref{10.1016/0550-3213(84)90322-5}{Nucl.~Phys. \textbf{B244}, 421
  (1984)\ignorespaces}\ignorespaces
\bibitem{Rosenberg:1962pp}
L.~Rosenberg,
\textit{``{Electromagnetic interactions of neutrinos}''},
\doiref{10.1103/PhysRev.129.2786}{Phys.~Rev. \textbf{129}, 2786
  (1963)\ignorespaces}\ignorespaces
\bibitem{Fox:2018ldq}
P.~J. Fox, I.~Low \& Y.~Zhang,
\textit{``{Top-philic $Z'$ forces at the LHC}''},
\doiref{10.1007/JHEP03(2018)074}{JHEP \textbf{1803}, 074
  (2018)\ignorespaces}\ignorespaces,
\normalsize{\texttt{\arxivref{1801.03505}{arXiv:1801.03505}}}\ignorespaces
\bibitem{Espinosa:2011ax}
J.~R. Espinosa, T.~Konstandin \& F.~Riva,
\textit{``{Strong Electroweak Phase Transitions in the Standard Model with a
  Singlet}''},
\doiref{10.1016/j.nuclphysb.2011.09.010}{Nucl.~Phys. \textbf{B854}, 592
  (2012)\ignorespaces}\ignorespaces,
\normalsize{\texttt{\arxivref{1107.5441}{arXiv:1107.5441}}}\ignorespaces
\bibitem{Patel:2013zla}
H.~H. Patel, M.~J. Ramsey-Musolf \& M.~B. Wise,
\textit{``{Color Breaking in the Early Universe}''},
\doiref{10.1103/PhysRevD.88.015003}{Phys.~Rev. \textbf{D88}, 015003
  (2013)\ignorespaces}\ignorespaces,
\normalsize{\texttt{\arxivref{1303.1140}{arXiv:1303.1140}}}\ignorespaces
\bibitem{Cheung:2013dca}
C.~Cheung \& Y.~Zhang,
\textit{``{Electroweak Cogenesis}''},
\doiref{10.1007/JHEP09(2013)002}{JHEP \textbf{1309}, 002
  (2013)\ignorespaces}\ignorespaces,
\normalsize{\texttt{\arxivref{1306.4321}{arXiv:1306.4321}}}\ignorespaces
\bibitem{Curtin:2014jma}
D.~Curtin, P.~Meade \& C.-T. Yu,
\textit{``{Testing Electroweak Baryogenesis with Future Colliders}''},
\doiref{10.1007/JHEP11(2014)127}{JHEP \textbf{1411}, 127
  (2014)\ignorespaces}\ignorespaces,
\normalsize{\texttt{\arxivref{1409.0005}{arXiv:1409.0005}}}\ignorespaces
\bibitem{deVries:2017ncy}
J.~de~Vries, M.~Postma, J.~van~de~Vis \& G.~White,
\textit{``{Electroweak Baryogenesis and the Standard Model Effective Field
  Theory}''},
\doiref{10.1007/JHEP01(2018)089}{JHEP \textbf{1801}, 089
  (2018)\ignorespaces}\ignorespaces,
\normalsize{\texttt{\arxivref{1710.04061}{arXiv:1710.04061}}}\ignorespaces
\bibitem{Dorsch:2018pat}
G.~C. Dorsch, S.~J. Huber \& T.~Konstandin,
\textit{``{Bubble wall velocities in the Standard Model and beyond}''},
\doiref{10.1088/1475-7516/2018/12/034}{JCAP \textbf{1812}, 034
  (2018)\ignorespaces}\ignorespaces,
\normalsize{\texttt{\arxivref{1809.04907}{arXiv:1809.04907}}}\ignorespaces
\bibitem{Cohen:1991iu}
A.~G. Cohen, D.~B. Kaplan \& A.~E. Nelson,
\textit{``{Spontaneous baryogenesis at the weak phase transition}''},
\doiref{10.1016/0370-2693(91)91711-4}{Phys.~Lett. \textbf{B263}, 86
  (1991)\ignorespaces}\ignorespaces
\bibitem{Kolb:1990vq}
E.~W. Kolb \& M.~S. Turner,
\textit{``{The Early Universe}''},
Front.~Phys. \textbf{69}, 1 (1990)\ignorespaces\ignorespaces
\bibitem{Davoudiasl:2004gf}
H.~Davoudiasl, R.~Kitano, G.~D. Kribs, H.~Murayama \& P.~J. Steinhardt,
\textit{``{Gravitational baryogenesis}''},
\doiref{10.1103/PhysRevLett.93.201301}{Phys.~Rev.~Lett. \textbf{93}, 201301
  (2004)\ignorespaces}\ignorespaces,
\normalsize{\texttt{\arxivref{hep-ph/0403019}{hep-ph/0403019}}}\ignorespaces
\bibitem{Davoudiasl:2013pda}
H.~Davoudiasl,
\textit{``{Gravitationally Induced Dark Matter Asymmetry and Dark Nucleon
  Decay}''},
\doiref{10.1103/PhysRevD.88.095004}{Phys.~Rev. \textbf{D88}, 095004
  (2013)\ignorespaces}\ignorespaces,
\normalsize{\texttt{\arxivref{1308.3473}{arXiv:1308.3473}}}\ignorespaces
\bibitem{Bodeker:1999gx}
D.~Bodeker, G.~D. Moore \& K.~Rummukainen,
\textit{``{Chern-Simons number diffusion and hard thermal loops on the
  lattice}''},
\doiref{10.1103/PhysRevD.61.056003}{Phys.~Rev. \textbf{D61}, 056003
  (2000)\ignorespaces}\ignorespaces,
\normalsize{\texttt{\arxivref{hep-ph/9907545}{hep-ph/9907545}}}\ignorespaces
\bibitem{Moore:1998swa}
G.~D. Moore,
\textit{``{Measuring the broken phase sphaleron rate nonperturbatively}''},
\doiref{10.1103/PhysRevD.59.014503}{Phys.~Rev. \textbf{D59}, 014503
  (1999)\ignorespaces}\ignorespaces,
\normalsize{\texttt{\arxivref{hep-ph/9805264}{hep-ph/9805264}}}\ignorespaces
\bibitem{Zhou:2019uzq}
R.~Zhou, L.~Bian \& H.-K. Guo,
\textit{``{Probing the Electroweak Sphaleron with Gravitational Waves}''},
\normalsize{\texttt{\arxivref{1910.00234}{arXiv:1910.00234}}}\ignorespaces
\bibitem{Aghanim:2018eyx}
Planck Collaboration, N.~Aghanim et~al.,
\textit{``{Planck 2018 results. VI. Cosmological parameters}''},
\normalsize{\texttt{\arxivref{1807.06209}{arXiv:1807.06209}}}\ignorespaces
\bibitem{Banerjee:2016tad}
NA64 Collaboration, D.~Banerjee et~al.,
\textit{``{Search for invisible decays of sub-GeV dark photons in
  missing-energy events at the CERN SPS}''},
\doiref{10.1103/PhysRevLett.118.011802}{Phys.~Rev.~Lett. \textbf{118}, 011802
  (2017)\ignorespaces}\ignorespaces,
\normalsize{\texttt{\arxivref{1610.02988}{arXiv:1610.02988}}}\ignorespaces
\bibitem{Lees:2017lec}
BaBar Collaboration, J.~P. Lees et~al.,
\textit{``{Search for Invisible Decays of a Dark Photon Produced in
  ${e}^{+}{e}^{-}$ Collisions at BaBar}''},
\doiref{10.1103/PhysRevLett.119.131804}{Phys.~Rev.~Lett. \textbf{119}, 131804
  (2017)\ignorespaces}\ignorespaces,
\normalsize{\texttt{\arxivref{1702.03327}{arXiv:1702.03327}}}\ignorespaces
\bibitem{Tanabashi:2018oca}
Particle Data Group Collaboration, M.~Tanabashi et~al.,
\textit{``{Review of Particle Physics}''},
\doiref{10.1103/PhysRevD.98.030001}{Phys.~Rev. \textbf{D98}, 030001
  (2018)\ignorespaces}\ignorespaces
\bibitem{Alcaraz:2006mx}
DELPHI, OPAL, ALEPH, LEP Electroweak Working Group, L3 Collaboration,
  J.~Alcaraz et~al.,
\textit{``{A Combination of preliminary electroweak measurements and
  constraints on the standard model}''},
\normalsize{\texttt{\arxivref{hep-ex/0612034}{hep-ex/0612034}}}\ignorespaces
\bibitem{Lees:2014xha}
BaBar Collaboration, J.~P. Lees et~al.,
\textit{``{Search for a Dark Photon in $e^+e^-$ Collisions at BaBar}''},
\doiref{10.1103/PhysRevLett.113.201801}{Phys.~Rev.~Lett. \textbf{113}, 201801
  (2014)\ignorespaces}\ignorespaces,
\normalsize{\texttt{\arxivref{1406.2980}{arXiv:1406.2980}}}\ignorespaces
\bibitem{Bjorken:2009mm}
J.~D. Bjorken, R.~Essig, P.~Schuster \& N.~Toro,
\textit{``{New Fixed-Target Experiments to Search for Dark Gauge Forces}''},
\doiref{10.1103/PhysRevD.80.075018}{Phys.~Rev. \textbf{D80}, 075018
  (2009)\ignorespaces}\ignorespaces,
\normalsize{\texttt{\arxivref{0906.0580}{arXiv:0906.0580}}}\ignorespaces
\bibitem{Bauer:2018onh}
M.~Bauer, P.~Foldenauer \& J.~Jaeckel,
\textit{``{Hunting All the Hidden Photons}''},
\doiref{10.1007/JHEP07(2018)094}{JHEP \textbf{1807}, 094
  (2018)\ignorespaces}\ignorespaces,
\normalsize{\texttt{\arxivref{1803.05466}{arXiv:1803.05466}}}\ignorespaces
\bibitem{Davoudiasl:2012qa}
H.~Davoudiasl, H.-S. Lee \& W.~J. Marciano,
\textit{``{Muon Anomaly and Dark Parity Violation}''},
\doiref{10.1103/PhysRevLett.109.031802}{Phys.~Rev.~Lett. \textbf{109}, 031802
  (2012)\ignorespaces}\ignorespaces,
\normalsize{\texttt{\arxivref{1205.2709}{arXiv:1205.2709}}}\ignorespaces
\bibitem{Battaglieri:2017aum}
M.~Battaglieri et~al.,
\textit{``{US Cosmic Visions: New Ideas in Dark Matter 2017: Community
  Report}''},
\normalsize{\texttt{\arxivref{1707.04591}{arXiv:1707.04591}}}\ignorespaces
\bibitem{Dror:2017ehi}
J.~A. Dror, R.~Lasenby \& M.~Pospelov,
\textit{``{New constraints on light vectors coupled to anomalous currents}''},
\doiref{10.1103/PhysRevLett.119.141803}{Phys.~Rev.~Lett. \textbf{119}, 141803
  (2017)\ignorespaces}\ignorespaces,
\normalsize{\texttt{\arxivref{1705.06726}{arXiv:1705.06726}}}\ignorespaces
\bibitem{Dror:2017nsg}
J.~A. Dror, R.~Lasenby \& M.~Pospelov,
\textit{``{Dark forces coupled to nonconserved currents}''},
\doiref{10.1103/PhysRevD.96.075036}{Phys.~Rev. \textbf{D96}, 075036
  (2017)\ignorespaces}\ignorespaces,
\normalsize{\texttt{\arxivref{1707.01503}{arXiv:1707.01503}}}\ignorespaces
\bibitem{dEnterria:2016fpc}
D.~d'Enterria,
\textit{``{Physics case of FCC-ee}''},
Frascati~Phys.~Ser. \textbf{61}, 17 (2016)\ignorespaces\ignorespaces,
\normalsize{\texttt{\arxivref{1601.06640}{arXiv:1601.06640}}}\ignorespaces,
in \textit{``{Proceedings, Physics Prospects for Linear and other Future
  Colliders after the Discovery of the Higgs (LFC15): Trento, Italy, September
  7-11, 2015}''},
17\ignorespaces
\bibitem{CEPC-SPPCStudyGroup:2015csa}
C.-S.~S. Group,
\textit{``{CEPC-SPPC Preliminary Conceptual Design Report. 1. Physics and
  Detector}''}
\bibitem{Carone:1995pu}
C.~D. Carone \& H.~Murayama,
\textit{``{Realistic models with a light U(1) gauge boson coupled to baryon
  number}''},
\doiref{10.1103/PhysRevD.52.484}{Phys.~Rev. \textbf{D52}, 484
  (1995)\ignorespaces}\ignorespaces,
\normalsize{\texttt{\arxivref{hep-ph/9501220}{hep-ph/9501220}}}\ignorespaces
\bibitem{Hook:2010tw}
A.~Hook, E.~Izaguirre \& J.~G. Wacker,
\textit{``{Model Independent Bounds on Kinetic Mixing}''},
\doiref{10.1155/2011/859762}{Adv.~High~Energy~Phys. \textbf{2011}, 859762
  (2011)\ignorespaces}\ignorespaces,
\normalsize{\texttt{\arxivref{1006.0973}{arXiv:1006.0973}}}\ignorespaces
\bibitem{Dobrescu:2014fca}
B.~A. Dobrescu \& C.~Frugiuele,
\textit{``{Hidden GeV-scale interactions of quarks}''},
\doiref{10.1103/PhysRevLett.113.061801}{Phys.~Rev.~Lett. \textbf{113}, 061801
  (2014)\ignorespaces}\ignorespaces,
\normalsize{\texttt{\arxivref{1404.3947}{arXiv:1404.3947}}}\ignorespaces
\bibitem{Barger:2003zh}
V.~Barger, P.~Langacker \& H.-S. Lee,
\textit{``{Primordial nucleosynthesis constraints on $Z^\prime$ properties}''},
\doiref{10.1103/PhysRevD.67.075009}{Phys.~Rev. \textbf{D67}, 075009
  (2003)\ignorespaces}\ignorespaces,
\normalsize{\texttt{\arxivref{hep-ph/0302066}{hep-ph/0302066}}}\ignorespaces
\bibitem{Ghosh:2010hy}
D.~K. Ghosh, G.~Senjanovic \& Y.~Zhang,
\textit{``{Naturally Light Sterile Neutrinos from Theory of R-parity}''},
\doiref{10.1016/j.physletb.2011.03.039}{Phys.~Lett. \textbf{B698}, 420
  (2011)\ignorespaces}\ignorespaces,
\normalsize{\texttt{\arxivref{1010.3968}{arXiv:1010.3968}}}\ignorespaces
\bibitem{Hamann:2011ge}
J.~Hamann, S.~Hannestad, G.~G. Raffelt \& Y.~Y.~Y. Wong,
\textit{``{Sterile neutrinos with eV masses in cosmology: How disfavoured
  exactly?}''},
\doiref{10.1088/1475-7516/2011/09/034}{JCAP \textbf{1109}, 034
  (2011)\ignorespaces}\ignorespaces,
\normalsize{\texttt{\arxivref{1108.4136}{arXiv:1108.4136}}}\ignorespaces
\bibitem{Abazajian:2019oqj}
K.~N. Abazajian \& J.~Heeck,
\textit{``{Observing Dirac neutrinos in the cosmic microwave background}''},
\normalsize{\texttt{\arxivref{1908.03286}{arXiv:1908.03286}}}\ignorespaces
\bibitem{Berryman:2017twh}
J.~M. Berryman, A.~de~Gouvêa, K.~J. Kelly \& Y.~Zhang,
\textit{``{Dark Matter and Neutrino Mass from the Smallest Non-Abelian Chiral
  Dark Sector}''},
\doiref{10.1103/PhysRevD.96.075010}{Phys.~Rev. \textbf{D96}, 075010
  (2017)\ignorespaces}\ignorespaces,
\normalsize{\texttt{\arxivref{1706.02722}{arXiv:1706.02722}}}\ignorespaces
\bibitem{Atre:2009rg}
A.~Atre, T.~Han, S.~Pascoli \& B.~Zhang,
\textit{``{The Search for Heavy Majorana Neutrinos}''},
\doiref{10.1088/1126-6708/2009/05/030}{JHEP \textbf{0905}, 030
  (2009)\ignorespaces}\ignorespaces,
\normalsize{\texttt{\arxivref{0901.3589}{arXiv:0901.3589}}}\ignorespaces
\bibitem{Krovi:2018fdr}
A.~Krovi, I.~Low \& Y.~Zhang,
\textit{``{Broadening Dark Matter Searches at the LHC: Mono-X versus Darkonium
  Channels}''},
\doiref{10.1007/JHEP10(2018)026}{JHEP \textbf{1810}, 026
  (2018)\ignorespaces}\ignorespaces,
\normalsize{\texttt{\arxivref{1807.07972}{arXiv:1807.07972}}}\ignorespaces
\bibitem{Fox:2011fx}
P.~J. Fox, R.~Harnik, J.~Kopp \& Y.~Tsai,
\textit{``{LEP Shines Light on Dark Matter}''},
\doiref{10.1103/PhysRevD.84.014028}{Phys.~Rev. \textbf{D84}, 014028
  (2011)\ignorespaces}\ignorespaces,
\normalsize{\texttt{\arxivref{1103.0240}{arXiv:1103.0240}}}\ignorespaces
\bibitem{Aprile:2018dbl}
XENON Collaboration, E.~Aprile et~al.,
\textit{``{Dark Matter Search Results from a One Tonne$\times$Year Exposure of
  XENON1T}''},
\normalsize{\texttt{\arxivref{1805.12562}{arXiv:1805.12562}}}\ignorespaces
\bibitem{Cui:2017nnn}
PandaX-II Collaboration, X.~Cui et~al.,
\textit{``{Dark Matter Results From 54-Ton-Day Exposure of PandaX-II
  Experiment}''},
\doiref{10.1103/PhysRevLett.119.181302}{Phys.~Rev.~Lett. \textbf{119}, 181302
  (2017)\ignorespaces}\ignorespaces,
\normalsize{\texttt{\arxivref{1708.06917}{arXiv:1708.06917}}}\ignorespaces
\bibitem{Akerib:2016vxi}
LUX Collaboration, D.~S. Akerib et~al.,
\textit{``{Results from a search for dark matter in the complete LUX
  exposure}''},
\doiref{10.1103/PhysRevLett.118.021303}{Phys.~Rev.~Lett. \textbf{118}, 021303
  (2017)\ignorespaces}\ignorespaces,
\normalsize{\texttt{\arxivref{1608.07648}{arXiv:1608.07648}}}\ignorespaces
\bibitem{Chowdhury:2011ga}
T.~A. Chowdhury, M.~Nemevsek, G.~Senjanovic \& Y.~Zhang,
\textit{``{Dark Matter as the Trigger of Strong Electroweak Phase
  Transition}''},
\doiref{10.1088/1475-7516/2012/02/029}{JCAP \textbf{1202}, 029
  (2012)\ignorespaces}\ignorespaces,
\normalsize{\texttt{\arxivref{1110.5334}{arXiv:1110.5334}}}\ignorespaces
\bibitem{Cline:2013gha}
J.~M. Cline, K.~Kainulainen, P.~Scott \& C.~Weniger,
\textit{``{Update on scalar singlet dark matter}''},
\doiref{10.1103/PhysRevD.92.039906, 10.1103/PhysRevD.88.055025}{Phys.~Rev.
  \textbf{D88}, 055025 (2013)\ignorespaces}\ignorespaces,
\normalsize{\texttt{\arxivref{1306.4710}{arXiv:1306.4710}}}\ignorespaces,
[Erratum: Phys. Rev.D92,no.3,039906(2015)]\ignorespaces
\bibitem{Wise:2014jva}
M.~B. Wise \& Y.~Zhang,
\textit{``{Stable Bound States of Asymmetric Dark Matter}''},
\doiref{10.1103/PhysRevD.90.055030, 10.1103/PhysRevD.91.039907}{Phys.~Rev.
  \textbf{D90}, 055030 (2014)\ignorespaces}\ignorespaces,
\normalsize{\texttt{\arxivref{1407.4121}{arXiv:1407.4121}}}\ignorespaces,
[Erratum: Phys. Rev.D91,no.3,039907(2015)]\ignorespaces
\bibitem{Zhang:2015era}
Y.~Zhang,
\textit{``{Long-lived Light Mediator to Dark Matter and Primordial Small Scale
  Spectrum}''},
\doiref{10.1088/1475-7516/2015/05/008}{JCAP \textbf{1505}, 008
  (2015)\ignorespaces}\ignorespaces,
\normalsize{\texttt{\arxivref{1502.06983}{arXiv:1502.06983}}}\ignorespaces
\bibitem{Kouvaris:2014uoa}
C.~Kouvaris, I.~M. Shoemaker \& K.~Tuominen,
\textit{``{Self-Interacting Dark Matter through the Higgs Portal}''},
\doiref{10.1103/PhysRevD.91.043519}{Phys.~Rev. \textbf{D91}, 043519
  (2015)\ignorespaces}\ignorespaces,
\normalsize{\texttt{\arxivref{1411.3730}{arXiv:1411.3730}}}\ignorespaces
\bibitem{Carena:2018vpt}
M.~Carena, Z.~Liu \& M.~Riembau,
\textit{``{Probing the electroweak phase transition via enhanced di-Higgs boson
  production}''},
\doiref{10.1103/PhysRevD.97.095032}{Phys.~Rev. \textbf{D97}, 095032
  (2018)\ignorespaces}\ignorespaces,
\normalsize{\texttt{\arxivref{1801.00794}{arXiv:1801.00794}}}\ignorespaces
\bibitem{Khachatryan:2016vau}
ATLAS, CMS Collaboration, G.~Aad et~al.,
\textit{``{Measurements of the Higgs boson production and decay rates and
  constraints on its couplings from a combined ATLAS and CMS analysis of the
  LHC pp collision data at $ \sqrt{s}=7 $ and 8 TeV}''},
\doiref{10.1007/JHEP08(2016)045}{JHEP \textbf{1608}, 045
  (2016)\ignorespaces}\ignorespaces,
\normalsize{\texttt{\arxivref{1606.02266}{arXiv:1606.02266}}}\ignorespaces
\bibitem{Accomando:2006ga}
E.~Accomando et~al.,
\textit{``{Workshop on CP Studies and Non-Standard Higgs Physics}''},
\normalsize{\texttt{\arxivref{hep-ph/0608079}{hep-ph/0608079}}}\ignorespaces
\bibitem{Chen:2014gka}
Y.~Chen, R.~Harnik \& R.~Vega-Morales,
\textit{``{Probing the Higgs Couplings to Photons in h->4l at the LHC}''},
\doiref{10.1103/PhysRevLett.113.191801}{Phys.~Rev.~Lett. \textbf{113}, 191801
  (2014)\ignorespaces}\ignorespaces,
\normalsize{\texttt{\arxivref{1404.1336}{arXiv:1404.1336}}}\ignorespaces
\bibitem{Izaguirre:2018atq}
E.~Izaguirre \& D.~Stolarski,
\textit{``{Searching for Higgs Decays to as Many as 8 Leptons}''},
\doiref{10.1103/PhysRevLett.121.221803}{Phys.~Rev.~Lett. \textbf{121}, 221803
  (2018)\ignorespaces}\ignorespaces,
\normalsize{\texttt{\arxivref{1805.12136}{arXiv:1805.12136}}}\ignorespaces
\bibitem{Sirunyan:2017lae}
CMS Collaboration, A.~M. Sirunyan et~al.,
\textit{``{Search for electroweak production of charginos and neutralinos in
  multilepton final states in proton-proton collisions at $\sqrt{s}=$ 13
  TeV}''},
\doiref{10.1007/JHEP03(2018)166}{JHEP \textbf{1803}, 166
  (2018)\ignorespaces}\ignorespaces,
\normalsize{\texttt{\arxivref{1709.05406}{arXiv:1709.05406}}}\ignorespaces
\bibitem{Aaboud:2018fvk}
ATLAS Collaboration, M.~Aaboud et~al.,
\textit{``{Search for Higgs boson decays to beyond-the-Standard-Model light
  bosons in four-lepton events with the ATLAS detector at $\sqrt{s}=13$
  TeV}''},
\doiref{10.1007/JHEP06(2018)166}{JHEP \textbf{1806}, 166
  (2018)\ignorespaces}\ignorespaces,
\normalsize{\texttt{\arxivref{1802.03388}{arXiv:1802.03388}}}\ignorespaces
\bibitem{Altmannshofer:2014pba}
W.~Altmannshofer, S.~Gori, M.~Pospelov \& I.~Yavin,
\textit{``{Neutrino Trident Production: A Powerful Probe of New Physics with
  Neutrino Beams}''},
\doiref{10.1103/PhysRevLett.113.091801}{Phys.~Rev.~Lett. \textbf{113}, 091801
  (2014)\ignorespaces}\ignorespaces,
\normalsize{\texttt{\arxivref{1406.2332}{arXiv:1406.2332}}}\ignorespaces
\bibitem{Araki:2017wyg}
T.~Araki, S.~Hoshino, T.~Ota, J.~Sato \& T.~Shimomura,
\textit{``{Detecting the $L_{\mu}-L_{\tau}$ gauge boson at Belle II}''},
\doiref{10.1103/PhysRevD.95.055006}{Phys.~Rev. \textbf{D95}, 055006
  (2017)\ignorespaces}\ignorespaces,
\normalsize{\texttt{\arxivref{1702.01497}{arXiv:1702.01497}}}\ignorespaces
\bibitem{TheBABAR:2016rlg}
BaBar Collaboration, J.~P. Lees et~al.,
\textit{``{Search for a muonic dark force at BABAR}''},
\doiref{10.1103/PhysRevD.94.011102}{Phys.~Rev. \textbf{D94}, 011102
  (2016)\ignorespaces}\ignorespaces,
\normalsize{\texttt{\arxivref{1606.03501}{arXiv:1606.03501}}}\ignorespaces
\bibitem{An:2012va}
H.~An, X.~Ji \& L.-T. Wang,
\textit{``{Light Dark Matter and $Z'$ Dark Force at Colliders}''},
\doiref{10.1007/JHEP07(2012)182}{JHEP \textbf{1207}, 182
  (2012)\ignorespaces}\ignorespaces,
\normalsize{\texttt{\arxivref{1202.2894}{arXiv:1202.2894}}}\ignorespaces
\bibitem{Dobrescu:2013coa}
B.~A. Dobrescu \& F.~Yu,
\textit{``{Coupling-Mass Mapping of Dijet Peak Searches}''},
\doiref{10.1103/PhysRevD.88.035021, 10.1103/PhysRevD.90.079901}{Phys.~Rev.
  \textbf{D88}, 035021 (2013)\ignorespaces}\ignorespaces,
\normalsize{\texttt{\arxivref{1306.2629}{arXiv:1306.2629}}}\ignorespaces,
[Erratum: Phys. Rev.D90,no.7,079901(2014)]\ignorespaces
\bibitem{FileviezPerez:2018jmr}
P.~Fileviez~Pérez, E.~Golias, R.-H. Li \& C.~Murgui,
\textit{``{Leptophobic Dark Matter and the Baryon Number Violation Scale}''},
\doiref{10.1103/PhysRevD.99.035009}{Phys.~Rev. \textbf{D99}, 035009
  (2019)\ignorespaces}\ignorespaces,
\normalsize{\texttt{\arxivref{1810.06646}{arXiv:1810.06646}}}\ignorespaces
\bibitem{Sirunyan:2017nvi}
CMS Collaboration, A.~M. Sirunyan et~al.,
\textit{``{Search for low mass vector resonances decaying into quark-antiquark
  pairs in proton-proton collisions at $ \sqrt{s}=13 $ TeV}''},
\doiref{10.1007/JHEP01(2018)097}{JHEP \textbf{1801}, 097
  (2018)\ignorespaces}\ignorespaces,
\normalsize{\texttt{\arxivref{1710.00159}{arXiv:1710.00159}}}\ignorespaces
\bibitem{ATLAS:2016xiv}
ATLAS Collaboration, T.~A. collaboration,
\textit{``{Search for light dijet resonances with the ATLAS detector using a
  Trigger-Level Analysis in LHC pp collisions at $\sqrt{s}=13$~TeV}''}
\bibitem{Sirunyan:2016iap}
CMS Collaboration, A.~M. Sirunyan et~al.,
\textit{``{Search for dijet resonances in proton–proton collisions at
  $\sqrt{s}$ = 13 TeV and constraints on dark matter and other models}''},
\doiref{10.1016/j.physletb.2017.09.029,
  10.1016/j.physletb.2017.02.012}{Phys.~Lett. \textbf{B769}, 520
  (2017)\ignorespaces}\ignorespaces,
\normalsize{\texttt{\arxivref{1611.03568}{arXiv:1611.03568}}}\ignorespaces,
[Erratum: Phys. Lett.B772,882(2017)]\ignorespaces
\bibitem{Aaboud:2017yvp}
ATLAS Collaboration, M.~Aaboud et~al.,
\textit{``{Search for new phenomena in dijet events using 37 fb$^{-1}$ of $pp$
  collision data collected at $\sqrt{s}=$13 TeV with the ATLAS detector}''},
\doiref{10.1103/PhysRevD.96.052004}{Phys.~Rev. \textbf{D96}, 052004
  (2017)\ignorespaces}\ignorespaces,
\normalsize{\texttt{\arxivref{1703.09127}{arXiv:1703.09127}}}\ignorespaces
\end{thebibliography}
\end{document}